\documentclass[]{emulateapj}
\begin{document}

\title{The c2d Spitzer spectroscopic survey of ices around low-mass young stellar objects II: CO$_2$ }

\author{Klaus M. Pontoppidan\altaffilmark{1}\altaffilmark{6}}
\altaffiltext{1}{California Institute of Technology, Division of Geological and Planetary Sciences, Mail Stop 150-21, Pasadena, CA 91125}
\email{pontoppi@gps.caltech.edu}

\author{A. C. A. Boogert\altaffilmark{2}}
\altaffiltext{2}{IPAC, NASA Herschel Science Center, Mail Code 100-22, California Institute of Technology, Pasadena, CA 91125, USA}

\author{Helen J. Fraser\altaffilmark{3}}
\altaffiltext{3}{University of Strathclyde, John Anderson Building, Glasgow G4 ONG, Scotland}

\author{Ewine F. van Dishoeck\altaffilmark{4}}

\author{Geoffrey A. Blake\altaffilmark{1}}

\author{Fred Lahuis\altaffilmark{4}}
\altaffiltext{4}{Leiden Observatory, P.O.Box 9513, NL-2300 RA Leiden, The Netherlands}

\author{Karin I. {\"O}berg\altaffilmark{4}}

\author{Neal J. Evans II\altaffilmark{5}}
\altaffiltext{5}{Department of Astronomy, University of Texas at Austin, 1 University Station, C1400, Austin, TX 78712-0259}

\altaffiltext{6}{Hubble Fellow}

\author{Colette Salyk\altaffilmark{1}}

\begin{abstract}
This paper presents Spitzer-IRS $\lambda/\Delta\lambda\sim 600$ spectroscopy of the CO$_2$ 15.2\,$\mu$m bending mode toward a
sample of 50 embedded low-mass stars in nearby star-forming clouds, taken mostly from the ``Cores to Disks (c2d)'' Legacy program. The average abundance 
of solid CO$_2$ relative to water in low-mass protostellar envelopes is $0.32\pm 0.02$, significantly higher than that found 
in quiescent molecular clouds and in massive star forming regions. It is found that a decomposition of all the observed CO$_2$ 
bending mode profiles requires a minimum of five unique components.
In general, roughly 2/3 of the CO$_2$ ice is found in a water-rich environment, while most of the remaining 1/3 is found in a CO environment
with strongly varying relative concentrations of CO$_2$ to CO along each line of sight. 
Ground-based observations of solid CO toward a large subset of the c2d sample are used to further constrain the CO$_2$:CO component and
suggest a model in which low-density clouds form the CO$_2$:H$_2$O component and higher density clouds form the CO$_2$:CO ice during and
after the freeze-out of gas-phase CO. The abundance of the CO$_2$:CO component 
is consistent with cosmic ray processing of the CO-rich part of the ice mantles, although a more quiescent formation mechanism is not ruled out.
It is suggested that the subsequent evolution of the CO$_2$ and CO profiles toward low-mass protostars, in particular the appearance of the
splitting of the CO$_2$ bending mode due to pure, crystalline CO$_2$, is first caused by distillation of the CO$_2$:CO component through evaporation of CO
due to thermal processing to $\sim 20-30\,$K in the inner regions of infalling envelopes. The formation of pure CO$_2$ via segregation 
from the H$_2$O rich mantle
may contribute to the band splitting at higher levels of thermal processing ($\gtrsim 50\,$K), but is harder to reconcile with the 
physical structure of protostellar envelopes around low-luminosity objects.

\end{abstract}

\keywords{astrochemistry --- circumstellar matter --- dust, extinction --- ISM: evolution}

\section{Introduction}
Although CO$_2$ is not an abundant gas-phase molecule in molecular clouds, 
it is one of a small number of molecular species consistently found in very high abundances inside ice mantles on dust grains 
\citep[$>10^{-5}$ with respect to H$_2$][]{Gerakines99,Whittet07}. 
The generally high abundance of solid CO$_2$ became apparent with the spectroscopic surveys conducted
with the Infrared Space Observatory (ISO) \citep{Gerakines99, Nummelin01}. Other species known to belong to this class of very abundant
molecules are CO and H$_2$O. In less than 10\% of dark cloud
regions surveyed, methanol 
(CH$_3$OH) is also found with similar abundances \citep{Pontoppidan03b,Boogert07}. Depending on the density and temperature of 
a cloud, the CO is found partly in the gas-phase and partly frozen onto grain surfaces, while the CO$_2$ and H$_2$O are
completely frozen as ice mantles \citep{Bergin95}, except in very hot or shocked regions \citep{Boonman03,Nomura04,Lahuis07}. 
The system of CO, CO$_2$, H$_2$O and, under some conditions, CH$_3$OH therefore 
represents the bulk of solid state volatiles in dense star forming clouds, and interactions between these four species 
can be expected to account for most of the solid state observables. Other species with abundances of less than 5\% relative to water,
such as CH$_4$, NH$_3$, OCN$^-$, HCOOH and OCS will be good tracers of chemistry and their local molecular environment, but
are unlikely to strongly affect the molecular environments, and therefore the band profiles, of the four major species.

The formation mechanism of solid CO$_2$ in the cold interstellar medium is still not understood, although a number of plausible scenarios
have been proposed. Since the direct surface route, $\rm CO+O\rightarrow CO_2$, is thought to possess a significant activation barrier, it was initially 
suggested that strong UV irradiation was needed to produce the observed CO$_2$ ice \citep{dhendecourt85}. 
Laboratory simulations of interstellar
ice mixtures of H$_2$O and CO confirmed that CO$_2$ is indeed readily formed during strong UV photolysis \citep{dhendecourt86}, and
initial detection of abundant CO$_2$ ice around UV-luminous massive young stars seemed to confirm this picture. 
However, recent 
detections of similar abundances of CO$_2$ in dark clouds observed along lines of sight toward background stars, far
away from any ionizing source \citep{Bergin05,Knez05,Whittet07}, argue against a UV irradiation route to CO$_2$, at least through enhanced UV from nearby protostars. 
\cite{Pontoppidan06} showed evidence for an increasing abundance of CO$_2$ ice with gas density in at 
least one low-mass star forming core. Furthermore, the
original premise of a barrier to oxygenation of CO is now in doubt \citep{Roser01}. 
Consequently, both theoretical and laboratory efforts to understand the 
formation of CO$_2$ are still very active.

Extensive surveys of the 3.1 and 4.67\,$\mu$m stretching mode of H$_2$O and CO ices
have been carried out in a range of different star-forming environments \citep{Whittet88, Chiar95, Pontoppidan03}. However, 
CO$_2$ can only be observed from space, and surveys have until recently been limited to a small sample of very luminous young stars 
\citep{Gerakines99}. While the 4.27\,$\mu$m stretching mode of CO$_2$ was detected toward a few background field stars
by ISO \citep{Whittet98,Nummelin01}, recent Spitzer observations of the 15.2\,$\mu$m bending mode of CO$_2$ have extended the sample
of CO$_2$ along quiescent lines of sight considerably \citep{Knez05, Bergin05, Whittet07}. 

The profiles of the CO$_2$ ice bands observed in quiescent regions and along lines of sight toward luminous protostars show
intriguing differences. In particular, ice in massive star-forming regions, which presumably traces more processed material, tends to 
show a splitting of the 15.2\,$\mu$m bending mode. This splitting has been identified as a general property of 
crystalline pure CO$_2$ and is readily reproduced in laboratory simulations of interstellar ices \citep[e.g.][]{Ehrenfreund97, Broekhuizen06}.
It seems implausible that this pure CO$_2$ layer would form directly through gas-phase deposition and subsequent surface reactions, and
it has been suggested that the CO$_2$ segregates from a CO$_2$:H$_2$O:CH$_3$OH=1:1:1 mixture upon strong heating \citep{Gerakines99}. Annealing in a laboratory
setting produces the same effect  \citep{Ehrenfreund97}, and strong heating is not unreasonable in the envelopes of massive young stars with luminosities in
excess of $10^3\,\rm L_{\odot}$.

In this paper, a survey of the 15.2\,$\mu$m CO$_2$ bending mode toward $\sim$50 young {\it low mass}
stars using the high resolution mode of the Infrared Spectrograph on board the Spitzer Space Telescope (Spitzer-IRS) is presented.
The sample stars have typical luminosities in the range $0.1-10\,\rm L_{\odot}$, thus bridging the observational
gap between the background stars and the massive protostars from the ISO sample.
The Spitzer spectra are complemented by ground-based observations of H$_2$O and CO ices, where available, as well as the archival
spectra from the Infrared Space Observatory used in \cite{Gerakines99}. 
While this paper concentrates on the region around 15\,$\mu$m, 
\cite{Boogert07} discusses the ices causing the 5-8\,$\mu$m absorption complex (henceforth referred to as Paper I). 
The 7.7\,$\rm\mu$m CH$_4$ and 9.0\,$\mu$m NH$_3$ bands are described in two separate papers ({\"O}berg et al. 2008, submitted 
and Bottinelli et al., in prep, respectively). 

The central questions that are addressed using the new Spitzer spectra of the CO$_2$ bending mode are:

\begin{itemize}
\item What are the differences, if any, between CO$_2$ ice in massive and low-mass star-forming environments?
\item What is the average abundance of CO$_2$ in low-mass protostellar envelopes compared to lower density quiescent clouds?
\item In which molecular environments can solid CO$_2$ be found, and what are their relative abundances?
\item Which process forms CO$_2$ in CO-dominated environments when the CO accretes from the gas-phase at high densities?
\item How does the component of pure CO$_2$ (as measured by the well-known splitting of the bending mode) form in low-mass protostellar envelopes?
\item What is the evolution of the CO$_2$ ice and what does it tell us about protostellar evolution?
\end{itemize}

\section{The infrared bands of solid CO$_2$}

The infrared vibrational modes of CO$_2$ are known to be very sensitive to the molecular environment. Observations of the band profiles
can determine whether the CO$_2$ molecules are embedded with water, CO and other CO$_2$ molecules. 
Solid CO$_2$ has two strong vibrational modes; the asymmetric stretching mode centered on 4.27\,$\mu$m, and the bending mode at 15.2\,$\mu$m.
The stretching mode is so strong that it is typically saturated along lines of sight through protostellar envelopes, and its $^{13}$CO$_2$
counterpart at 4.38\,$\mu$m is often used for profile analysis instead. 
However, neither stretching modes are covered by the spectral range of Spitzer-IRS.

Fortunately, the CO$_2$ bending mode is an excellent diagnostic of molecular environments. 
For instance, pure CO$_2$ will
typically produce a split band in the bending mode due to Davydov splitting -- a long range interaction in crystalline materials. Conversely, 
CO$_2$ embedded in a hydrogen-bonding matrix will produce a broad, smooth profile. While these differences are relatively well known, 
it is noted that the different formation scenarios from gas-grain chemical models make distinct predictions for the molecular 
environment of the CO$_2$. Thus models favoring a formation route via OH predict that the CO$_2$ will be found in a water-dominated
matrix. 

Indeed, observations of the CO$_2$ stretching and bending modes toward young massive stars with ISO
\citep{Gerakines99} have shown that the CO$_2$ ice is dominated by a band consistent with CO$_2$ in a hydrogen-bonding environment.
They also found that a double-peaked component, consistent with a relatively pure, crystalline CO$_2$ is generally present at a lower level toward their
sample of massive stars.

\section{Observations}

The spectra of the 15.2\,$\mu$m CO$_2$ bending mode have been obtained using the short-high (SH) module of Spitzer-IRS, with a 
spectral resolving power of $\lambda/\Delta\lambda \sim 600$, covering 10-19.5\,$\mu$m, corresponding to 1\,$\rm cm^{-1}$. 
The Spitzer spectra were obtained as part of the ``Cores to Disks'' Legacy program (PID 172,179) as well as a few archival spectra observed as
part of the GTO programs (PID 2). All SH spectra from the c2d database that show clear detections of the CO$_2$ bending mode have been included.
The spectra have been reduced using the c2d pipeline from basic calibrated data (BCD) products version S13.0.2. For each spectrum, clearly deviant points 
were removed and individual orders were 
scaled by small factors to align the overlapping regions between orders.
The overlapping regions between these two orders usually match very well, which lends support to the reality
of small features in the spectra for most sources. A small number of sources show evidence for
absorption by the Q-branch of gas-phase CO$_2$ at 15.0 micron. Since this survey is concerned with solid CO$_2$, pixels affected by gas-phase 
absorption have been removed from the fits of IRS 46, WL 12 and DG Tau B. 
Some of the most embedded sources have saturated or nearly saturated bending mode bands. For these sources, special
care has to be taken to ensure that the background level is well-determined. 

In addition to the Spitzer data, the five highest-quality spectra of massive YSOs observed with the ISO-SWS \citep{Gerakines99} have been included. 
The ISO spectra provide a useful comparison of the structure of CO$_2$ ices in the warmer, more energetic envelopes of young massive stars to
the comparatively quiescent envelopes of low-mass YSOs.

Finally, to study the relation of the CO$_2$ ices with CO and water, ground-based spectra of the 4.67\,$\mu$m stretching mode of solid CO and the 3.08\,$\mu$m stretching mode
of water ice have been collected using the Infrared Spectrometer and Array Camera (ISAAC) on the Very Large Telescope 
(VLT)\footnote{Based on observations made with ESO Telescopes at the Paranal Observatory under programme 
ID 164.C-0605}. 
Most of the 4.67\,$\mu$m VLT-ISAAC spectra are published in \cite{Pontoppidan03}, but a significant fraction are previously unpublished spectra obtained with NIRSPEC
at the Keck Telescope. Sources with no CO ice data are typically too faint for useful ground-based 4.67\,$\mu$m spectroscopy. The column
densities of water ice are derived using the 3.08\,$\mu$m ground-based spectra or taken from Paper I. 
The data set is summarized in Table \ref{sources}. The new observations of the CO ice bands not covered in \cite{Pontoppidan03} are summarized
in Table \ref{newCO}.  

\begin{table*}[h]
\caption{Observed sample of embedded young stars}
\begin{tabular}{lllllll}
\hline
\hline
Source & CO$_2$ & CO and H$_2$O & RA [J2000] & DEC [J2000] & Observation ID \\
\hline
W3 IRS5           &ISO-SWS  &NIRSPEC &02 25 40.8  &+62 05 52.8  & 42701302\\
L1448 IRS1        &Spitzer  &NIRSPEC &03 25 09.4  &+30 46 21.7  & 5656832 \\
L1448 NA          &Spitzer  &--      &03 25 36.5  &+30 45 21.4  & 5828096 \\
L1455 SMM1        &Spitzer  &--      &03 27 43.2  &+30 12 28.8  & 15917056\\
RNO 15            &Spitzer  &NIRSPEC &03 27 47.7  &+30 12 04.3  & 5633280 \\
IRAS 03254+3050   &Spitzer  &NIRSPEC &03 28 34.5  &+31 00 51.2  & 13460480\\
IRAS 03271+3013   &Spitzer  &NIRSPEC &03 30 15.2  &+30 23 48.8  & 5634304 \\
B1 a              &Spitzer  &NIRSPEC &03 33 16.7  &+31 07 55.1  & 15918080\\
B1 c              &Spitzer  &--      &03 33 17.9  &+31 09 31.0  & 15916544\\
IRAS 03439+3233   &Spitzer  &NIRSPEC &03 47 05.4  &+32 43 08.5  & 5635072 \\ %L1455 IRS3
IRAS 03445+3242   &Spitzer  &NIRSPEC &03 47 41.6  &+32 51 43.8  & 5635328 \\
L1489 IRS         &Spitzer  &ISAAC   &04 04 42.6  &+26 18 56.8  & 3528960 \\
DG Tau B          &Spitzer  &NIRSPEC &04 27 02.7  &+26 05 30.5  & 3540992 \\
GL 989	          &ISO-SWS  &NIRSPEC &06 41 10.2  &+09 29 33.7  & 72602619\\
HH46 IR           &Spitzer  &ISAAC   &08 25 43.8  &--51 00 35.6 & 7130112 \\
CED 110 IRS4      &Spitzer  &--      &11 06 46.6  &--77 22 32.4 & 5639680 \\
B 35              &Spitzer  &--      &11 07 21.5  &--77 22 11.8 & 5639680 \\
CED 110 IRS6      &Spitzer  &ISAAC   &11 07 09.2  &--77 23 04.3 & 5639680 \\
IRAS 12553-7651   &Spitzer  &--      &12 59 06.6  &--77 07 40.0 & 9830912 \\
ISO ChaII 54      &Spitzer  &--      &13 00 59.2  &--77 14 02.7 & 15735040\\
IRAS 13546-3941   &Spitzer  &--      &13 57 38.9  &--39 56 00.2 & 5642752 \\
IRAS 15398-3359   &Spitzer  &--      &15 43 02.3  &--34 09 06.7 & 5828864 \\
GSS 30 IRS1       &Spitzer  &ISAAC   &16 26 21.4  &--24 23 04.1 & 5647616 \\
WL 12             &Spitzer  &ISAAC   &16 26 44.2  &--24 34 48.4 & 5647616 \\
GY 224            &Spitzer  &NIRSPEC &16 27 11.2  &--24 40 46.7 & 9829888 \\
WL 20             &Spitzer  &--      &16 27 15.7  &--24 38 45.6 & 9829888 \\
IRS 37            &Spitzer  &ISAAC   &16 27 17.6  &--24 28 56.5 & 5647616 \\
IRS 42            &Spitzer  &ISAAC   &16 27 21.5  &--24 41 43.1 & 5647616 \\
WL 6              &Spitzer  &ISAAC   &16 27 21.8  &--24 29 53.3 & 5647616 \\
CRBR 2422.8-3423  &Spitzer  &ISAAC   &16 27 24.6  &--24 41 03.3 & 9346048 \\
IRS 43            &Spitzer  &ISAAC   &16 27 27.0  &--24 40 52.0 & 12699648\\
IRS 44            &Spitzer  &ISAAC   &16 27 28.1  &--24 39 35.0 & 12699648\\
Elias 32/IRS 45   &Spitzer  &ISAAC   &16 27 28.4  &--24 27 21.4 & 12664320\\
IRS 46            &Spitzer  &ISAAC   &16 27 29.4  &--24 39 16.3 & 9829888 \\
VSSG 17/ IRS 47   &Spitzer  &ISAAC   &16 27 30.2  &--24 27 43.4 & 5647616 \\
IRS 51            &Spitzer  &ISAAC   &16 27 39.8  &--24 43 15.1 & 9829888 \\
IRS 63            &Spitzer  &ISAAC   &16 31 35.7  &--24 01 29.5 & 9827840 \\
L1689 IRS5        &Spitzer  &--      &16 31 52.1  &--24 56 15.2 & 12664064\\
RNO 91            &Spitzer  &ISAAC   &16 34 29.3  &--15 47 01.4 & 5650432 \\
W33 A             &ISO-SWS  &ISO-SWS &18 14 39.7  &--17 52 02.0 & 32900920\\
GL 2136           &ISO-SWS  &ISO-SWS &18 22 27.0  &--13 30 10.0 & 33000222\\
Serp S68          &Spitzer  &--      &18 29 48.1  &+01 16 42.5  & 9828608 \\
EC 74             &Spitzer  &NIRSPEC &18 29 55.7  &+01 14 31.6  & 9407232 \\
SVS 4-5           &Spitzer  &ISAAC   &18 29 57.6  &+01 13 00.6  & 9407232 \\
EC 82             &Spitzer  &ISAAC   &18 29 56.9  &+01 14 46.5  & 9407232 \\
EC 90             &Spitzer  &ISAAC   &18 29 57.8  &+01 14 05.9  & 9828352 \\
SVS 4-10          &Spitzer  &ISAAC   &18 29 57.9  &+01 12 51.6  & 9407232 \\
CK 4              &Spitzer  &NIRSPEC &18 29 58.2  &+01 15 21.7  & 9407232 \\
CK 2              &Spitzer  &ISAAC   &18 30 00.6  &+01 15 20.1  & 11828224\\
RCrA IRS 5        &Spitzer  &ISAAC   &19 01 48.0  &--36 57 21.6 & 9835264 \\
RCrA IRS 7A       &Spitzer  &ISAAC   &19 01 55.3  &--36 57 22.0 & 9835008 \\
RCrA IRS 7B       &Spitzer  &ISAAC   &19 01 56.4  &--36 57 28.0 & 9835008 \\
CrA IRAS 32       &Spitzer  &--      &19 02 58.7  &--37 07 34.5 & 9832192 \\ 
S140 IRS1         &ISO-SWS  &NIRSPEC &22 19 18.4  &+63 18 45.0  & 22002135\\
NGC 7538 IRS 9    &ISO-SWS  &NIRSPEC &23 14 01.7  &+61 27 20.0  & 09801532\\
IRAS 23238+7401   &Spitzer  &--      &23 25 46.7  &+74 17 37.2  & 9833728 \\

\hline
\end{tabular}
\label{sources}
\end{table*}

\begin{table*}[ht]
\centering
\caption{New CO ice band observations}
\begin{tabular}{llll}
\hline
\hline
Source & $\tau$(CO:H$_2$O) (red)$^a\,^b$ & $\tau$(Pure CO) (middle)$^c$ & $\tau$(CO:CO$_2$) (blue)$^d$ \\
\hline
W3 IRS5            &   0.11$\pm$ 0.05     &   0.27$\pm$ 0.06          & 0.06$\pm$  0.05 \\
L1448 IRS1         &   0.10$\pm$ 0.05     &   0.18$\pm$ 0.06          & 0.09$\pm$  0.03 \\    
RNO 15             &   0.12$\pm$ 0.03     &   0.27$\pm$ 0.03          & 0.09$\pm$  0.03 \\   
IRAS 03254+3050    &   0.12$\pm$ 0.06     &   0.19$\pm$ 0.10          & 0.11$\pm$  0.05 \\ 
IRAS 03271+3013    &   $<$0.5             &   1.0$\pm$  0.5            & $<$0.5     \\ 
B1 a               &   $<$0.5             &   4.0$\pm$ 2.0             & $<$0.3     \\
IRAS 03439+3233    &   0.09$\pm$ 0.05     &   0.25$\pm$ 0.10           & 0.08$\pm$ 0.04 \\ 
IRAS 03445+3242    &   0.30$\pm$ 0.05     &   1.16$\pm$ 0.20           & 0.33$\pm$ 0.05  \\ 
DG Tau B           &   0.13$\pm$ 0.05     &   0.12$\pm$ 0.10           & 0.11$\pm$ 0.05  \\
GL 989	           &   0.19$\pm$ 0.03     &   0.37$\pm$ 0.05           & 0.16$\pm$ 0.05  \\
HH46 IR            &   0.80$\pm$ 0.08     &   0.68$\pm$ 0.06           & 0.16$\pm$ 0.05  \\
GY 224             &   $<$0.2             &   $<$0.2                   & $<$0.2     \\
IRS 37             &   0.08$\pm$ 0.02     &   0.45$\pm$ 0.04           & 0.10$\pm$ 0.05  \\
GL 2136            &   0.20$\pm$ 0.05     &   0.10$\pm$ 0.05           & $<$0.05    \\
EC 74              &   $<$0.10            &   0.75$\pm$ 0.06           & 0.10$\pm$ 0.06  \\
CK 4               &   $<$0.03            &   0.48$\pm$ 0.05           & 0.07$\pm$ 0.03  \\
S140 IRS1          &   0.05$\pm$ 0.02     &   0.04$\pm$ 0.02           & $<$0.03    \\
NGC 7538 IRS 9     &   0.52$\pm$ 0.05     &   3.17$\pm$ 0.10           & 0.30$\pm$ 0.03  \\

\hline
\end{tabular}
\begin{itemize}
\item[$^a$] Decomposition as described in \cite{Pontoppidan03}.
\item[$^b$] Adopted conversion to column density: $N_{\rm CO:H_2O} = 16.0\,{\rm cm^{-1}} \tau_{\rm CO:H_2O} A^{-1}$, where $A=1.1\,10^{-17}\,\rm cm\,molecule^{-1}$ is the CO band strength.
\item[$^c$] Adopted conversion to column density, using a continuous distribution of ellipsoids (CDE): $N_{\rm CO} = 6.0\,{\rm cm^{-1}} \tau_{\rm CO} A^{-1}$.
\item[$^d$] $N_{\rm CO:CO_2} = 3.0\,{\rm cm^{-1}} \tau_{\rm CO:CO_2} A^{-1}$.
\end{itemize}
\label{newCO}
\end{table*}

\section{Profile decomposition}

\subsection{Continuum determination}

To directly compare with dust models, the spectra of ice absorption bands have to be converted to an optical depth scale. This
requires that an appropriate continuum be defined, a process somewhat complicated for CO$_2$ by the location of the bending mode
on the blue side of the broad silicate bending mode and on the red side of the H$_2$O libration band. Unfortunately, not knowing the 
shape of the underlying continuum, this is a problem with no unique solution. In this work, continua for each spectrum are constructed by fitting
a third-order polynomial to the spectral ranges: 13\,$\mu$m -- 14.7\,$\mu$m and 18.2\,$\mu$m -- 19.5\,$\mu$m. The shape of the blue wing
of the silicate bending mode is simulated by a Gaussian in frequency space with center at 608\,cm$^{-1}$ and a FWHM of 73\,cm$^{-1}$. The
aim is to construct a shape of the continuum that has a negative second derivative under the CO$_2$ band. The same procedure has been used
for all the Spitzer spectra, and the resulting continua are shown for each spectrum in Figures \ref{decomp-1} through \ref{decomp-13}.

\subsection{Laboratory data}

A number of laboratory spectra of CO$_2$ ices have been taken from the literature. Ice inventories of envelopes around young low-mass have shown
that the ices are dominated by H$_2$O, CO and CO$_2$, so this study concentrates on systems involving these three species. In some regions of low-mass star formation, 
CH$_3$OH is found in large amounts (up to 25\% relative to H$_2$O), but this seems to affect only a small subset of the sample presented here.

The available laboratory spectra are divided into CO$_2$:H$_2$O mixtures and CO$_2$:CO mixtures, each set with distinct characteristics. As in the 
case of all solid state features due to abundant molecules, the band shapes can be strongly modified by surface modes, depending on the shape distribution of the 
dust grains \citep{Tielens91}. Astronomical spectra can therefore not be directly compared to absorbance laboratory spectra. Rather, complex refractive 
indices must be combined with a dust model
to calculate opacities relevant for the small irregular dust grains of the interstellar medium. This study is consequently restricted to laboratory
experiments for which optical constants have been calculated. 

\cite{Ehrenfreund97,Dartois99} present optical constants for a wide range of CO$_2$:CO 
mixtures, as well as a few CO$_2$:H$_2$O mixtures, obtained under high vacuum (HV) conditions and 2.0\,cm$^{-1}$ resolution. 
More recently, a number of detailed studies of relevant CO$_2$-rich
ices were performed by \cite{Broekhuizen06} and \cite{Oberg07}, also under HV conditions. While these studies do not
provide optical constants directly, they do report absorbance spectra as 
well as approximate ice thicknesses, making it possible to derive optical constants using the Kramers-Kronig relations. 

For the H$_2$O-rich ices, the CO$_2$:H$_2$O=14:100 mixture at 10\,K from \cite{Ehrenfreund97} was chosen. The CO-rich ices show a CO$_2$ bending model
profile that is dependent on the mixing ratio. A function is therefore constructed that returns an ice spectrum for any relative concentration between CO$_2$:CO=1:4 and 
CO$_2$:CO=1:1 by interpolating between the available 10\,K laboratory spectra within this range. At very low concentrations of CO$_2$ relative to CO 
(CO$_2$:CO$<$1:10), the bending mode becomes
quite narrow, and the shape becomes independent of concentration. At very high concentrations of CO$_2$, the bending mode exhibits the well-known
split, characteristic of a crystalline structure of the ice. The peaks are very narrow - of order 1\,cm$^{-1}$, and so the higher resolution data from 
\cite{Broekhuizen06} are used for pure CO$_2$ ice. This spectrum, however, suffers
from a misalignment in the spectrometer optics at 0.5\,cm$^{-1}$ resolution
so that the bending mode is too weak by a factor of 3 relative to the
stretching mode and the noise is relatively high. As a result, it was necessary to scale the absorbance of the bending mode to fit with the band strength of 
$1.1\times 10^{-17}\,\rm cm^{-1}$ reported
by \cite{Gerakines95} before calculating the corresponding set of optical constants. 
The noise in the spectrum is reduced by fitting a number of Gaussians, rather than smoothing it, which would reduce the resolution. 
Because all the laboratory spectra used were obtained under high-vacuum conditions, they may be contaminated by H$_2$O.

\begin{figure}
  \includegraphics[width=8cm]{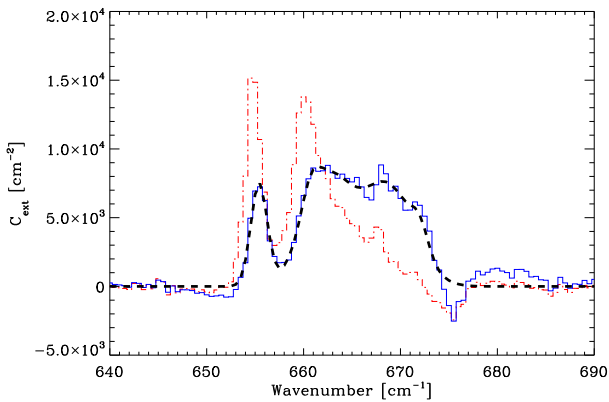}
  \caption{Comparison of the absorbance spectrum of \cite{Broekhuizen06} relevant for pure CO$_2$ ice (red, dash-dotted curve) with that calculated for a 
CDE distribution of particles in the Rayleigh limit (blue, solid curve). The dashed curve shows the component fit used to eliminate the noise. }
  \label{CDE}
\end{figure}

At 15\,$\mu$m the dust grains are most likely well into the Rayleigh limit of $2\pi a\ll 15\,\mu$m, where $a$ is the radius of the largest grains. This
means that scattering  of light out of the line of sight is unimportant, and the opacities can be treated as pure absorption coefficients.  
A continuous distribution of ellipsoids (CDE) is used to convert the optical constants to opacities. This is a convenient method of simulating
the effect of irregularly shaped grains, found to work well for ice bands in the Rayleigh limit 
and it has been used successfully for solid CO \citep{Pontoppidan03} and CO$_2$ \citep{Gerakines99}. Figure \ref{CDE} illustrates the process of converting the absorbance spectrum
of pure CO$_2$ into an opacity that can be used for comparing with the Spitzer spectra. 
The laboratory spectra are summarized in Table \ref{labspecs}.

\begin{table}
\caption{Laboratory spectra}
\begin{tabular}{llll}
\hline
\hline
Mixture & T [K] &  Resolution [cm$^{-1}$] & Reference \\
\hline
CO$_2$:H$_2$O=14:100 &10 &2 &1\\
CO$_2$:CO=4:100      &10 &2 &1\\
CO$_2$:CO=26:100     &10 &2 &1\\
CO$_2$:CO=70:100     &10 &2 &1\\
CO$_2$:CO=112:100    &10 &2 &1\\
Pure CO$_2$          &15 &0.5&2\\
\hline
\end{tabular}
\begin{itemize}
\item[1] \cite{Ehrenfreund97}
\item[2] \cite{Broekhuizen06}
\end{itemize}
\label{labspecs}
\end{table}

\subsection{Component analysis}
\label{CompAnalysis}
The strategy adopted here for analyzing the general shape of the CO$_2$ ice bending mode in low-mass young stellar envelopes is to 
determine the minimum number of unique components required to fit all the observed bands. In this context, a unique
component is a band that only changes its relative depth, but not its shape from source to source. This approach was used in \cite{Pontoppidan03}
to determine that only three unique components could be used to fit the 4.67\,$\mu$m stretching mode of solid CO, and is also
used in Paper I to decompose the 5-8\,$\mu$m complex. The three unique CO components were 1) A broad, red-shifted component
associated with CO in a water-rich mantle, 2) a component indistinguishable from pure CO and 3) a narrow, blue-shifted component due
to either CO in a CO$_2$ environment, or CO in a crystalline form. The three CO components were named ``red'', ``middle''
and ``blue'', and is seen below, the ``red'' and ``blue'' components have counterparts in the CO$_2$ bending modes. 
Note that the ``red'' and ``middle'' components are also often referred to in the literature as ``polar'' and ``nonpolar'', respectively. 
The bending mode of CO$_2$ probes ice structures that are
somewhat more complicated than those probed by CO. While most of the CO ice desorbs efficiently at temperatures higher than 20\,K, the less volatile CO$_2$ ice
goes through several additional structural changes. The most characteristic is the appearance of the double peak seen in pure CO$_2$ (see Figure \ref{CDE}). 

\begin{figure}
    
  \includegraphics[width=8cm]{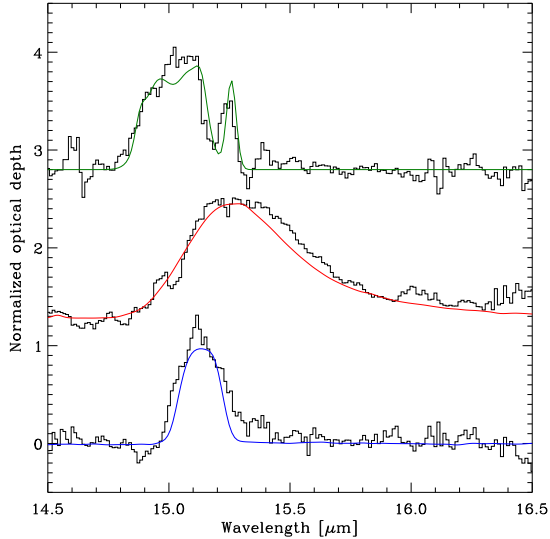}
  \caption{Empirically determined profiles for the three dominant components of the CO$_2$ bending mode. The data are
as follows. Top: RNO 91 - IRS 42, fit using pure CO$_2$. Middle: CRBR 2422.8-3423 - IRS 51, fit using CO$_2$:H$_2$O=14:100. Bottom:
IRS 63 - IRS 42, fit using CO$_2$:CO = 26:100.}
  \label{subtract_extreme}
\end{figure}

In order to empirically derive the shape of the components of the CO$_2$ bending mode, pairs of 
spectra are subtracted. If the spectra are superpositions of a small number of components with varying relative contributions, 
it will be possible to isolate each component. The three dominant components determined this way are shown in Figure \ref{subtract_extreme}
where they are compared to laboratory simulations.

Consequently, it is found that the minimum number of unique components required to fit all the observed CO$_2$ bending mode profiles is five:

\begin{figure*}
    
  \includegraphics[angle=90,width=16cm]{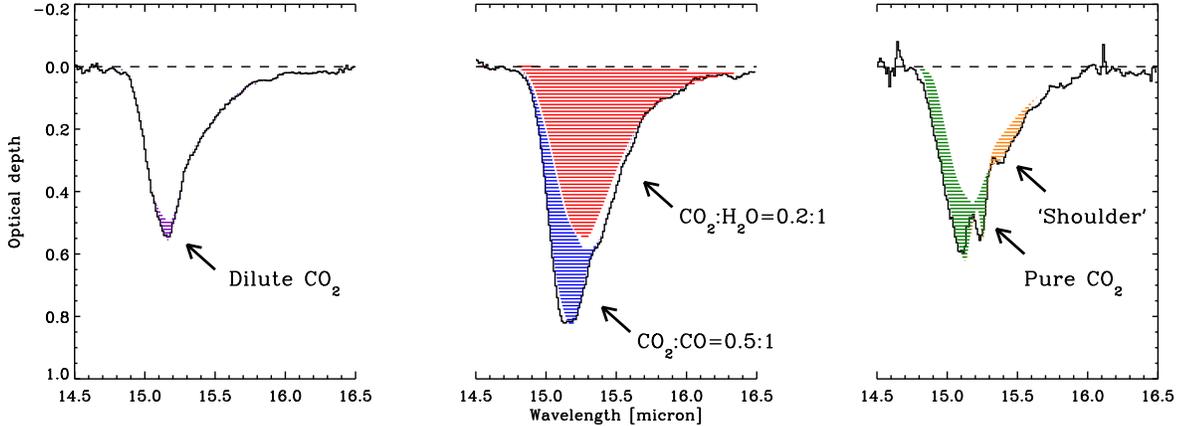}
  \caption{Sketch of the 5 different components used to fit the CO$_2$ band. The spectra used to illustrate the components are from left to right:
 IRS 51, SVS 4-5 and RNO 91.}
  \label{co2_components_sketch}
\end{figure*}

\begin{itemize}
\item An H$_2$O rich component, modeled with a laboratory spectrum with a concentration of CO$_2$:H$_2$O=14:100. This is required to fit the red wing of almost
all the observed bands. This is referred to as the ``red'' component.
\item A component with a roughly equal mixture of CO$_2$ and CO. Strictly speaking, this is not constructed as a unique component since 
the CO$_2$:CO mixing ratio is included as a free parameter. This is required because
the empirical profiles of the blue component, one of which is shown in Figure \ref{subtract_extreme}, have varying widths. This
behavior can be reproduced by varying the concentration of CO$_2$ in CO. While the band profiles are only available for a set of discrete 
mixing ratios, profiles are constructed with arbitrary mixing ratios by linearly interpolating the available laboratory profiles at each frequency point. 
In effect, this allows a measurement of the CO$_2$:CO mixing ratio for each observed CO$_2$ bending mode. This is referred to as the ``blue'' component.
\item A component in which CO$_2$ is very dilute in an otherwise pure CO ice. This is a very narrow component centered on 15.15\,$\mu$m (660\,cm$^{-1}$).  
In practice, this band is modeled by a CO$_2$:CO=4:100 laboratory spectrum. Note that at such high dilution, the shape of the band is not
sensitive to the exact mixing ratio. 
\item A component of pure CO$_2$, producing the characteristic double-peaked structure often seen in protostellar sources.
\item An additional, relatively narrow component on the red side of the main band. This component is unambiguously identified in only a few sources, most
of them the massive young stars included from the ISO sample. The component has in the past been identified as an interaction with CH$_3$OH 
in strongly annealed ices with the mixing ratio H$_2$O:CO$_2$:CH$_3$OH=1:1:1 \citep{Gerakines99}. Since there is 
no laboratory spectrum of the shoulder in isolation, the component is modeled empirically using a superposition of two Gaussians:

\begin{eqnarray*}
\lefteqn{\tau_{\rm shoulder}(\nu)/\tau_0 =} \\
 & & { } \exp\left[\frac{-(\nu-645\,{\rm cm^{-1}})^2}{2 (2.1\,{\rm cm^{-1}})^2}\right] +      \\ 
 & & { } 1.85\exp\left[\frac{-(\nu-650\,{\rm cm^{-1}})^2}{2 (2.8\,{\rm cm^{-1}})^2}\right],
\end{eqnarray*}
where $\tau_0$ is a scale factor. 

\end{itemize} 

The ``red'' and ``blue'' components generally dominate the bending mode profiles, and the total CO$_2$ abundance. The remaining three
components represent subtle differences due to trace constituents. It is stressed that all the components correspond to distinct and plausible molecular environments. 
The relative contributions to three typical CO$_2$ bending mode profiles are sketched in Figure \ref{co2_components_sketch}.

\begin{figure*}[h]
    
  \includegraphics[width=15cm]{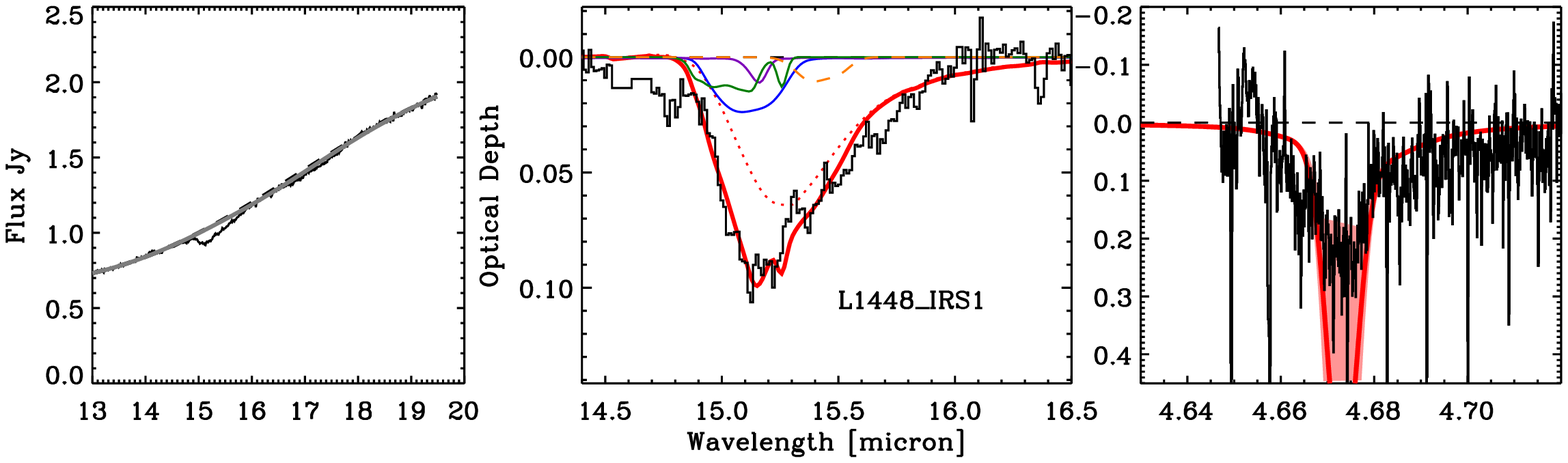}
  \includegraphics[width=15cm]{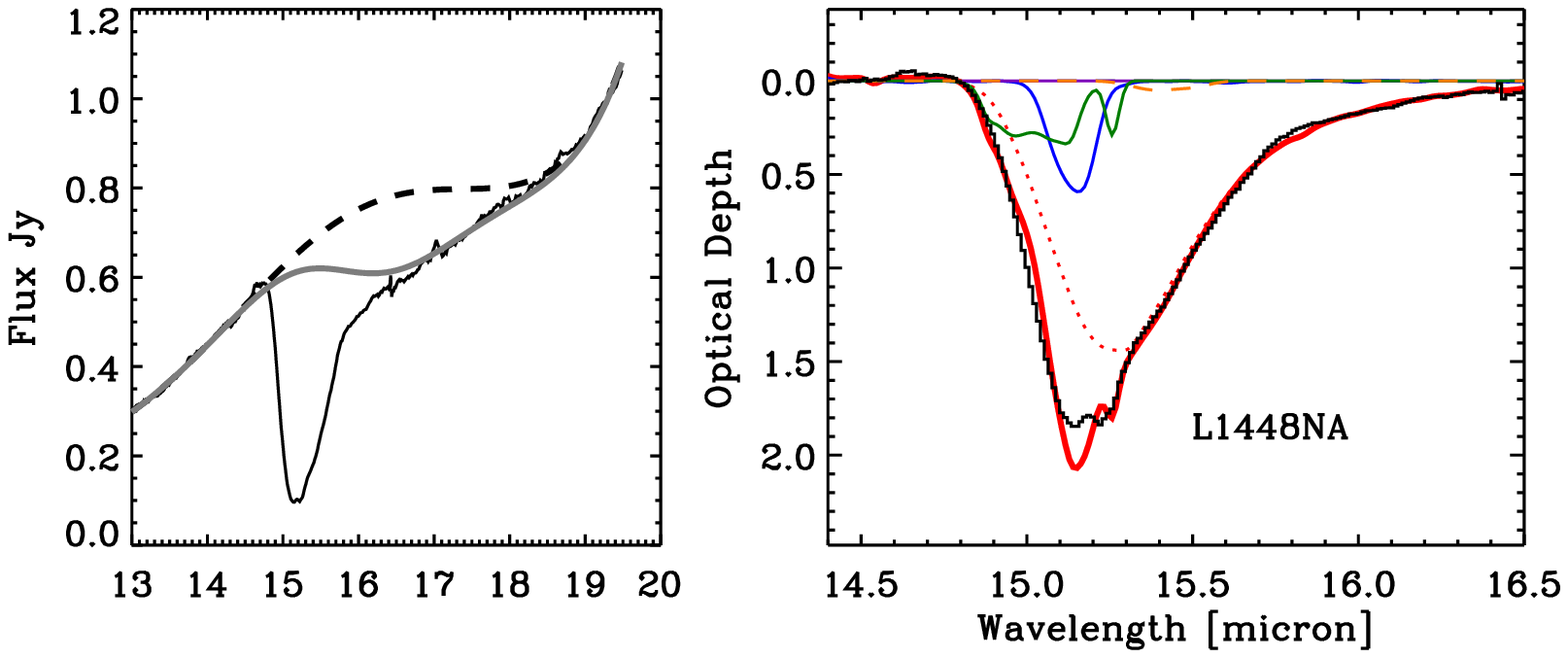}
  \includegraphics[width=15cm]{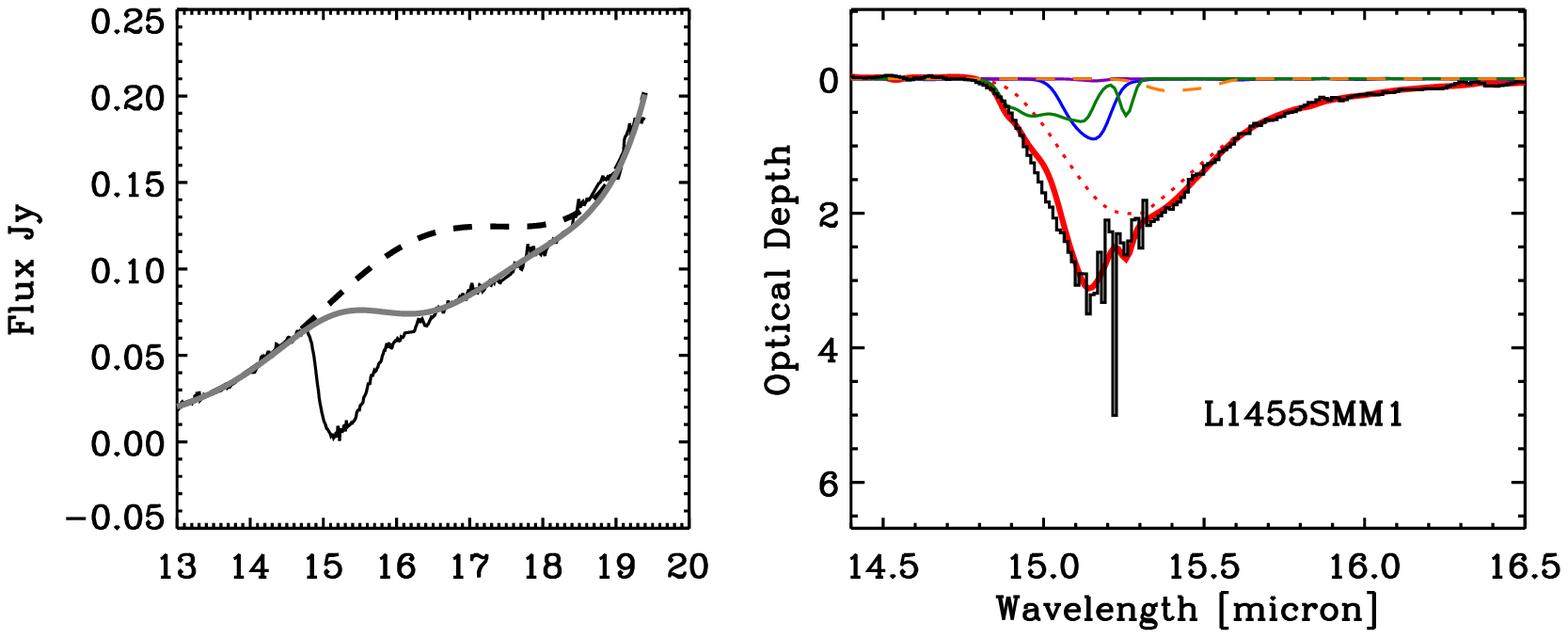}
  \includegraphics[width=15cm]{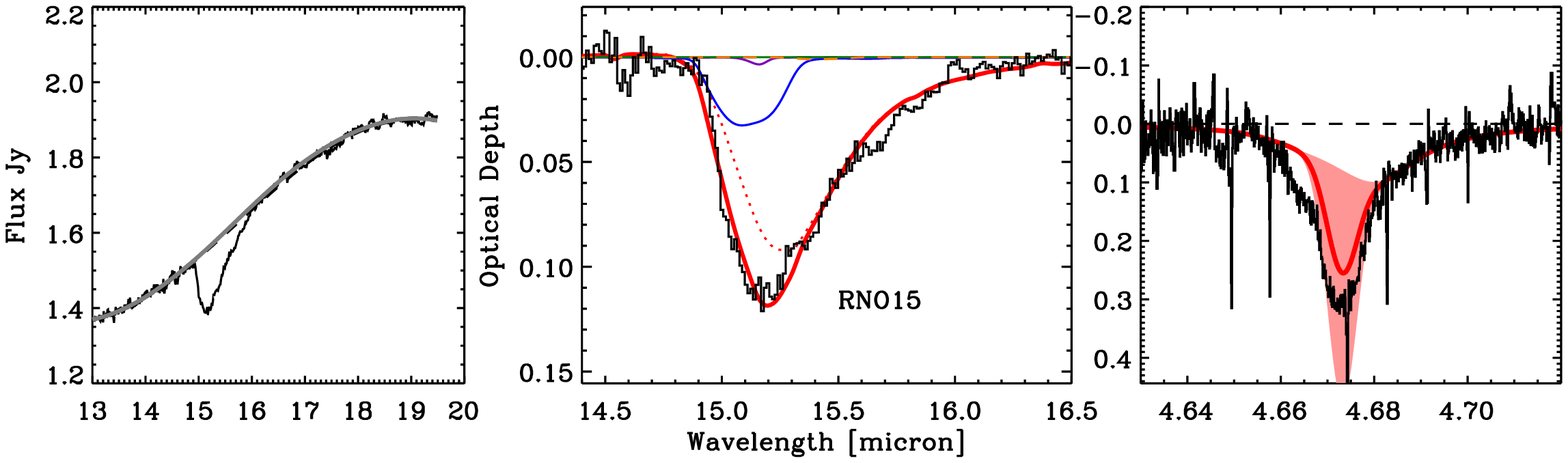}
  \caption{{\it Left panels}: Continuum determinations for the SH spectra. The dashed curves are low-order polynomial fits, while
    the solid curves are the fits with a Gaussian included to simulate the blue wing of the 18\,$\mu$m silicate bending mode. {\it Middle panels}: 
    Component fits of the CO$_2$ bending mode profiles. 
    {\it Right panels}: The CO stretching mode profiles predicted by the fit to the CO$_2$ bending mode (where available). 
    The components are discussed in Section \ref{CompAnalysis}. The shaded areas indicate the profiles allowed by the uncertainties
in the fit to the CO$_2$ bending modes. }
  \label{decomp-1}
\end{figure*}

\begin{figure*}[h]
    
  \includegraphics[width=15cm]{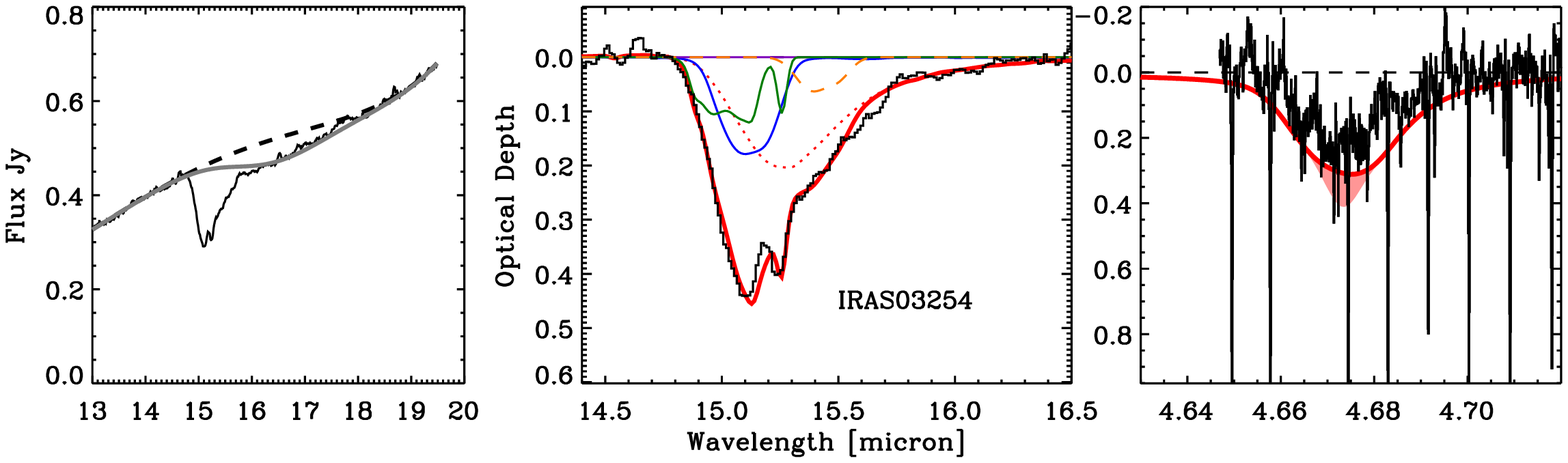}
  \includegraphics[width=15cm]{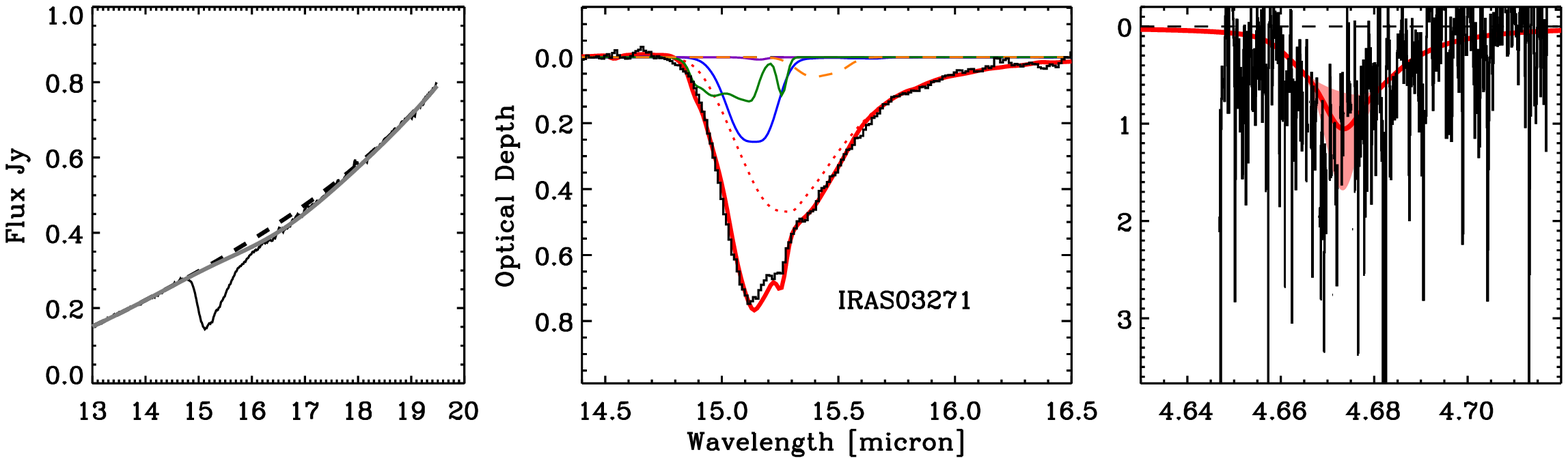}
  \includegraphics[width=15cm]{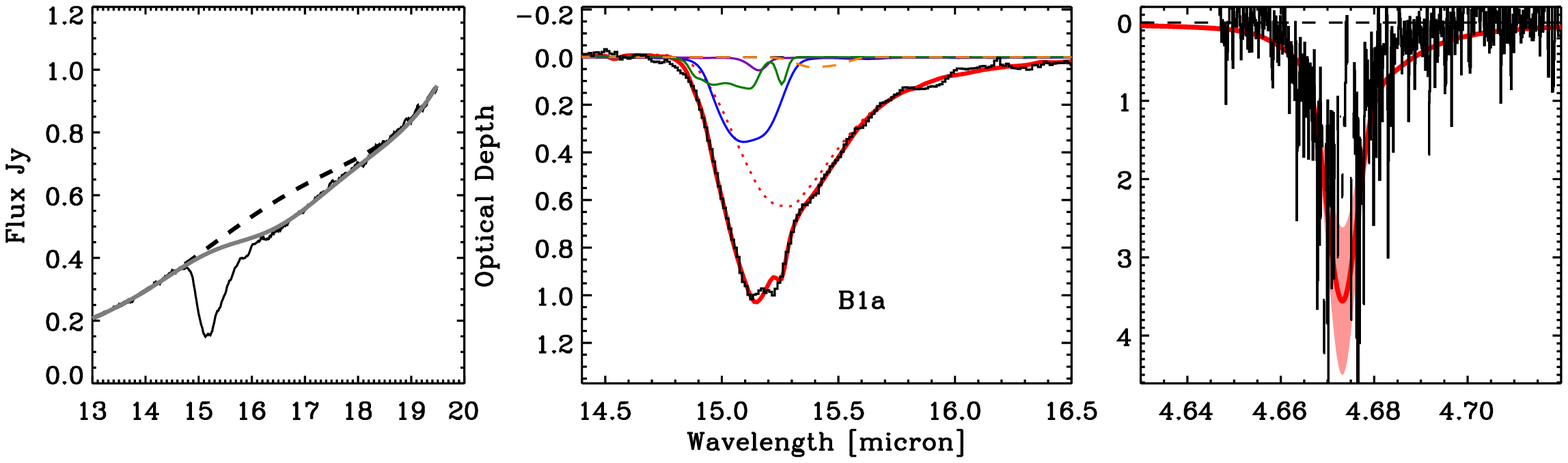}
  \includegraphics[width=15cm]{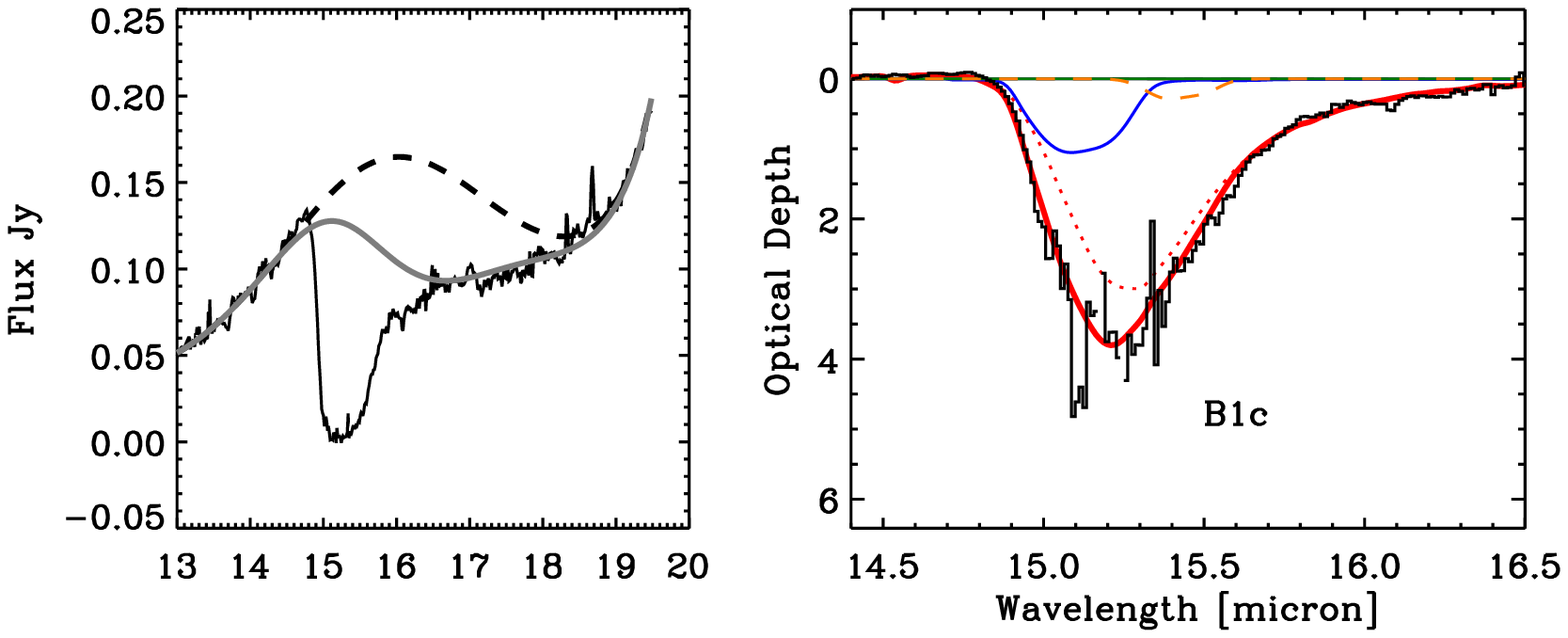}
  \caption{As Figure \ref{decomp-1}.}
  \label{decomp-2}
\end{figure*}

\begin{figure*}[h]
    
  \includegraphics[width=15cm]{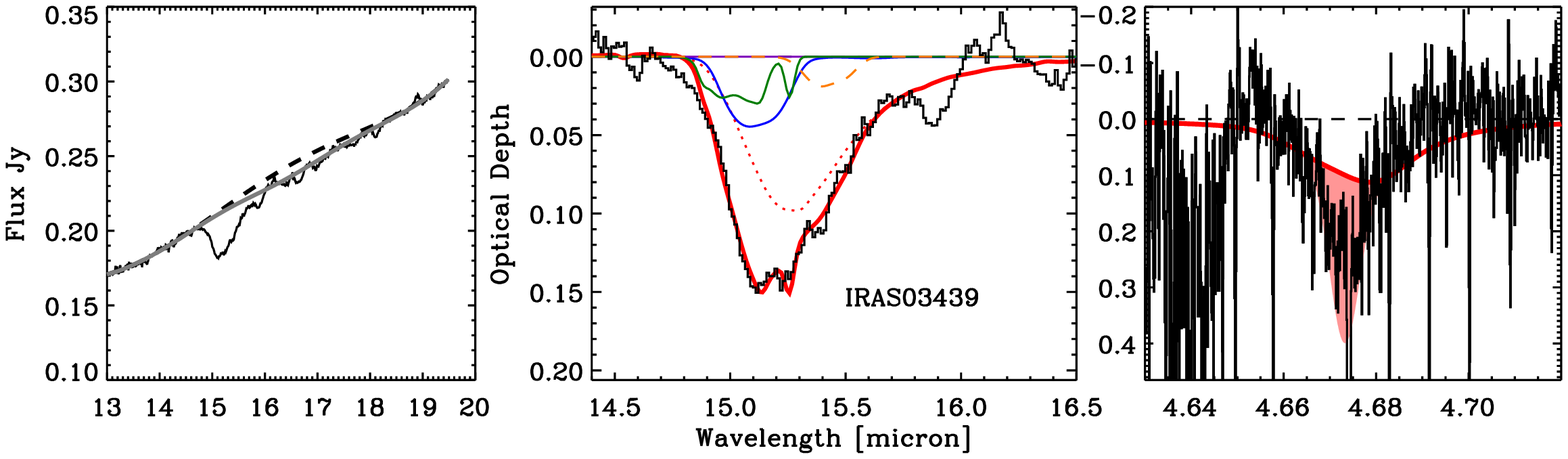}
  \includegraphics[width=15cm]{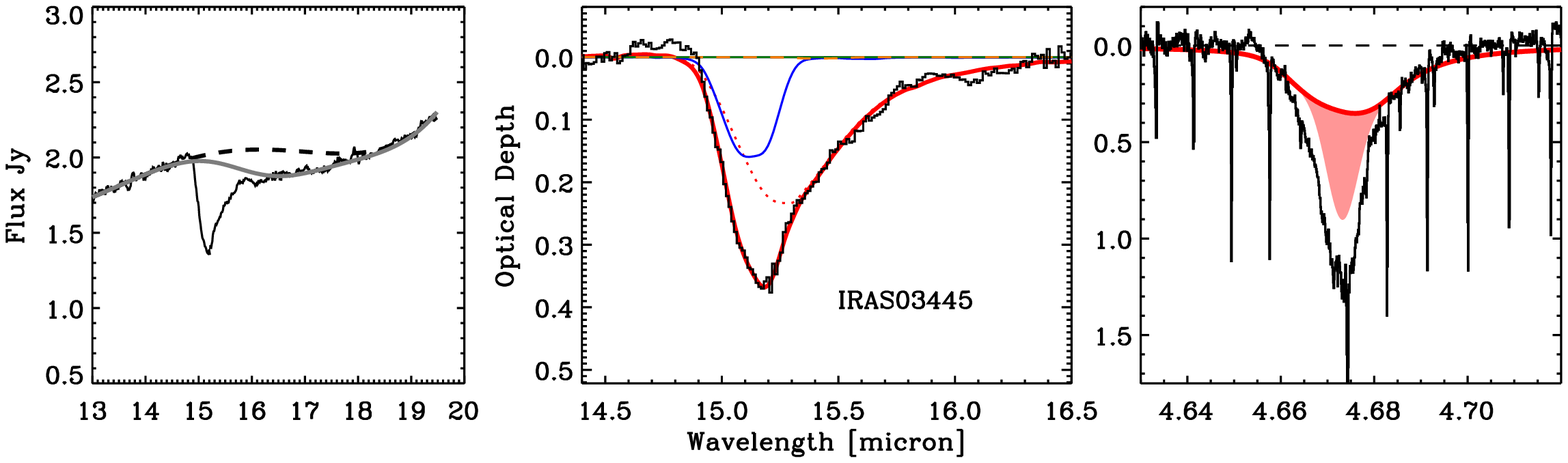}
  \includegraphics[width=15cm]{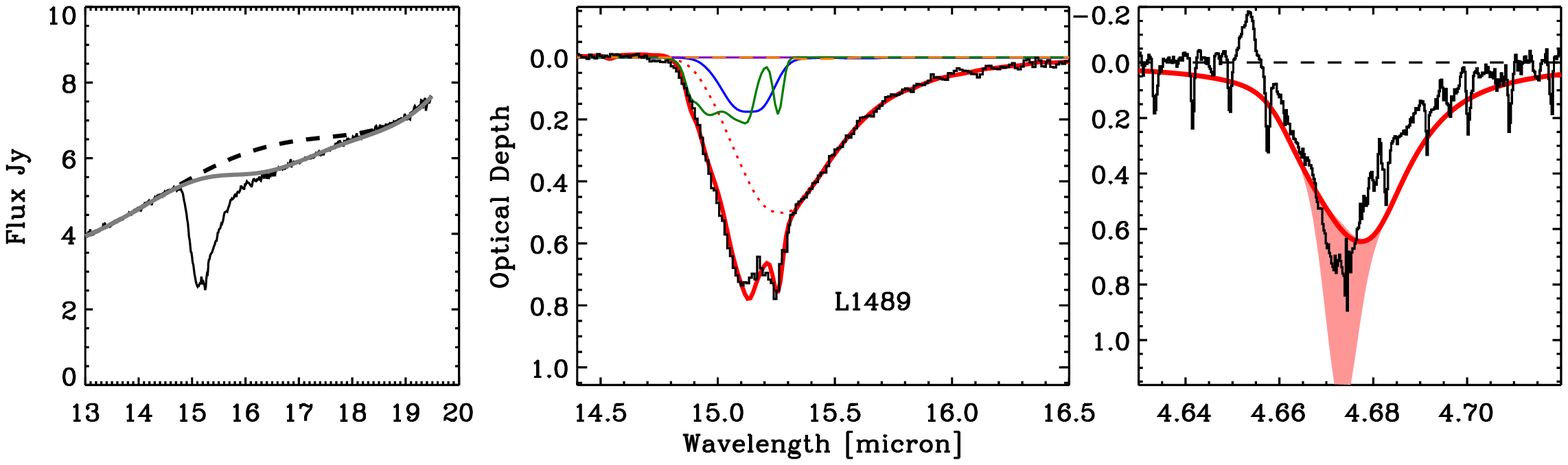}
  \includegraphics[width=15cm]{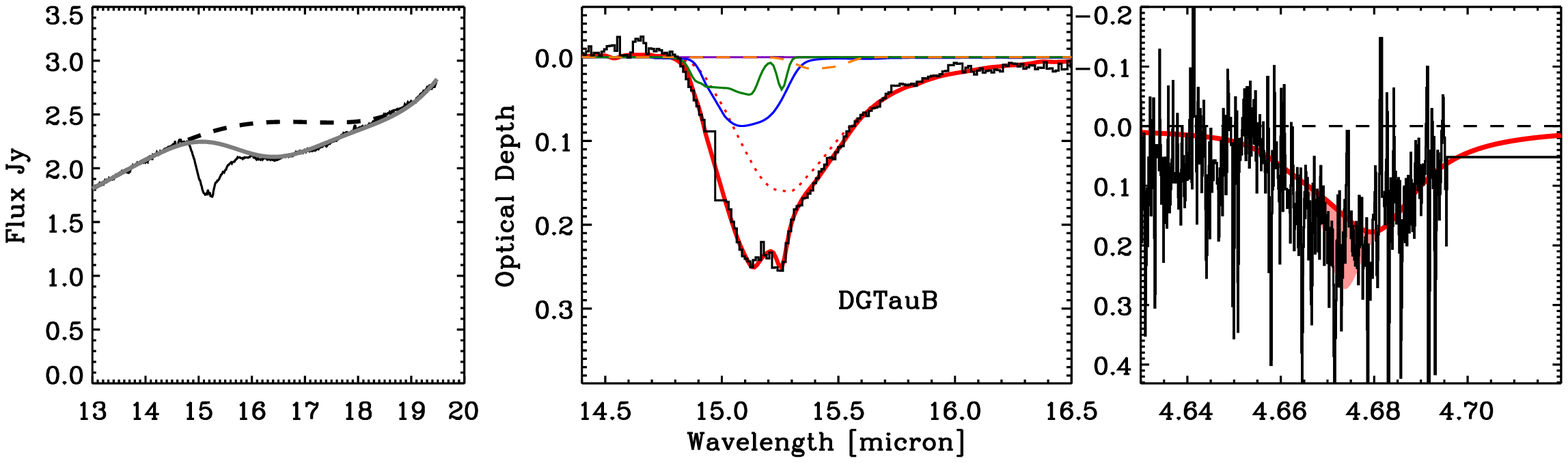}
  \caption{As Figure \ref{decomp-1}.}
  \label{decomp-3}
\end{figure*}

\begin{figure*}[h]
    
  \includegraphics[width=15cm]{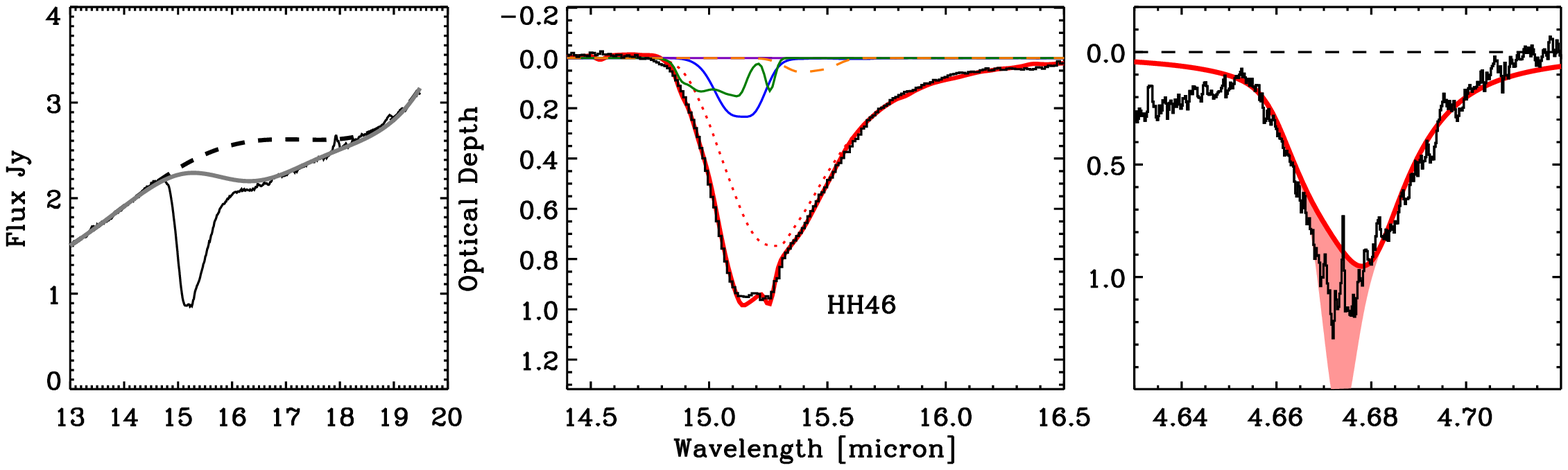}
  \includegraphics[width=15cm]{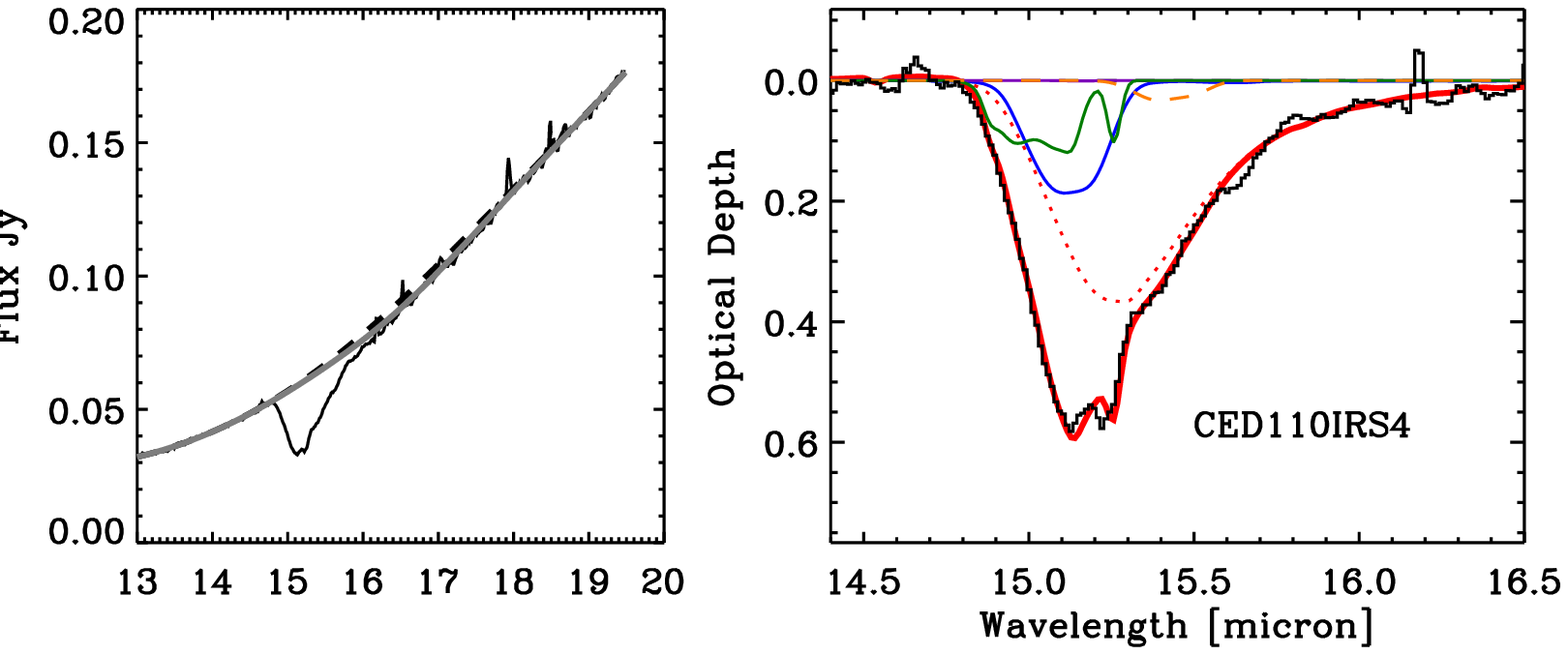}
  \includegraphics[width=15cm]{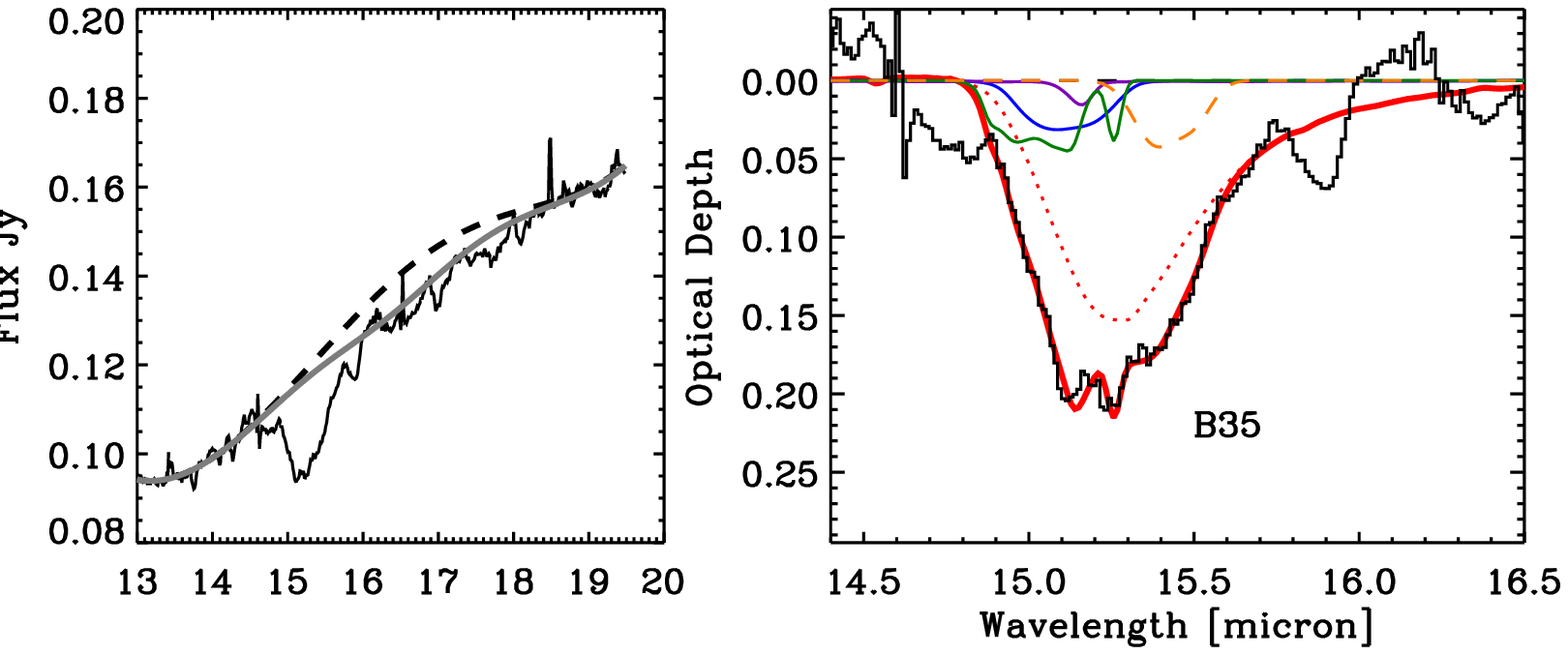}
  \includegraphics[width=15cm]{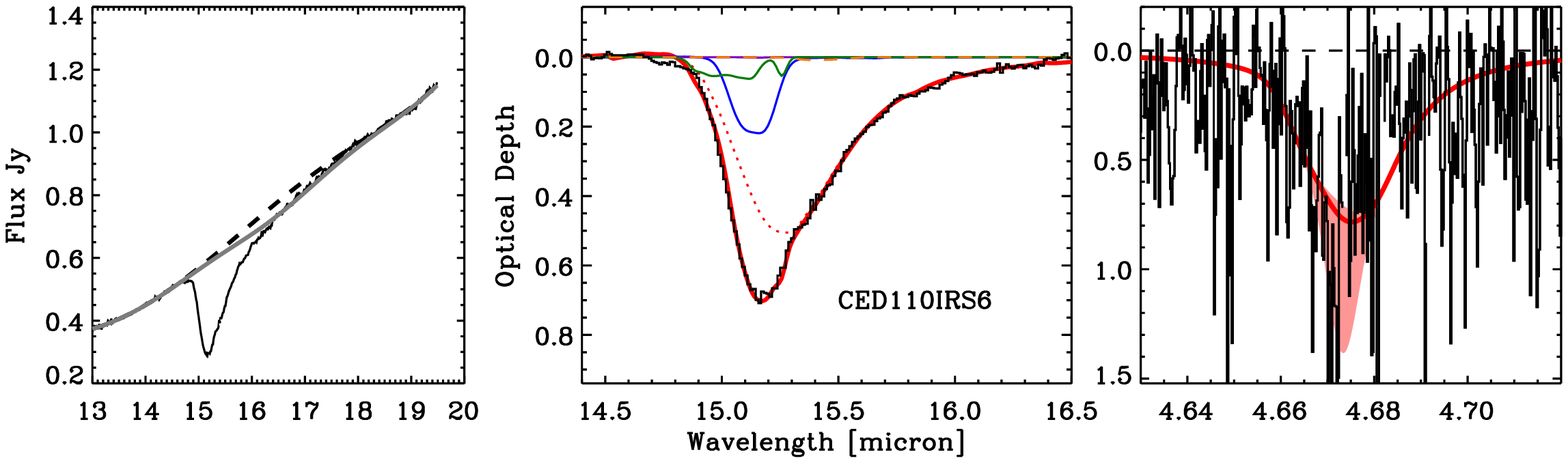}
  \caption{As Figure \ref{decomp-1}.}
  \label{decomp-4}
\end{figure*}

\begin{figure*}[h]
    
  \includegraphics[width=15cm]{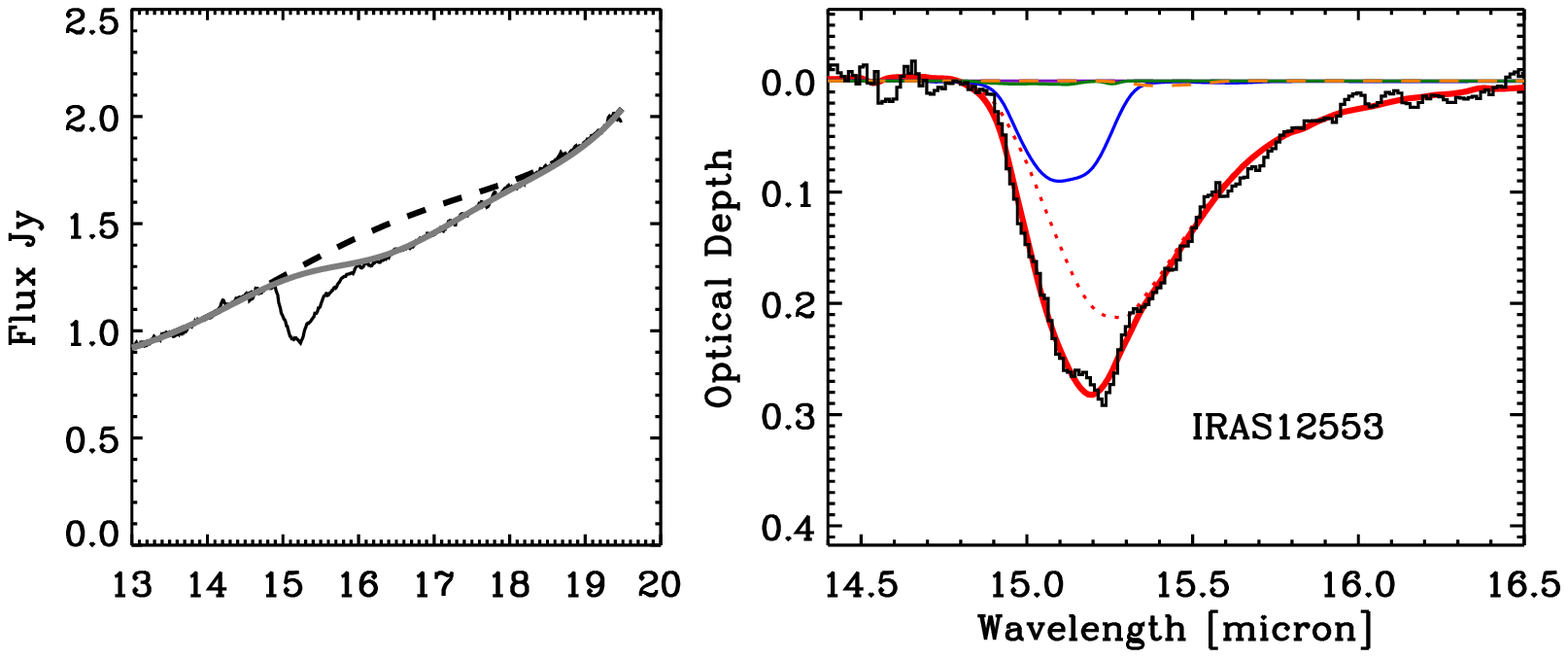}
  \includegraphics[width=15cm]{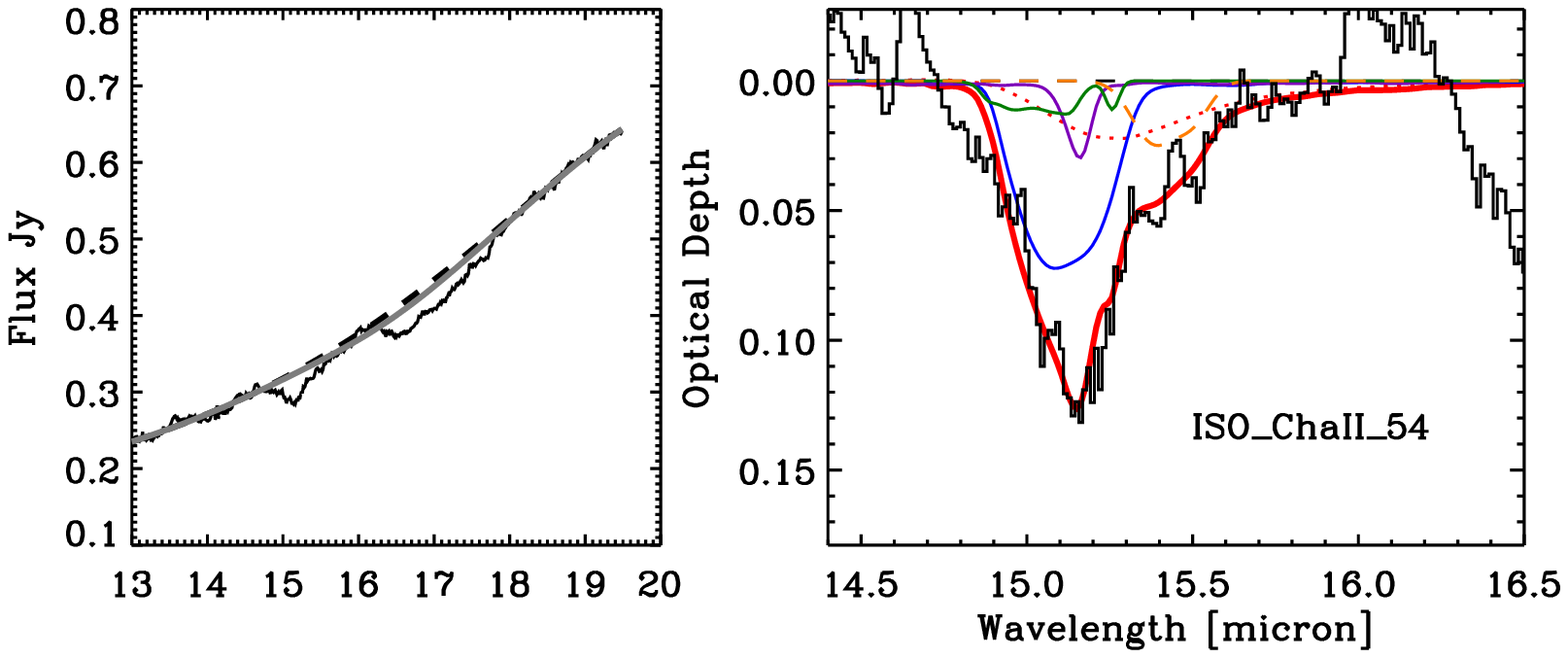}
  \includegraphics[width=15cm]{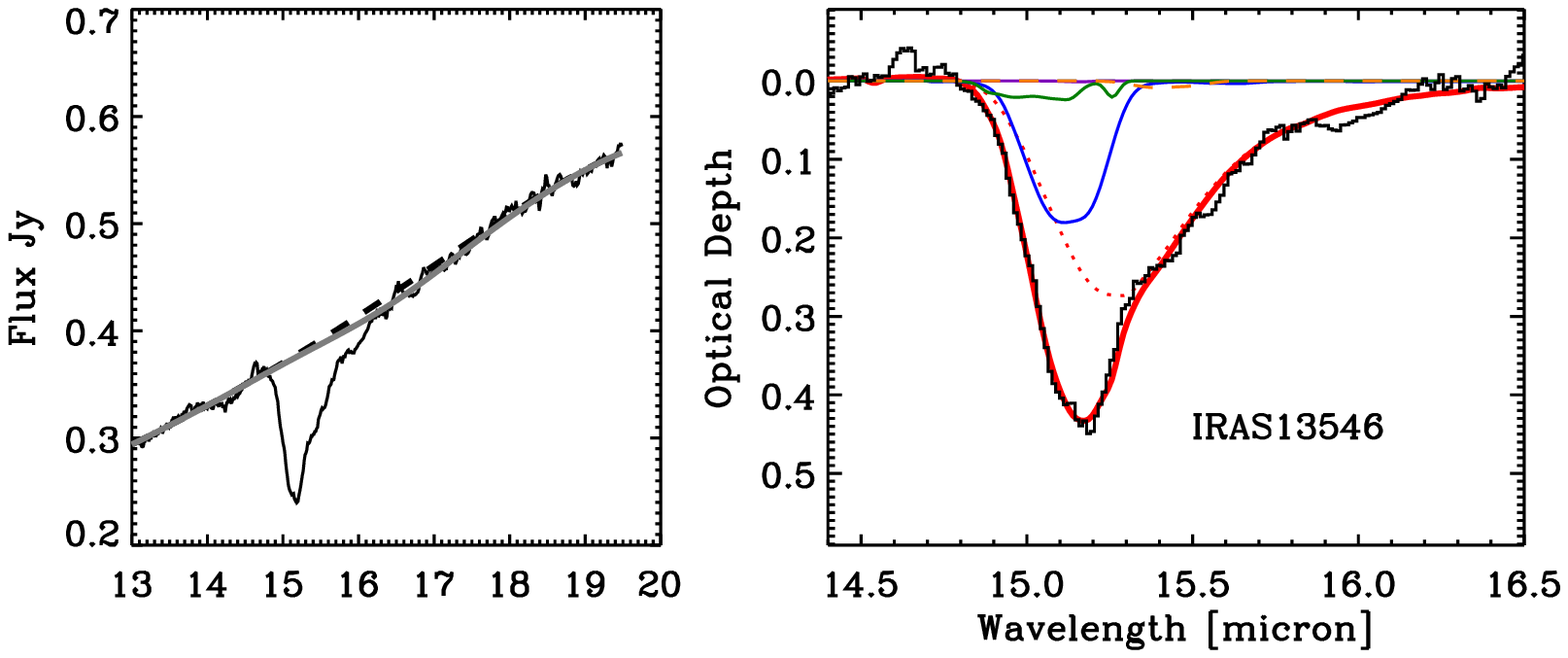}
  \includegraphics[width=15cm]{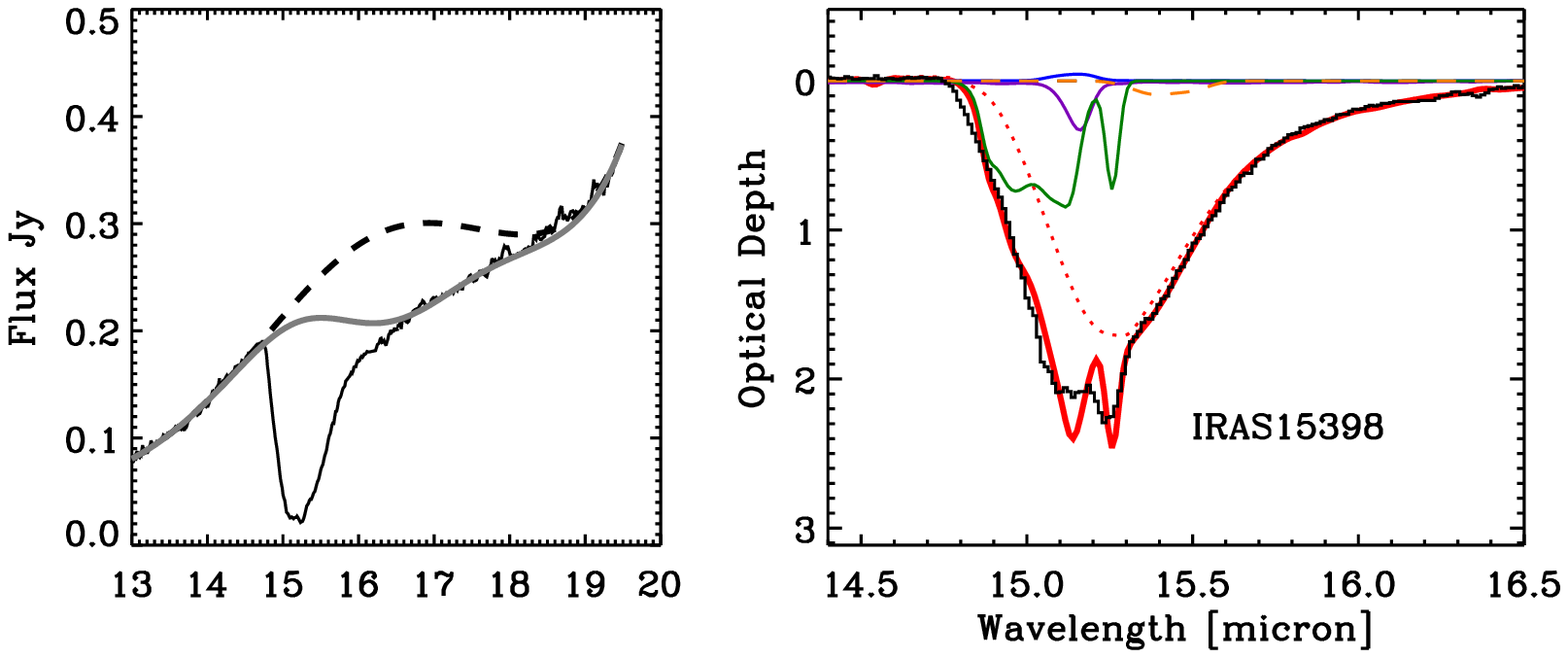}
  \caption{As Figure \ref{decomp-1}.}
  \label{decomp-5}
\end{figure*}

\begin{figure*}[h]
    
  \includegraphics[width=15cm]{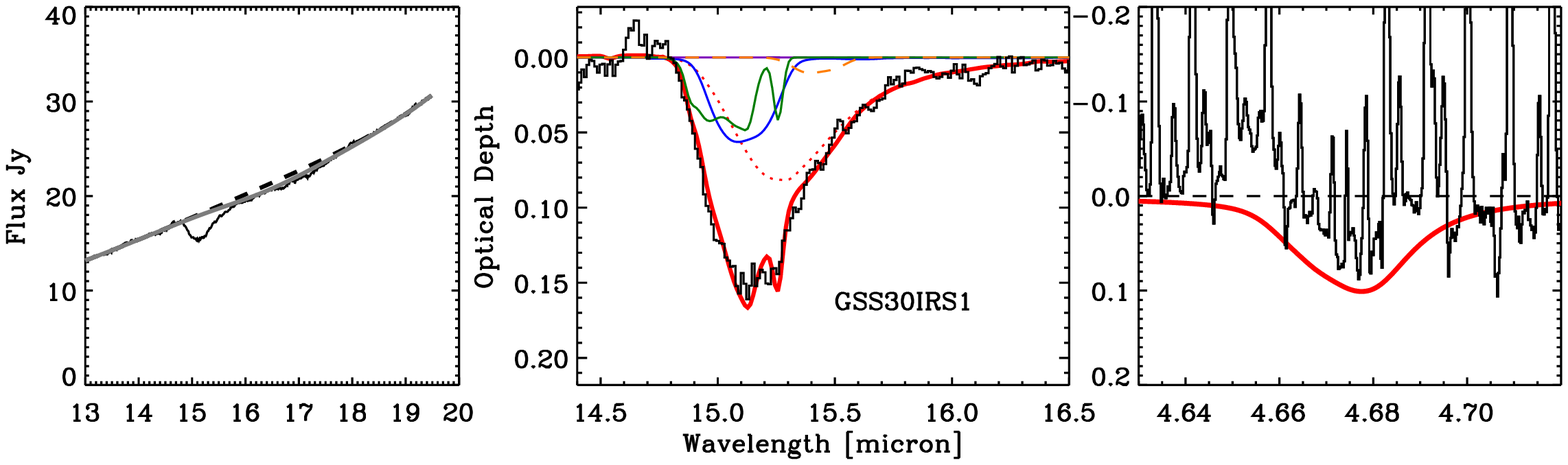}
  \includegraphics[width=15cm]{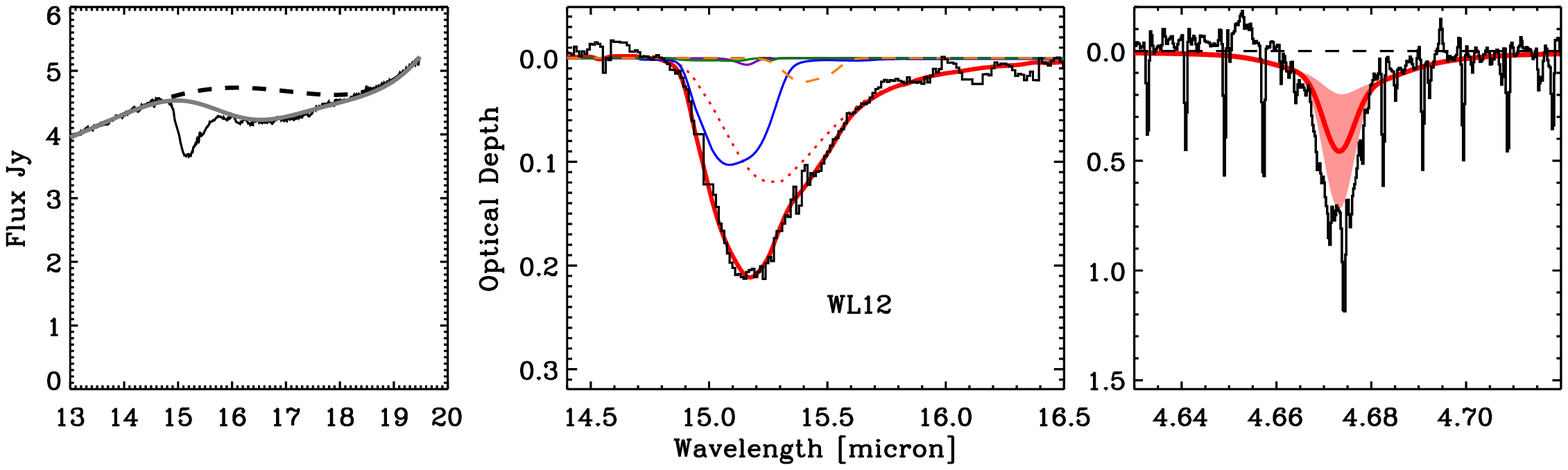}
  \includegraphics[width=15cm]{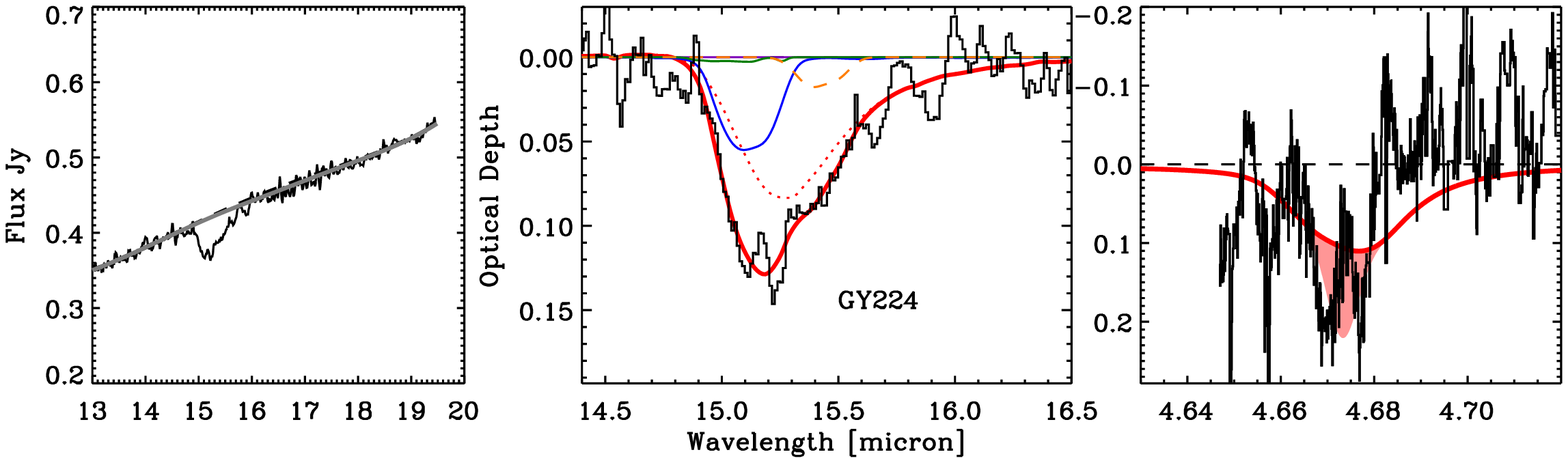}
  \includegraphics[width=15cm]{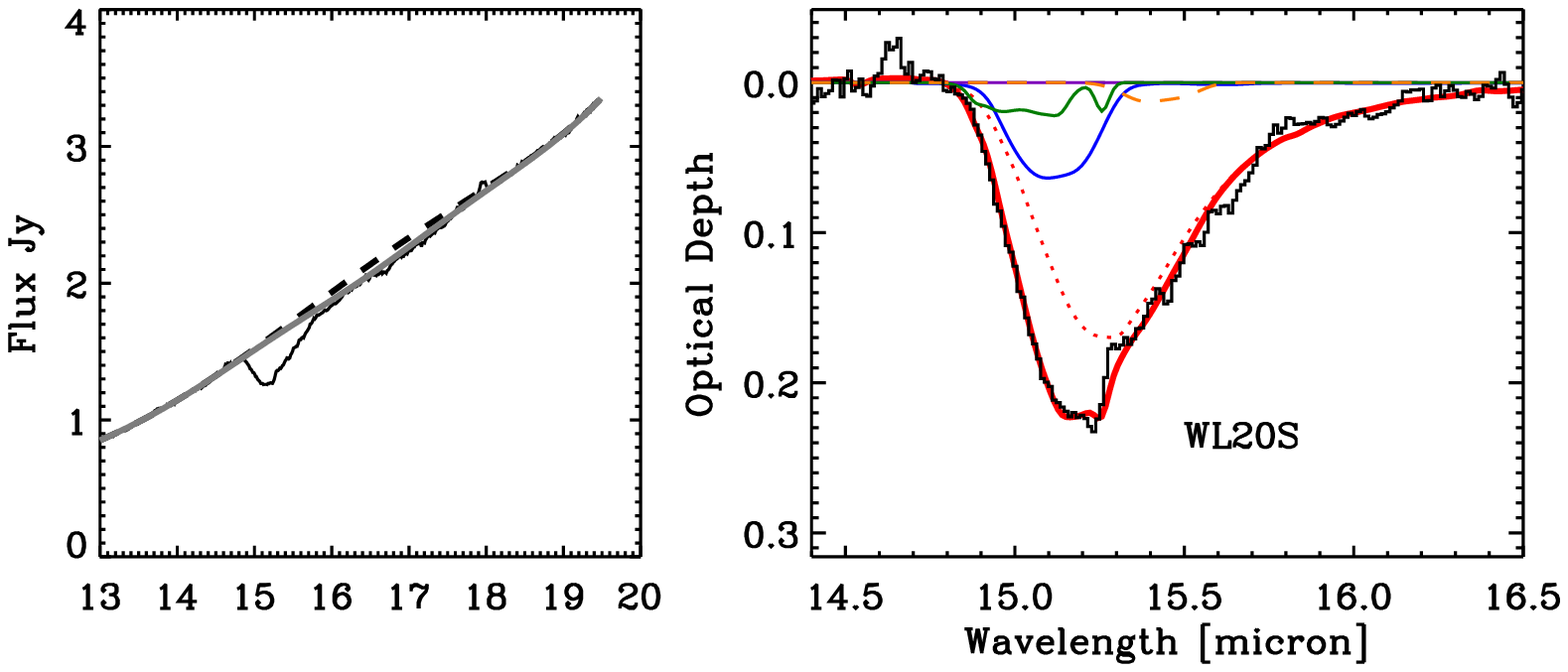}
  \caption{As Figure \ref{decomp-1}.}
  \label{decomp-6}
\end{figure*}

\begin{figure*}[h]
    
  \includegraphics[width=15cm]{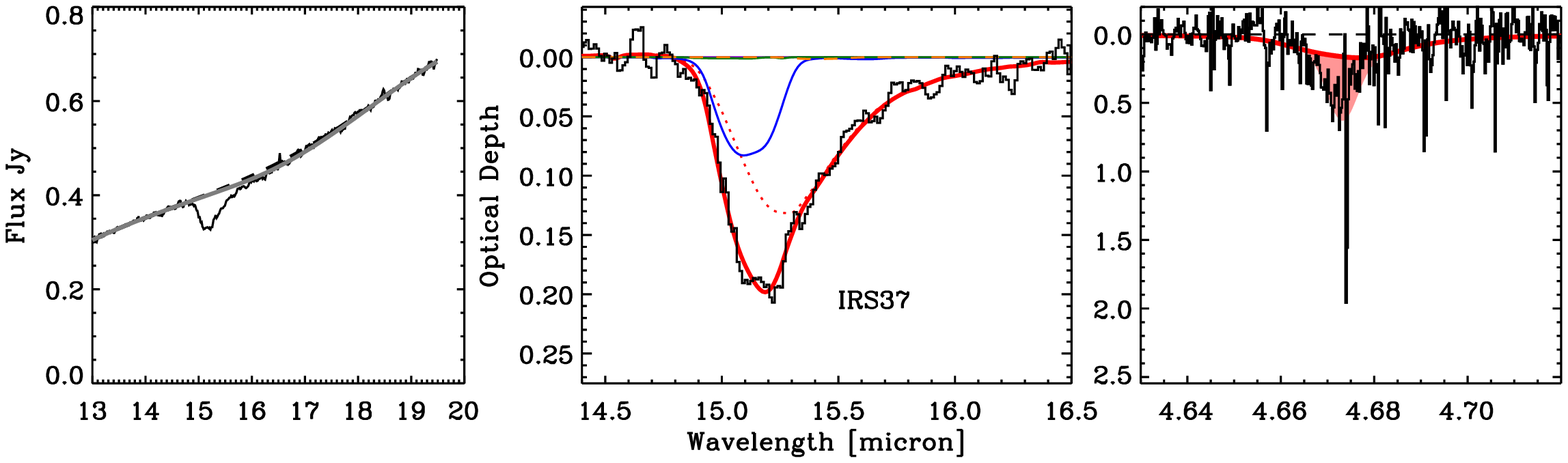}
  \includegraphics[width=15cm]{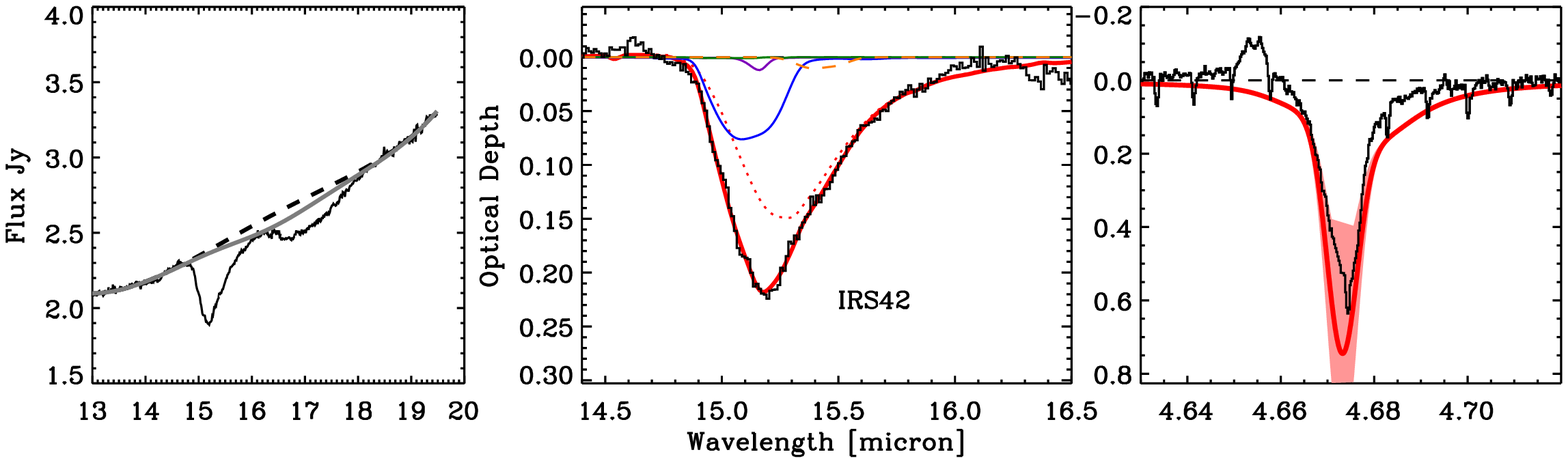}
  \includegraphics[width=15cm]{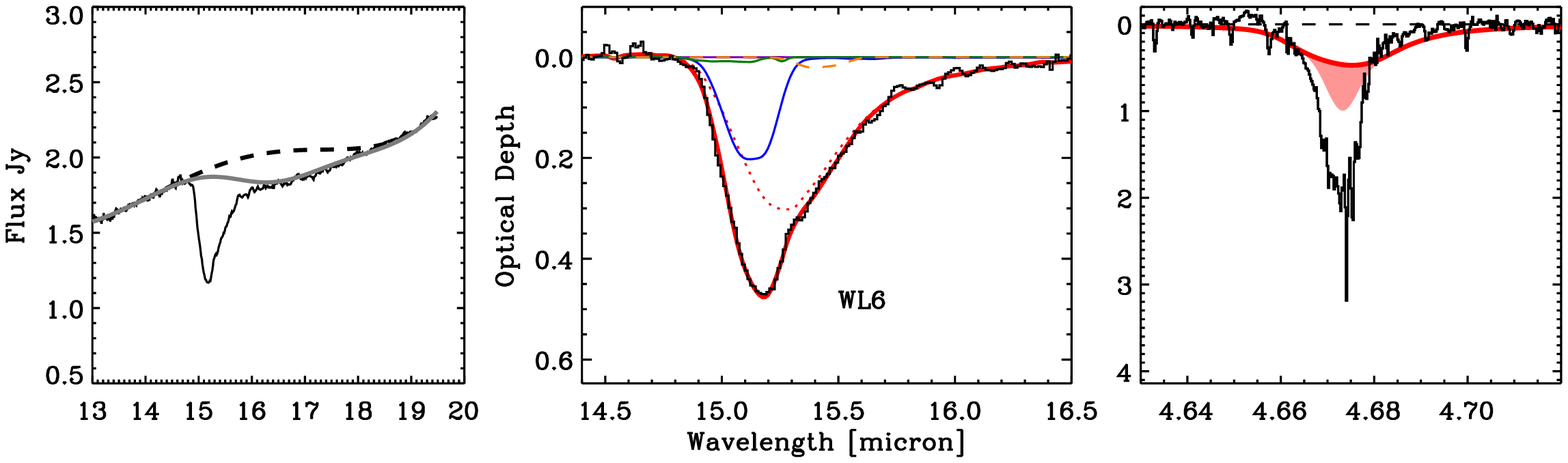}
  \includegraphics[width=15cm]{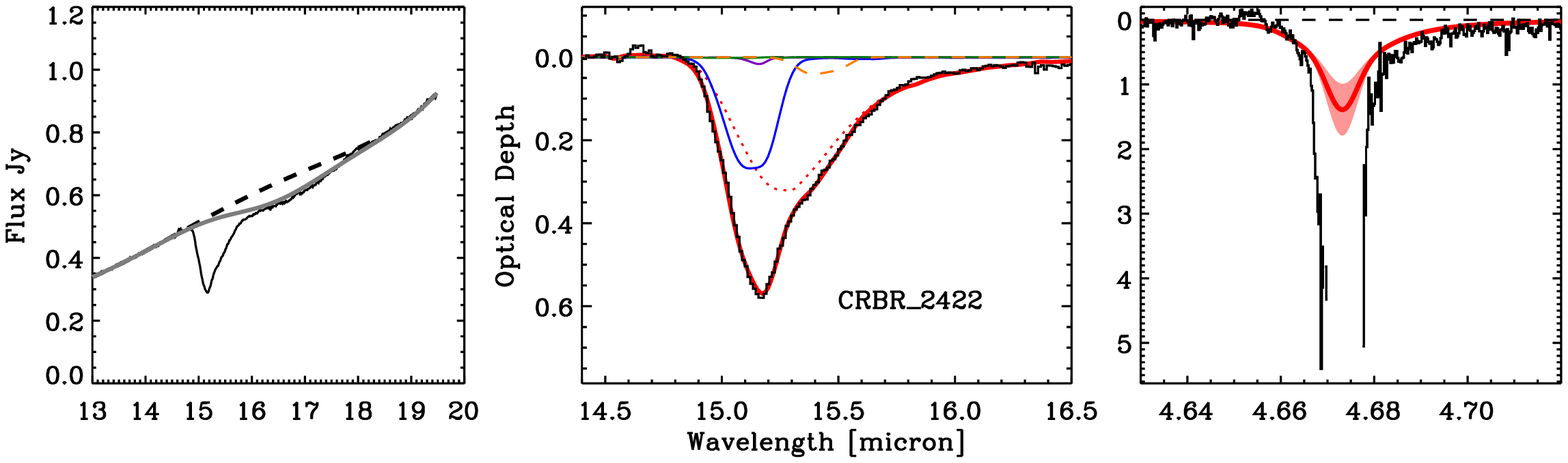}
  \caption{As Figure \ref{decomp-1}.}
  \label{decomp-7}
\end{figure*}

\begin{figure*}[h]
    
  \includegraphics[width=15cm]{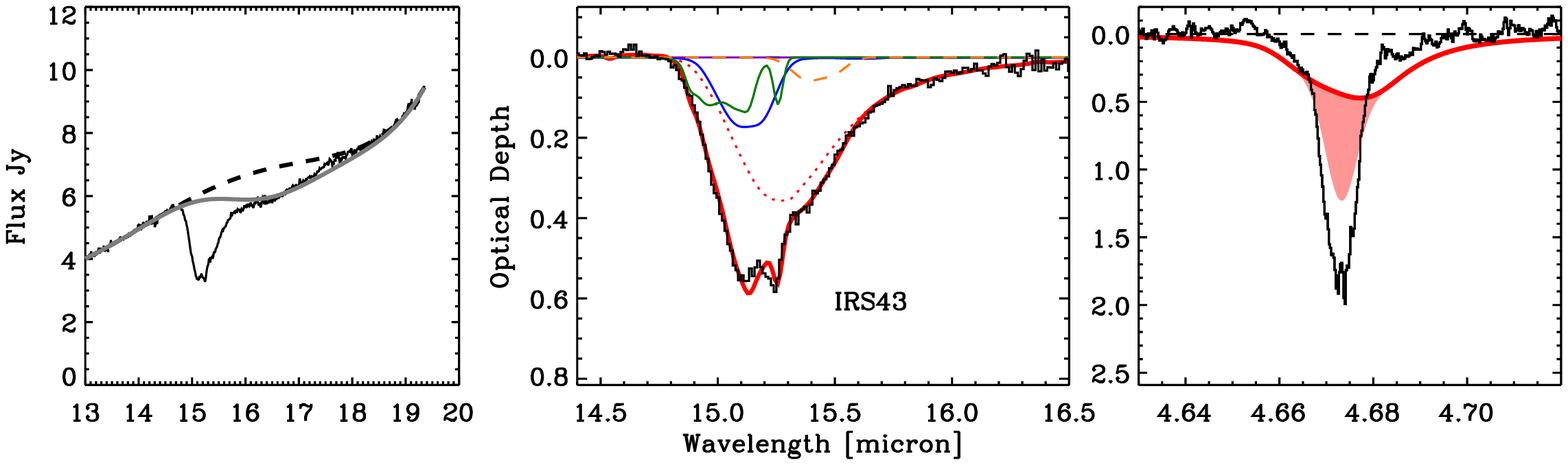}
  \includegraphics[width=15cm]{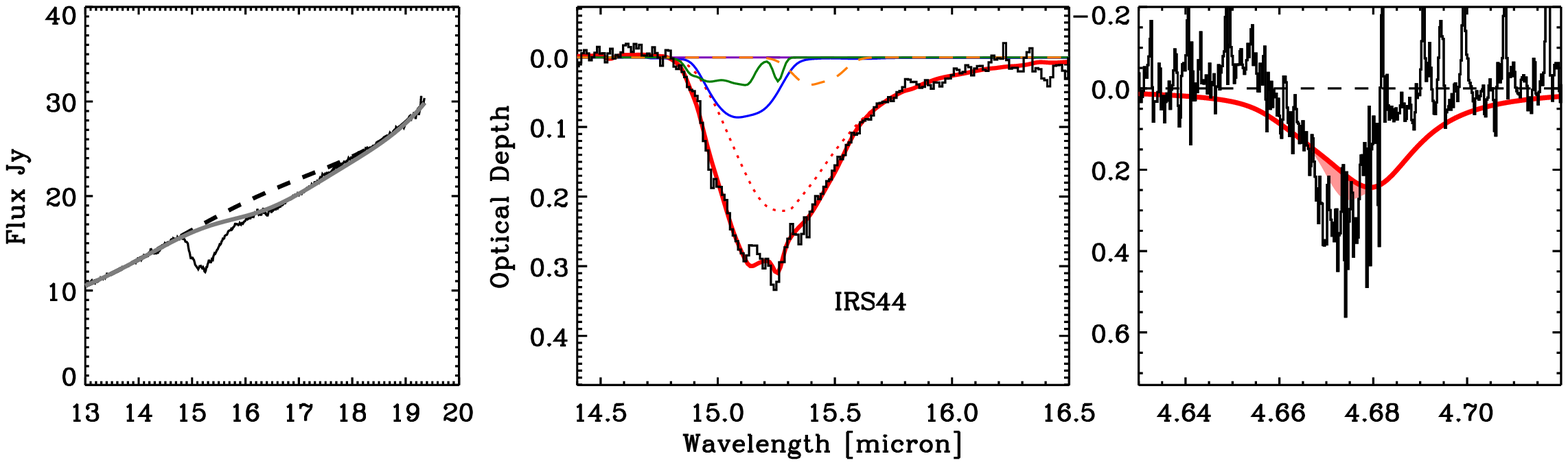}
  \includegraphics[width=15cm]{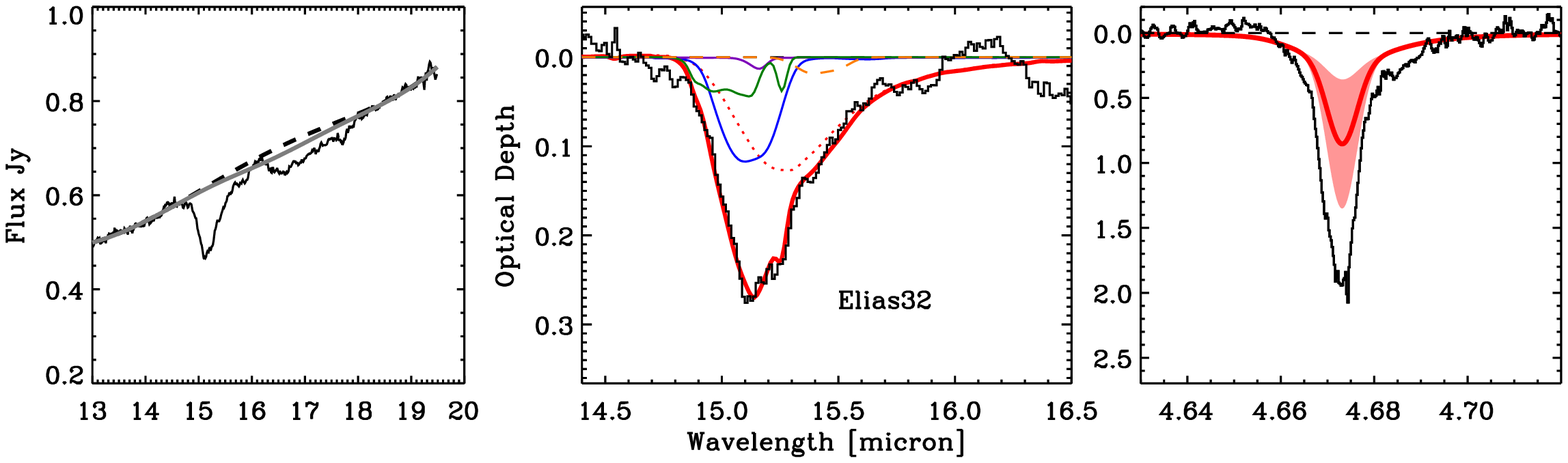}
  \includegraphics[width=15cm]{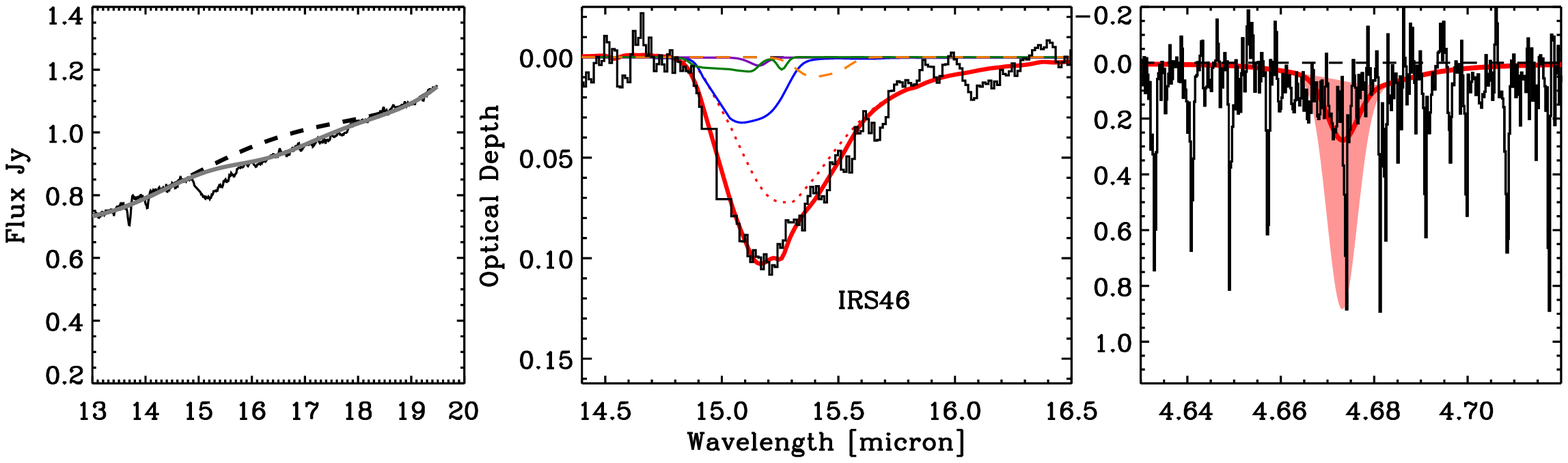}
  \caption{As Figure \ref{decomp-1}.}
  \label{decomp-8}
\end{figure*}

\begin{figure*}[h]
    
  \includegraphics[width=15cm]{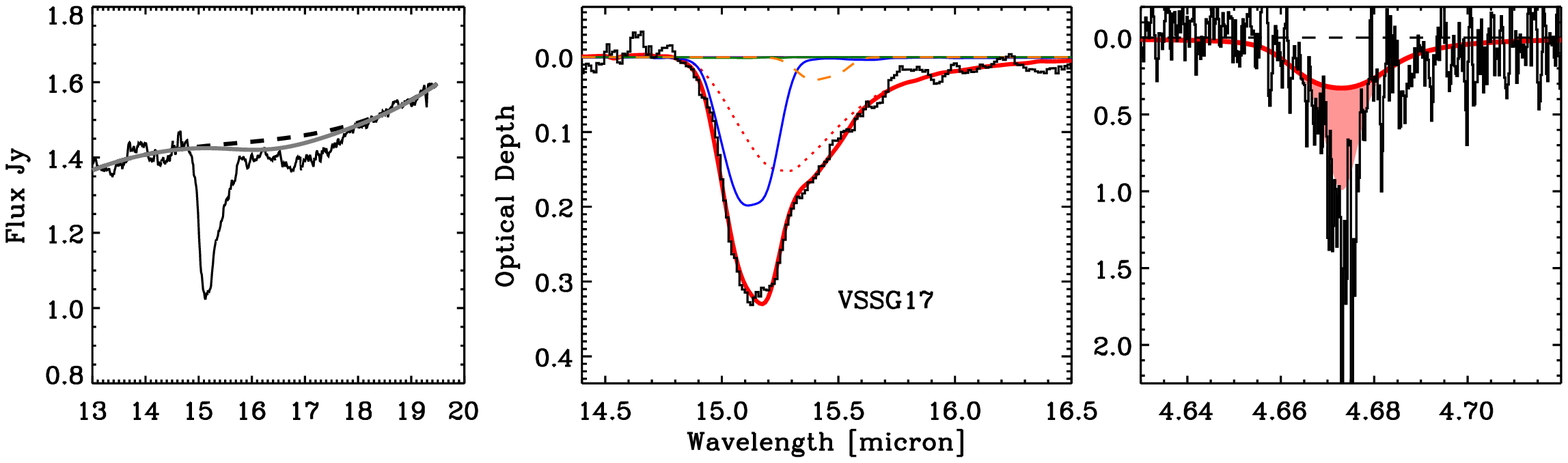}
  \includegraphics[width=15cm]{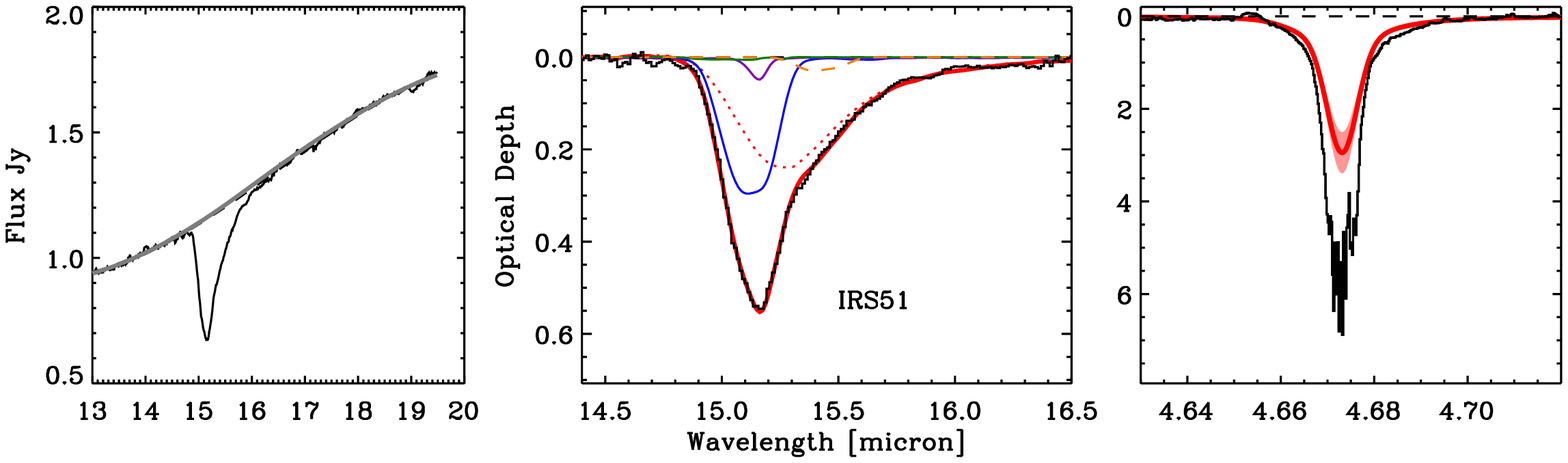}
  \includegraphics[width=15cm]{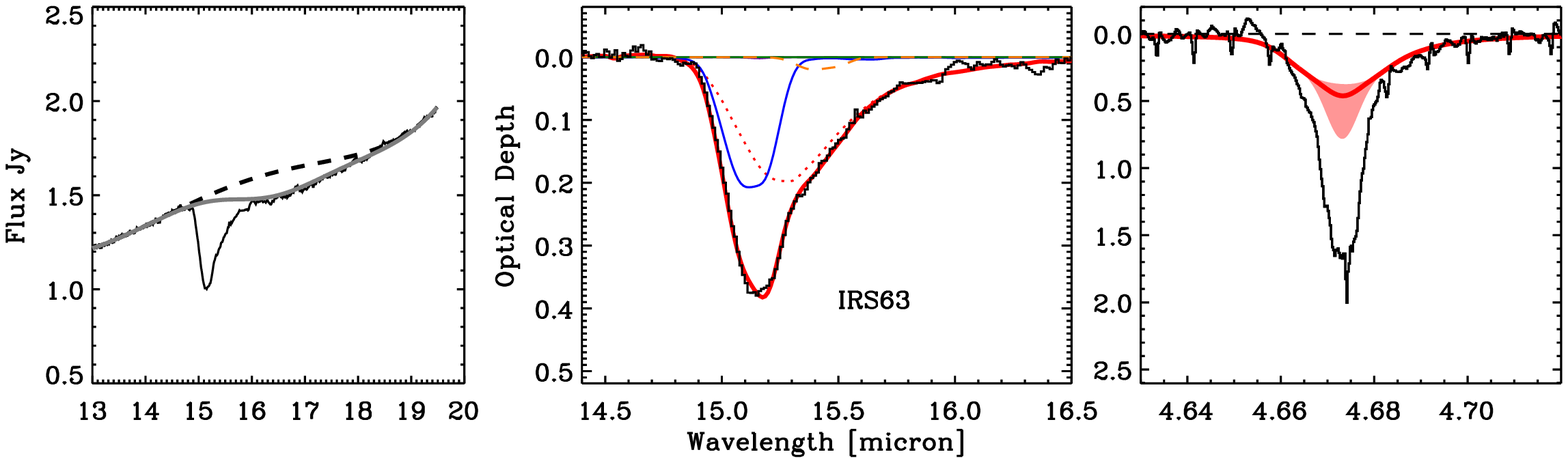}
  \includegraphics[width=15cm]{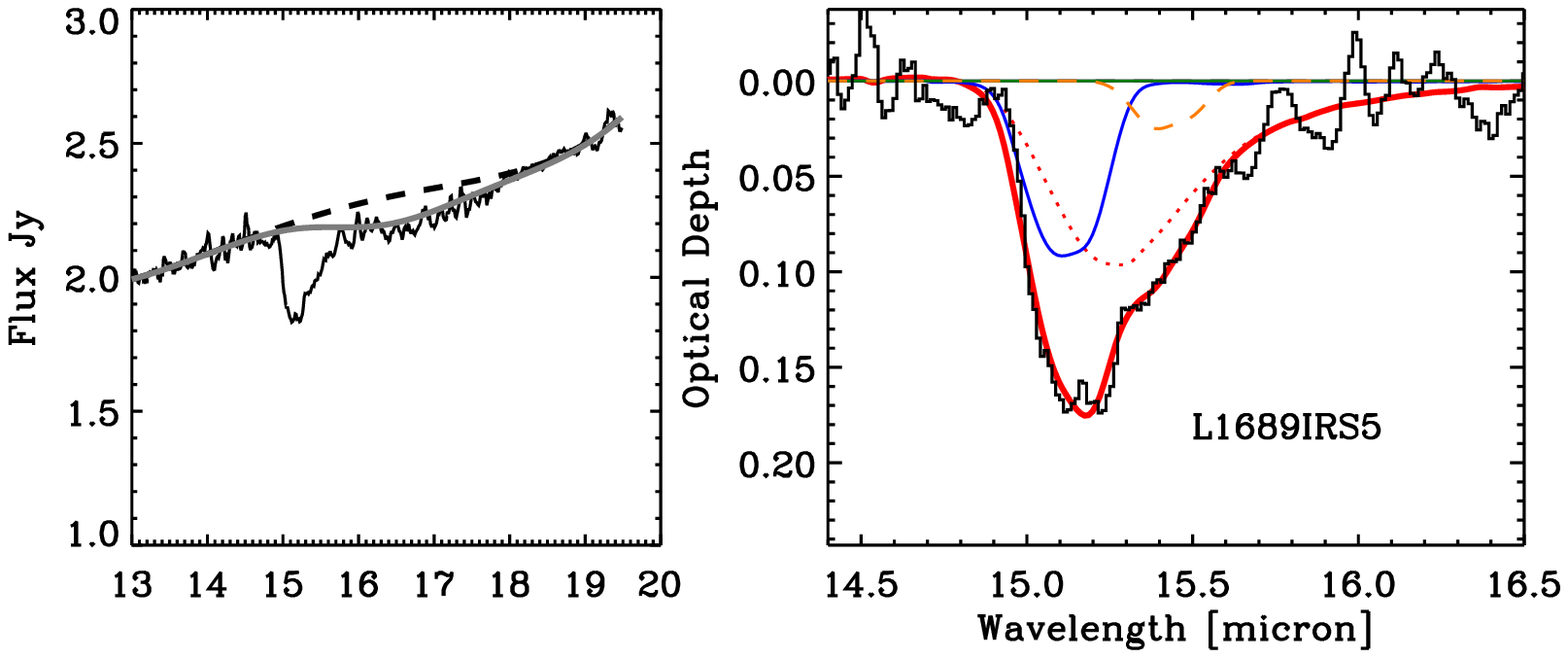}
  \caption{As Figure \ref{decomp-1}.}
  \label{decomp-9}
\end{figure*}

\begin{figure*}[h]
    
  \includegraphics[width=15cm]{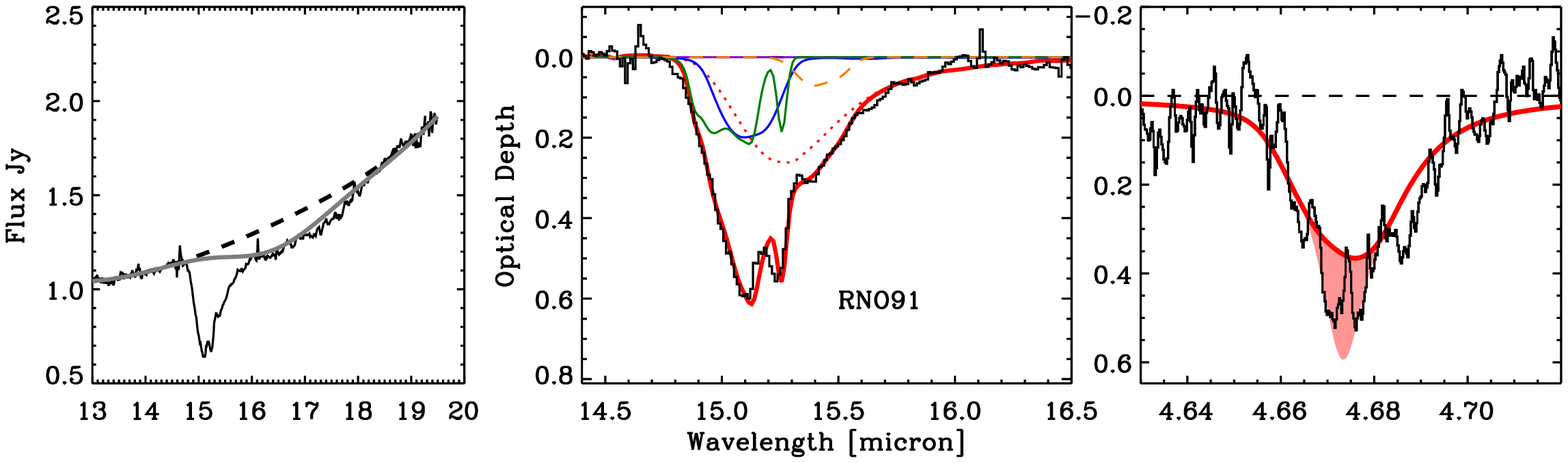}
  \includegraphics[width=15cm]{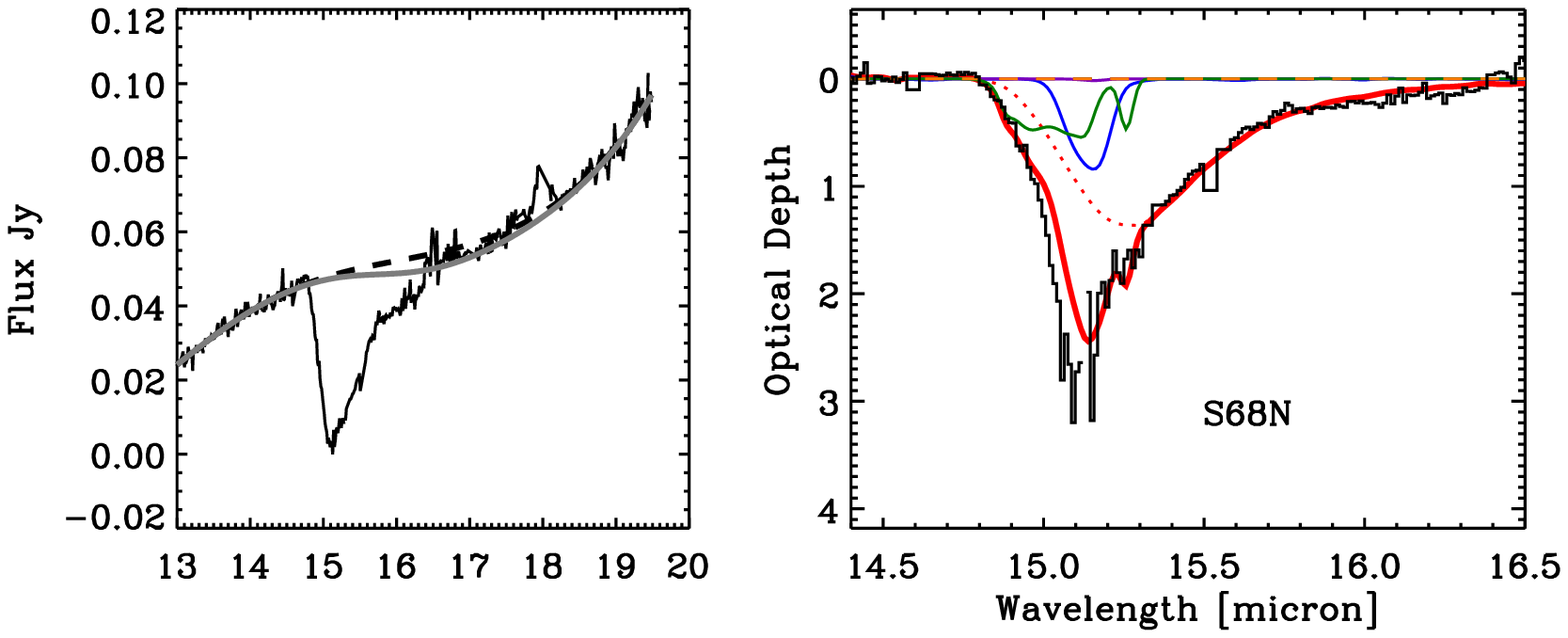}
  \includegraphics[width=15cm]{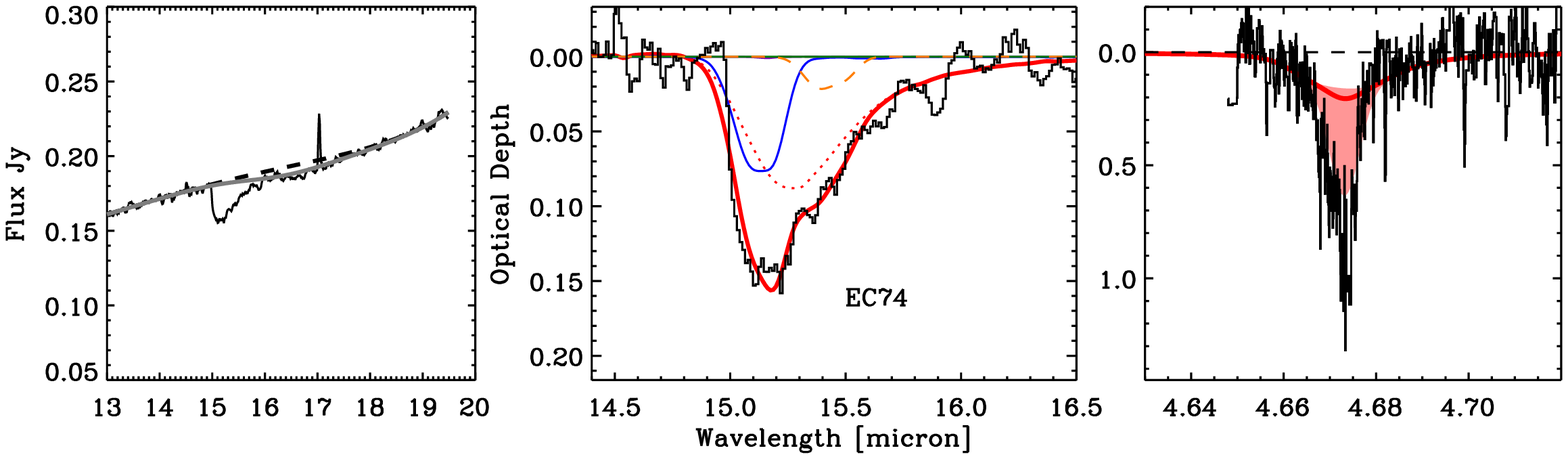}
  \includegraphics[width=15cm]{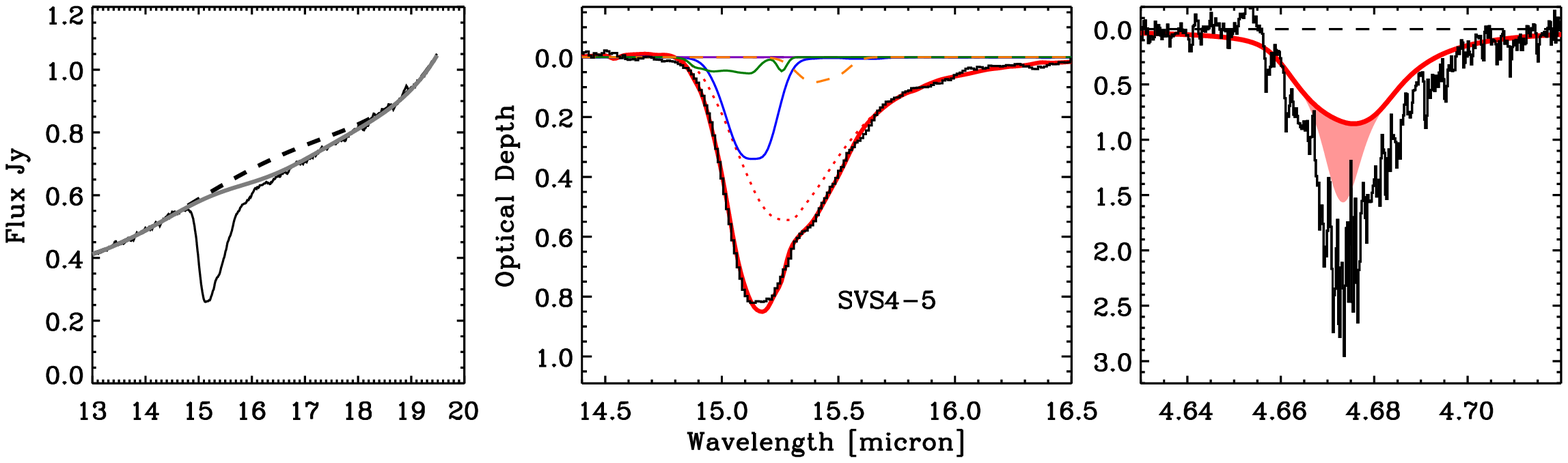}
  \caption{As Figure \ref{decomp-1}.}
  \label{decomp-10}
\end{figure*}

\begin{figure*}[h]
    
  \includegraphics[width=15cm]{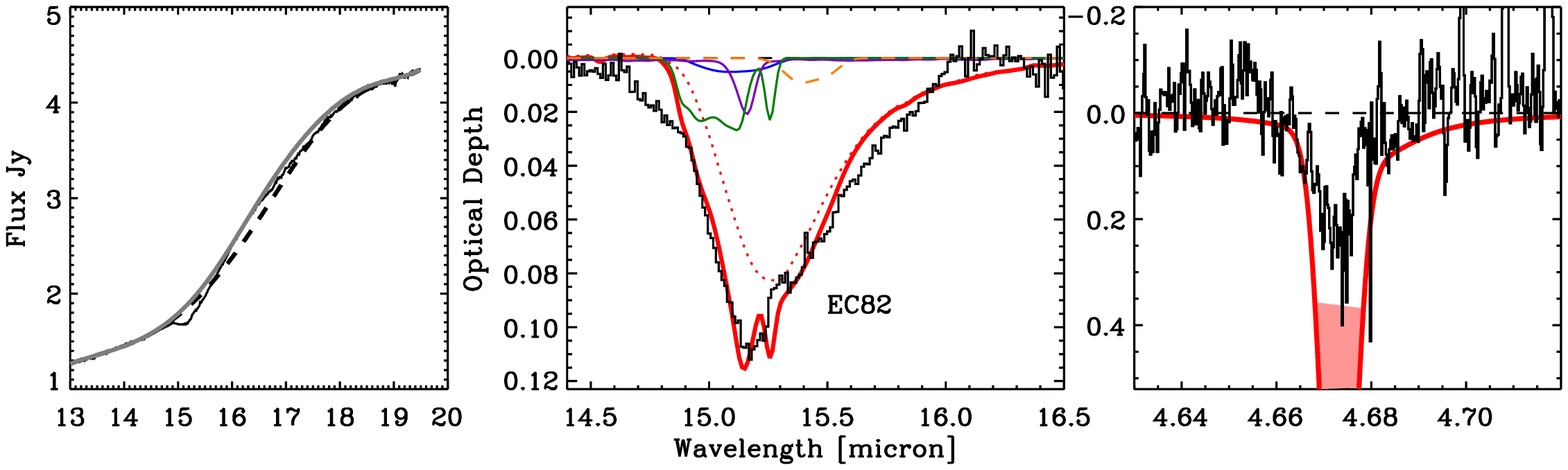}
  \includegraphics[width=15cm]{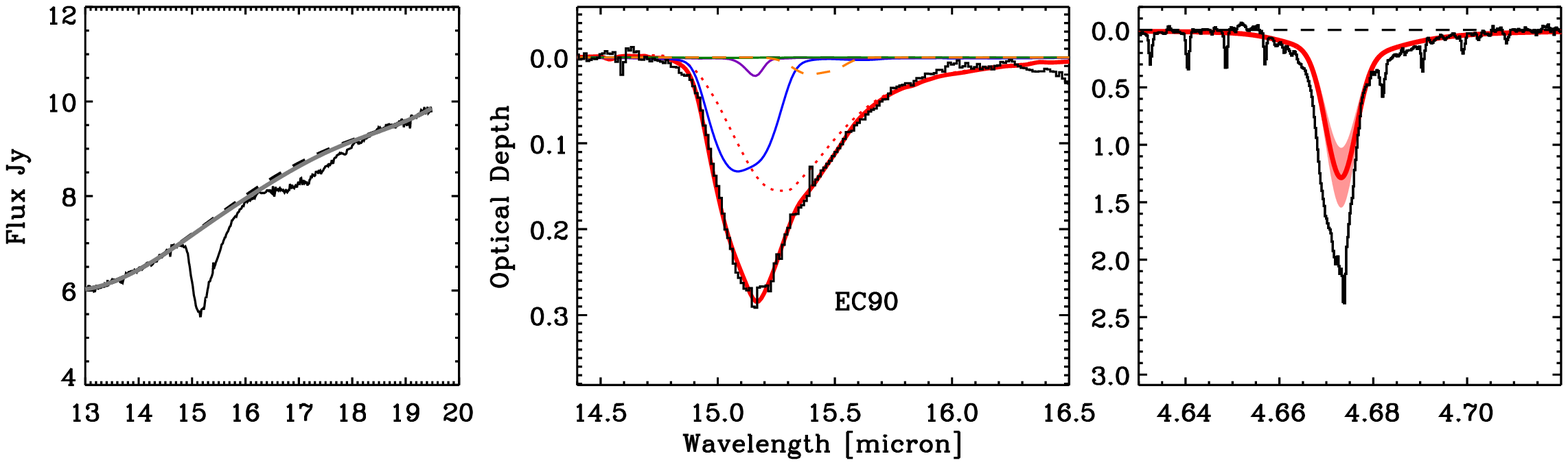}
  \includegraphics[width=15cm]{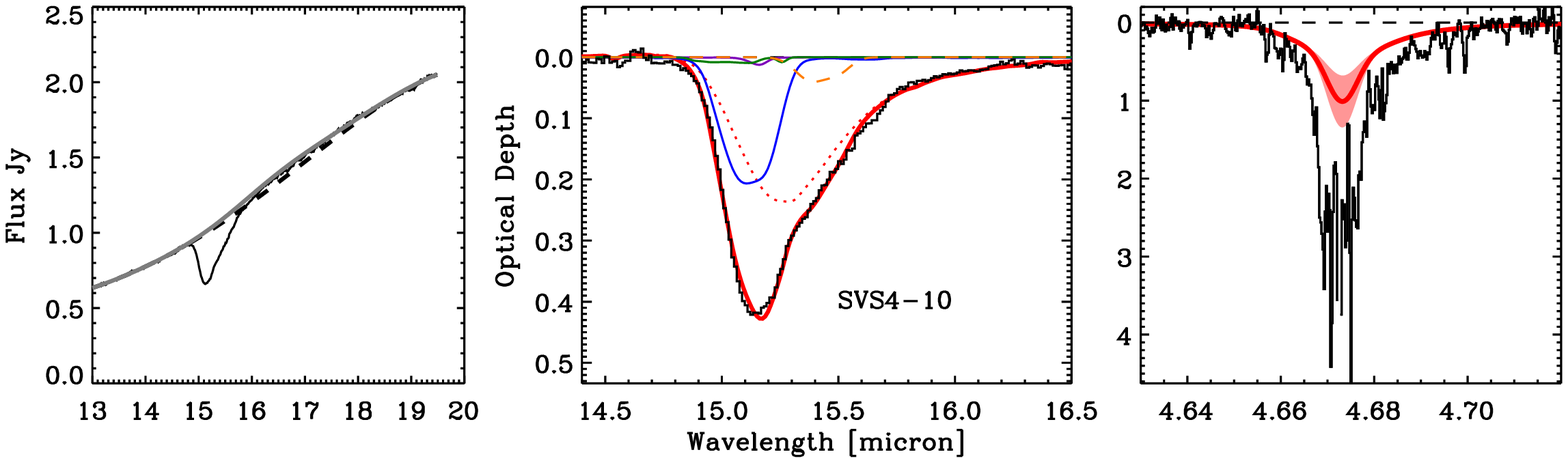}
  \includegraphics[width=15cm]{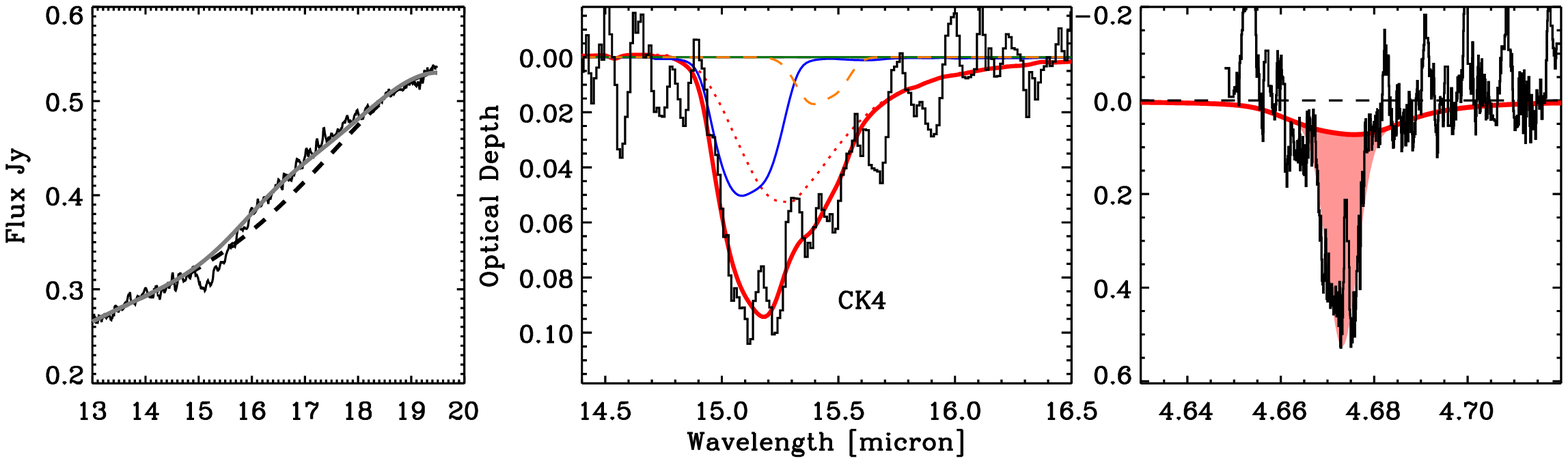}
  \caption{As Figure \ref{decomp-1}.}
  \label{decomp-11}
\end{figure*}

\begin{figure*}[h]
    
  \includegraphics[width=15cm]{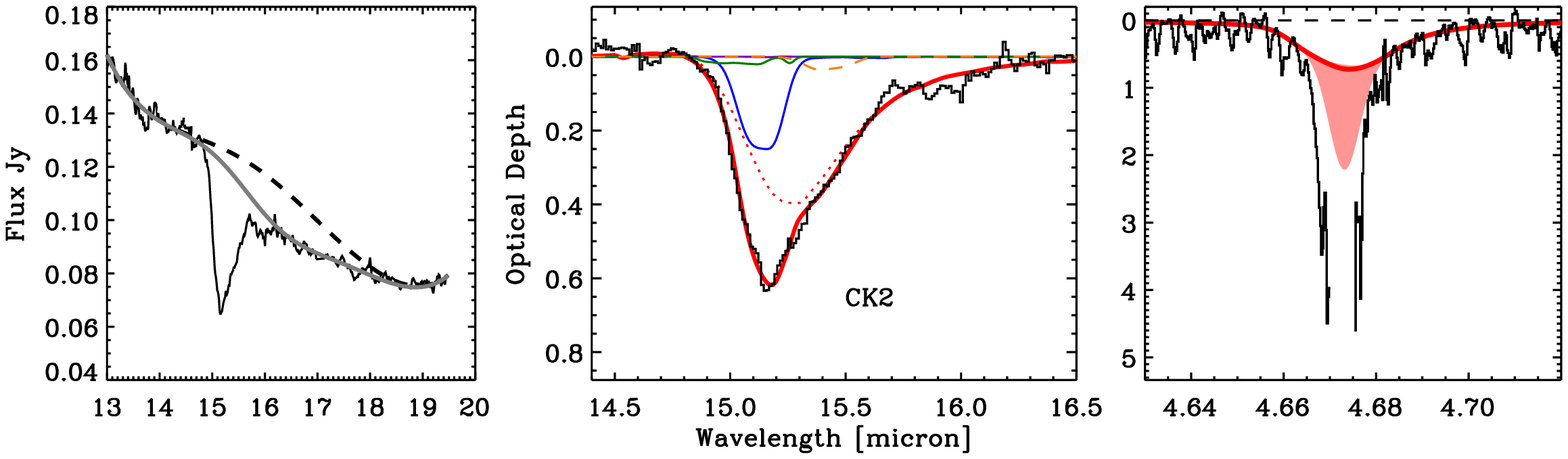}
  \includegraphics[width=15cm]{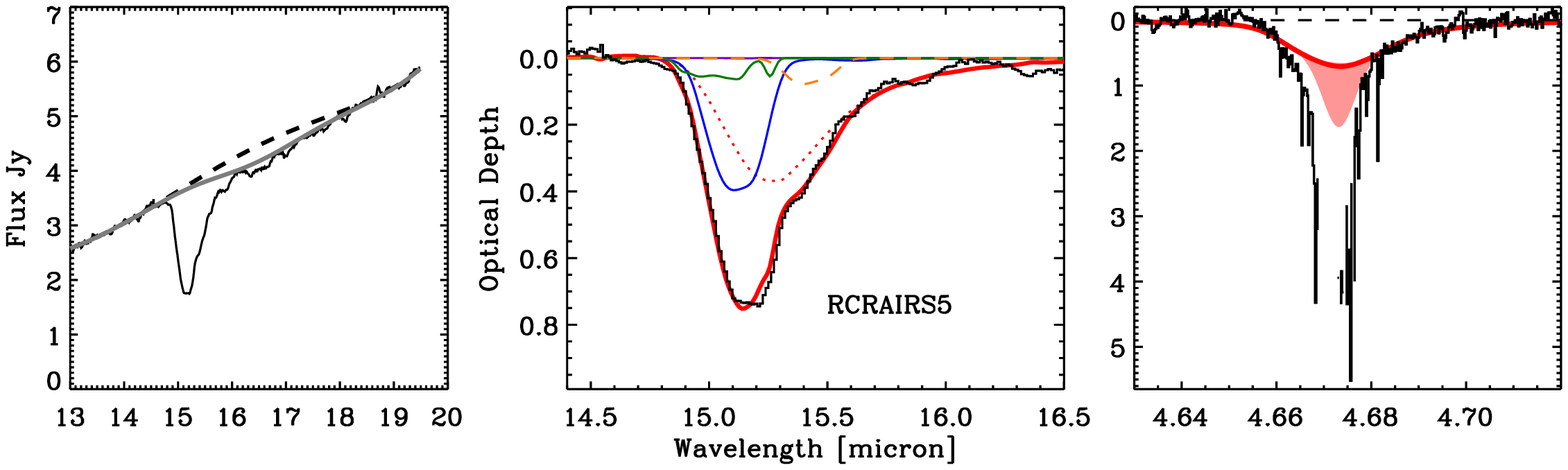}
  \includegraphics[width=15cm]{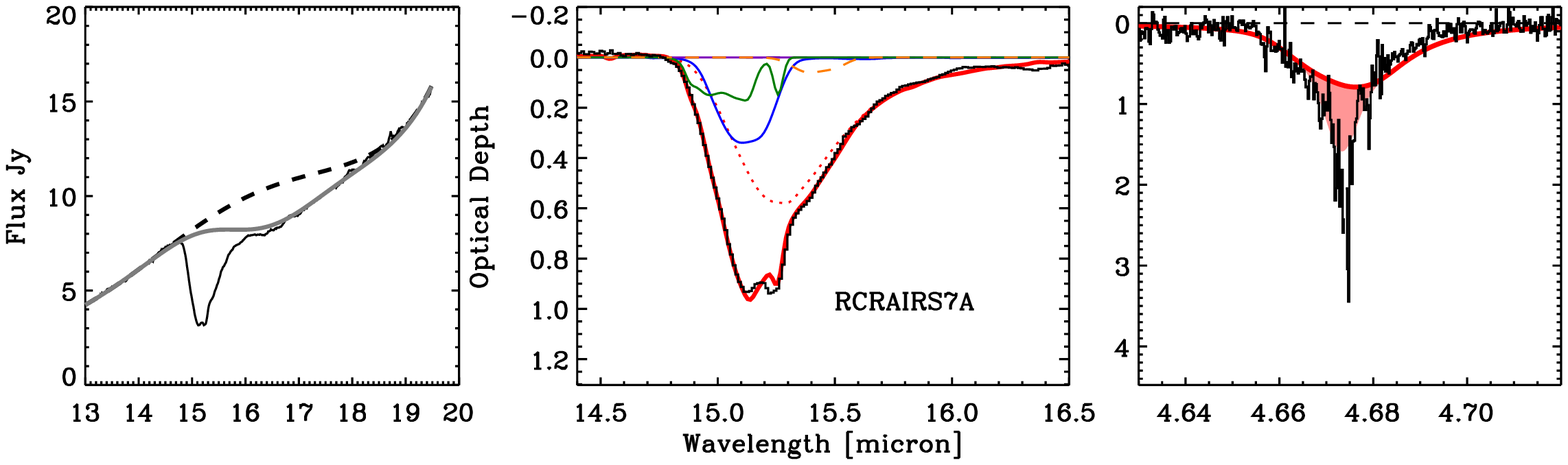}
  \includegraphics[width=15cm]{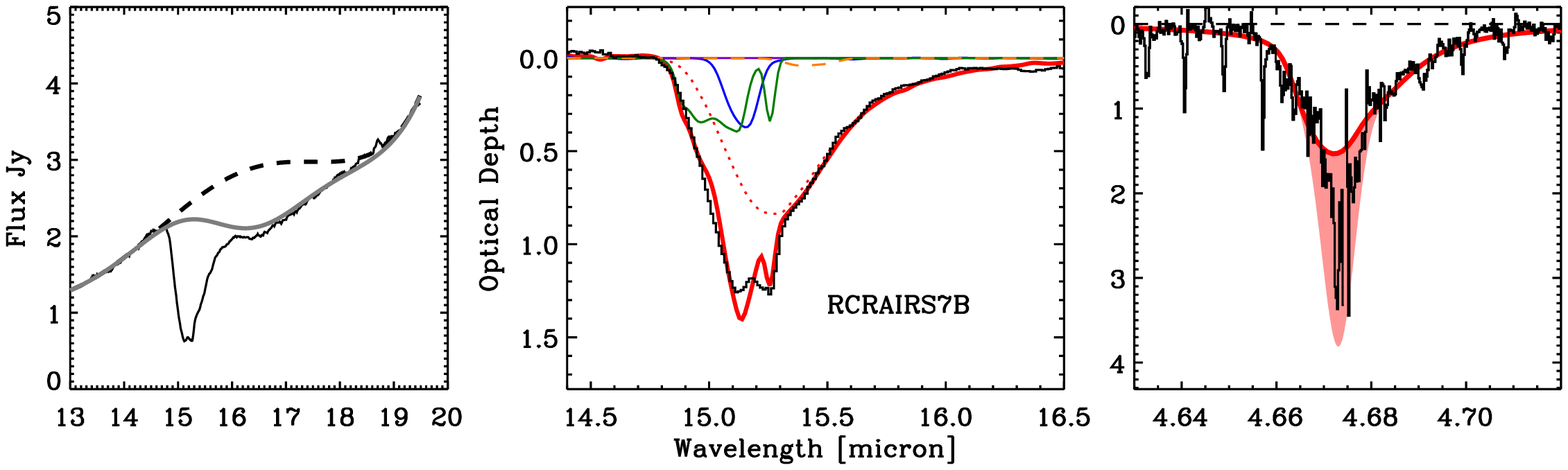}
  \caption{As Figure \ref{decomp-1}.}
  \label{decomp-12}
\end{figure*}

\begin{figure*}[h]
    
  \includegraphics[width=15cm]{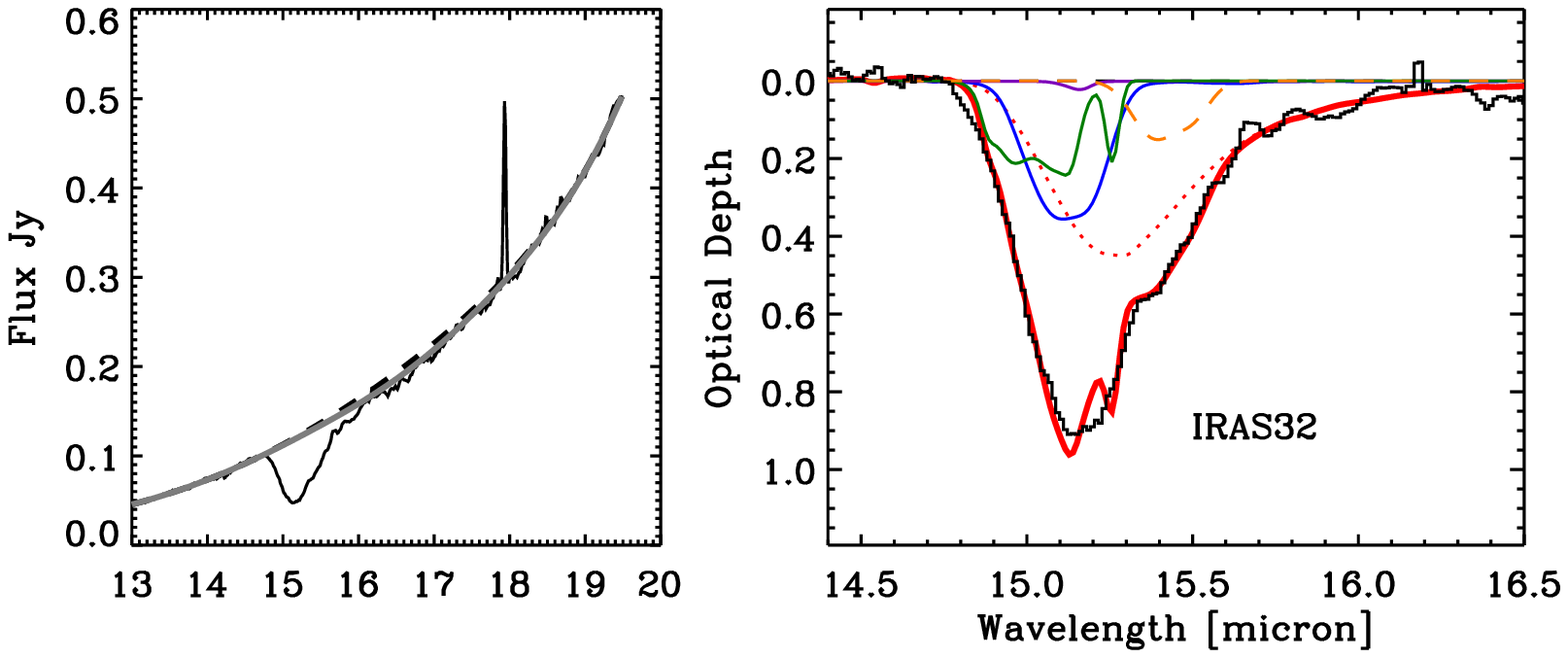}
  \includegraphics[width=15cm]{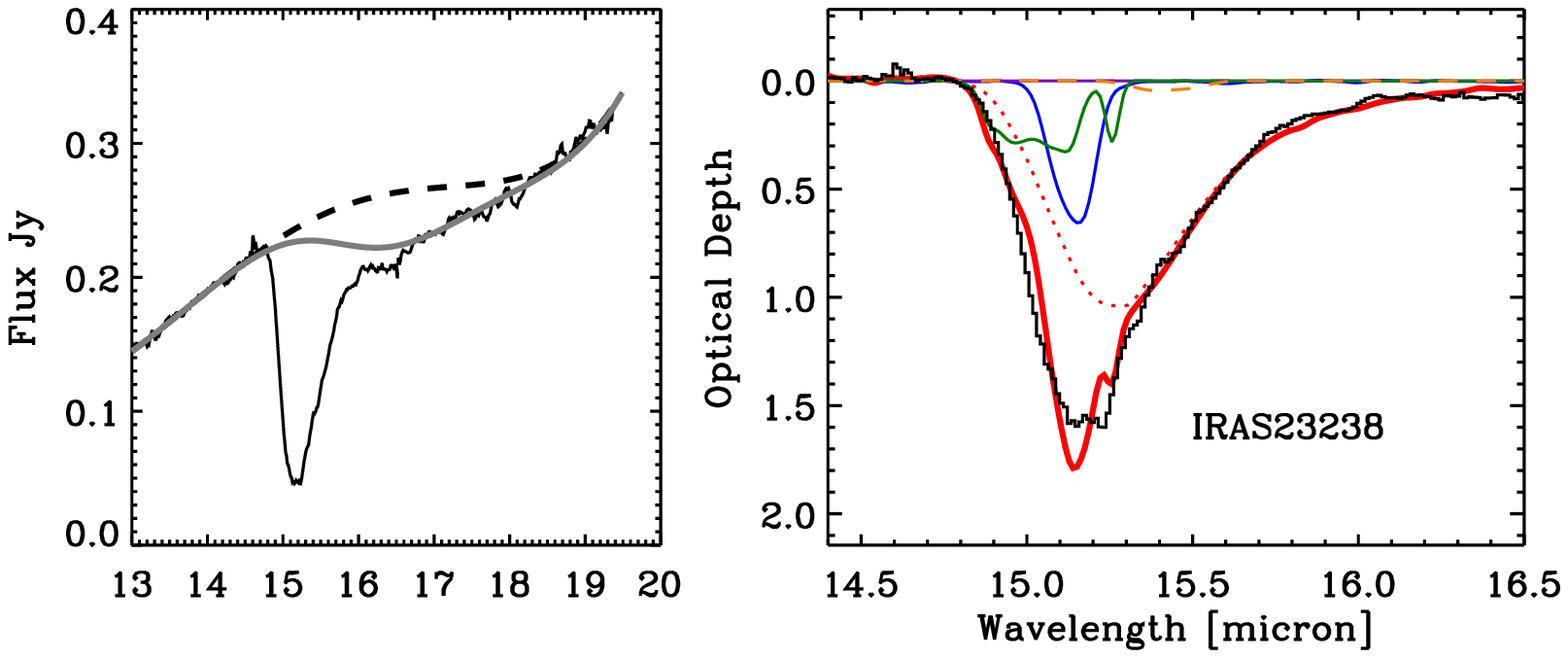}
  \caption{As Figure \ref{decomp-1}.}
  \label{decomp-13}
\end{figure*}

\begin{figure*}[h]
    
  \includegraphics[width=15cm]{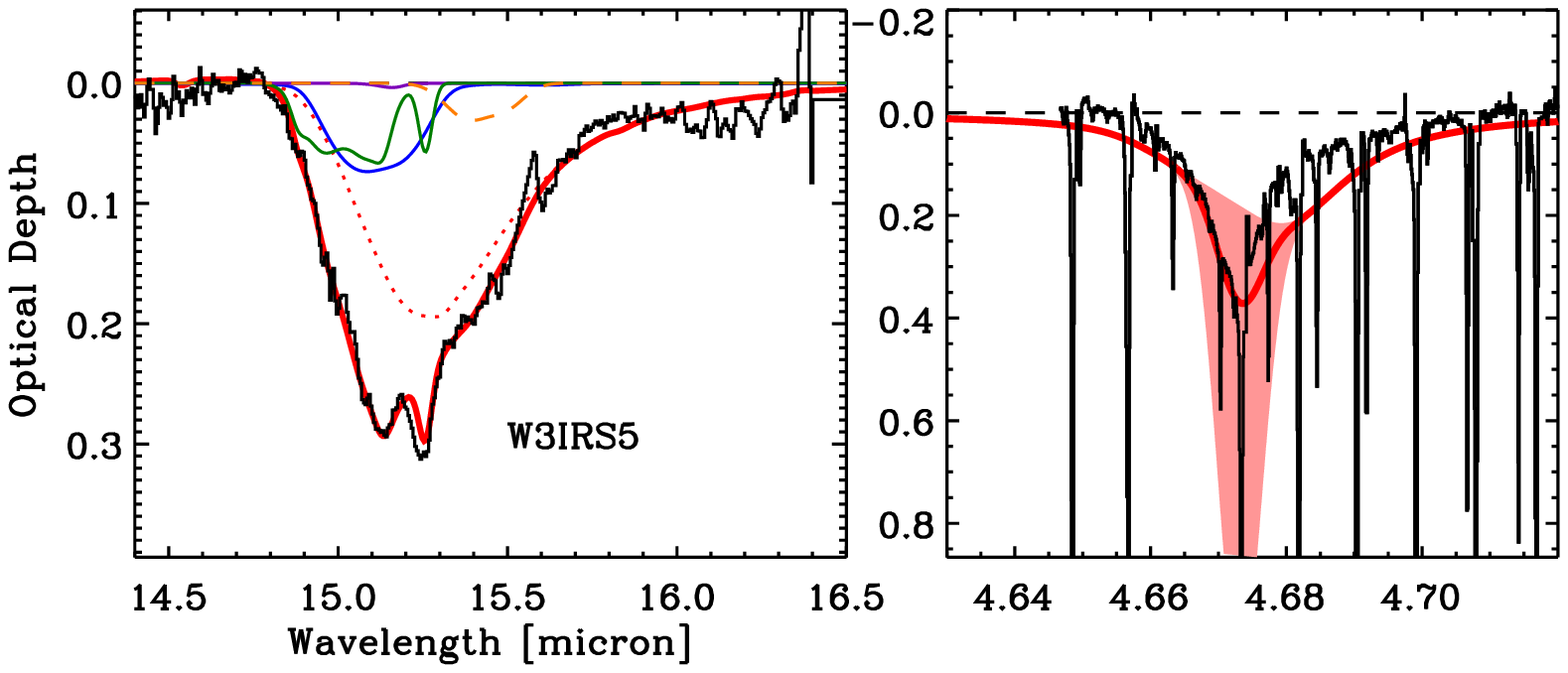}
  \includegraphics[width=15cm]{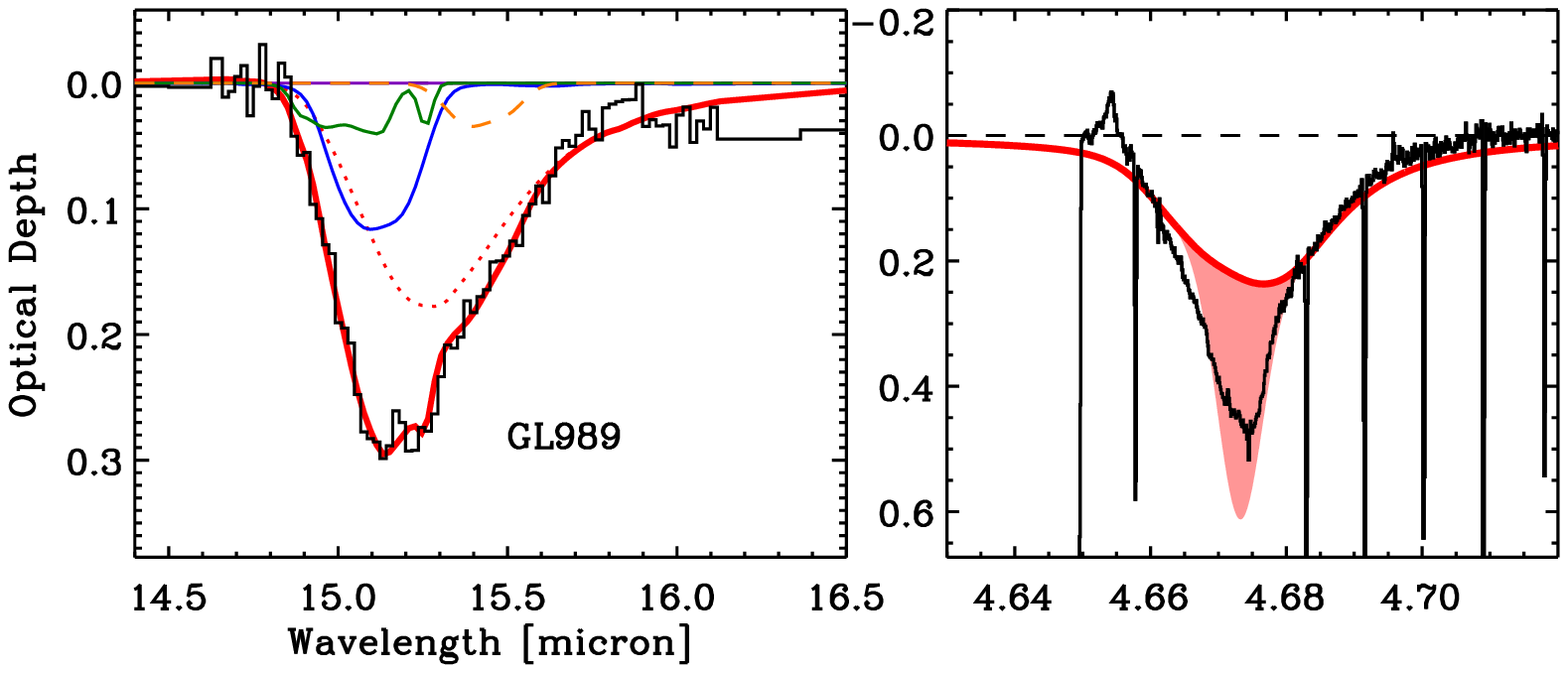}
  \includegraphics[width=15cm]{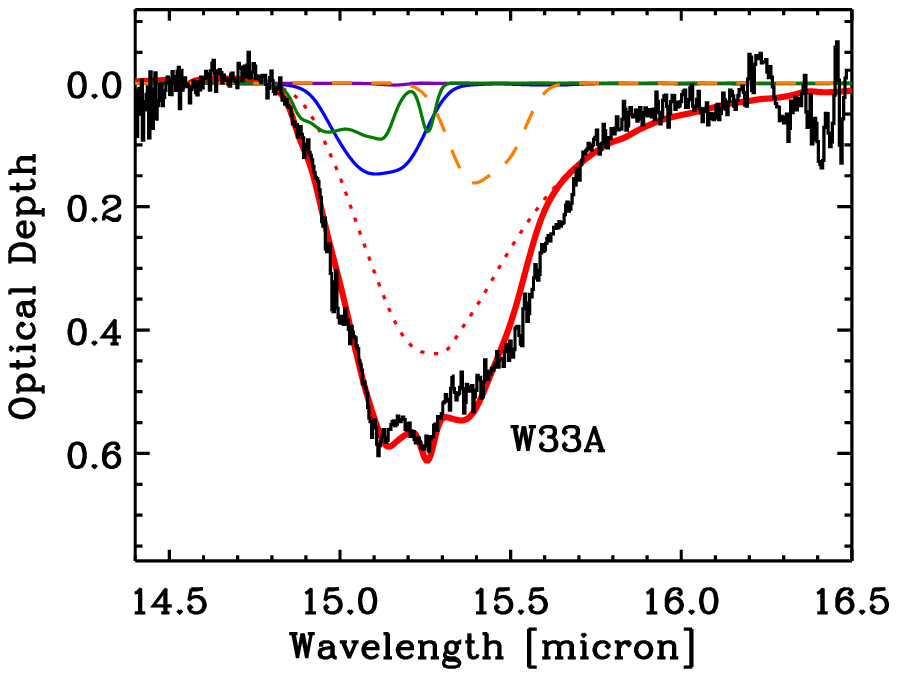}
  \caption{As Figure \ref{decomp-1}, but for archival ISO spectra of massive young stars from \cite{Gerakines99}.}
  \label{decomp-14}
\end{figure*}

\begin{figure*}[h]
    
  \includegraphics[width=15cm]{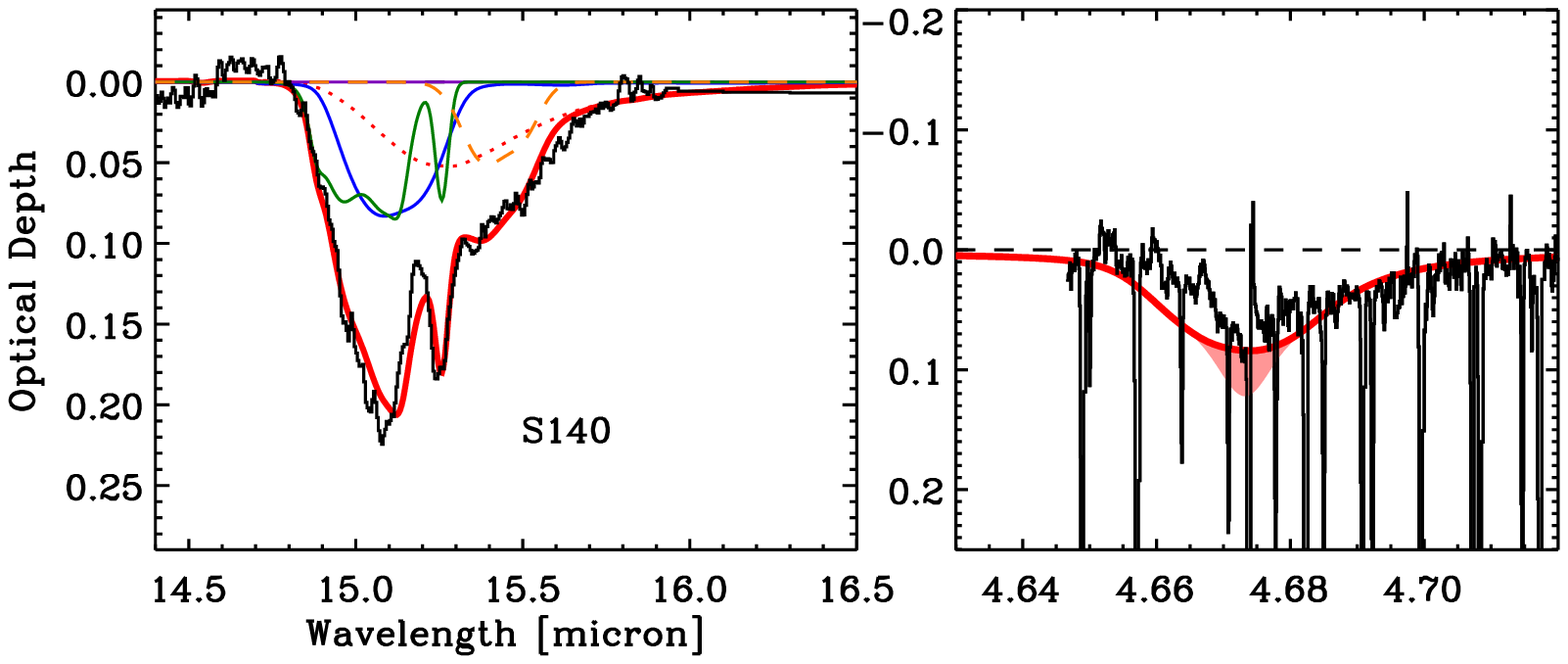}
  \includegraphics[width=15cm]{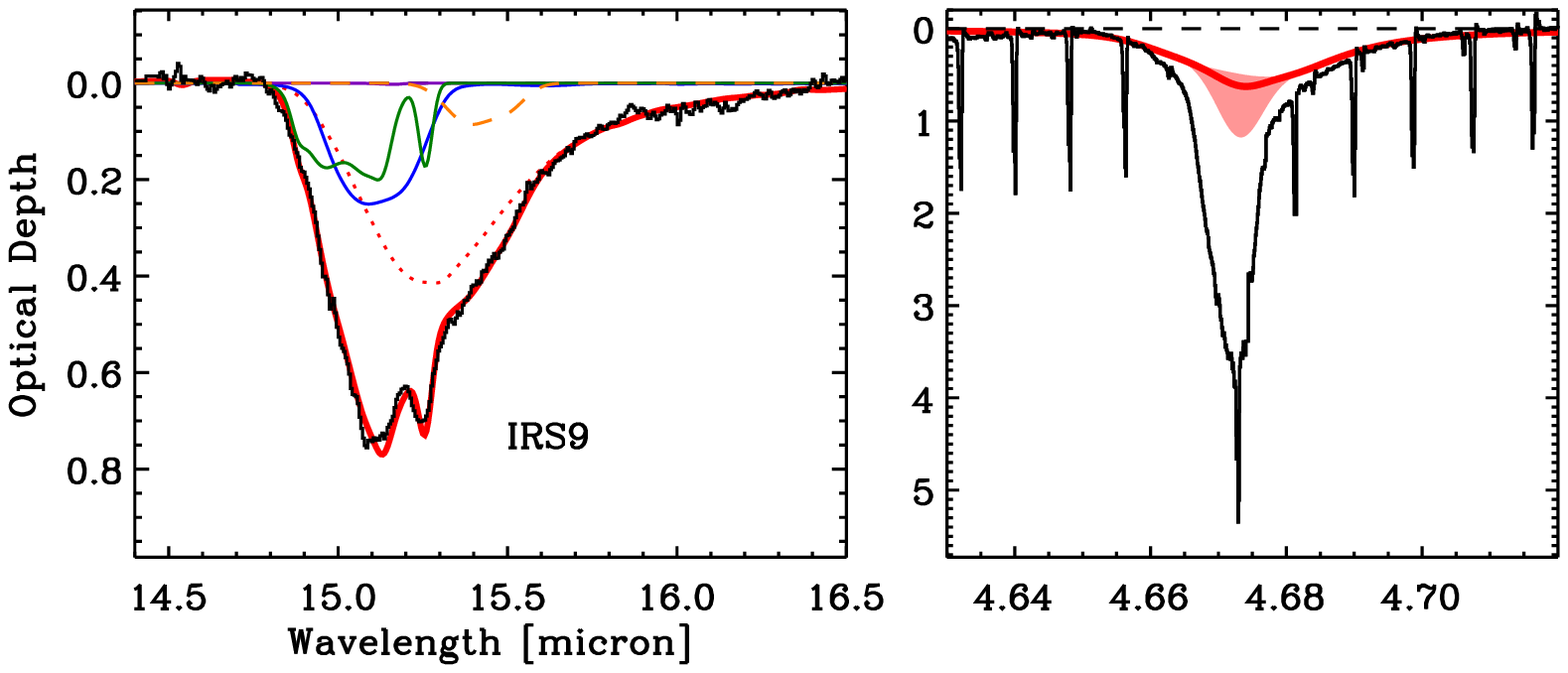}
  \includegraphics[width=15cm]{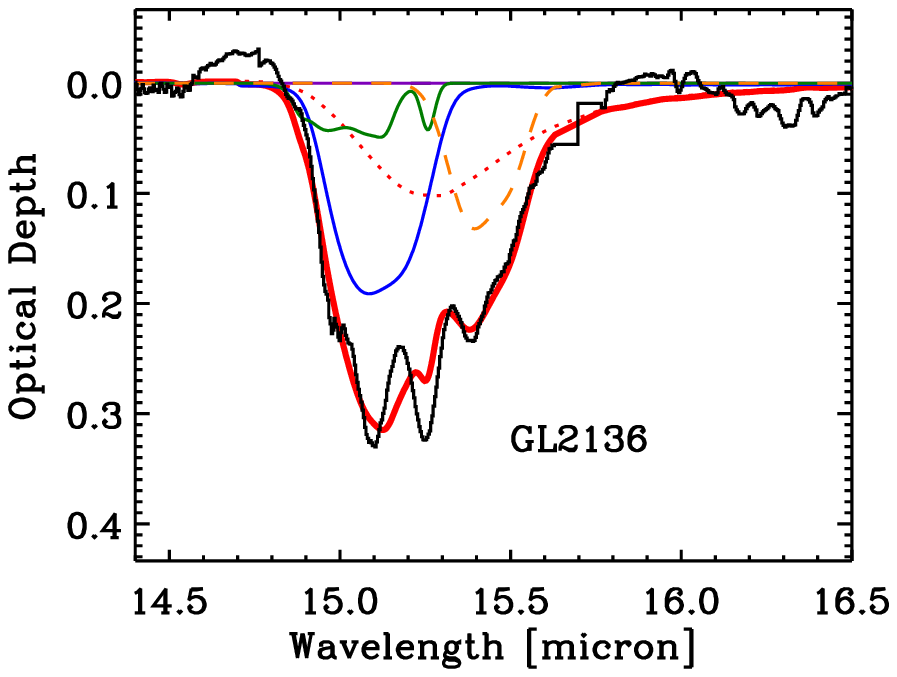}
   \caption{As Figure \ref{decomp-1}, but for archival ISO spectra of massive young stars from \cite{Gerakines99}.}
  \label{decomp-15}
\end{figure*}

A non-linear least squares fitting routine from the IDL library of C. Markwardt \footnote{http://cow.physics.wisc.edu/$\sim$craigm/idl/idl.html} 
is used to find the best fit to each bending mode spectrum. Since three of the five components used to interpret the CO$_2$ spectrum include CO, 
each fit in part predicts a corresponding CO stretching mode spectrum. This information is used to construct a model CO ice spectrum that can
be compared directly with the observed 4.67\,$\mu$m CO bands. The model CO profiles are therefore only fitted to the CO$_2$ bending mode 
profiles and are plotted on the observed
CO profiles for comparison only. In summary, each 15.2\,$\mu$m CO$_2$ bending mode
is fitted with a function with 6 free parameters; the depth of the 5 components and the mixing ratio of CO$_2$ to CO for the blue component.
The spectra and best fits are shown in Figures \ref{decomp-1} to \ref{decomp-15}, along
with the CO stretching mode bands, where available. The CO$_2$ column densities are given in Table \ref{bigtable}.

It is found that the ``blue'' CO stretching mode band corresponding to the ``blue'' CO$_2$:CO 
mixture was in general much too deep to fit the observed 
CO bands and it has consequently been reduced by a factor of 3 relative to the CO$_2$ bending mode in the comparison plots for all the sources. This discrepancy is
discussed in Section \ref{COCO2system}. In this paper, the same band strength of 1.1$\times 10^{-17}\,\rm cm\,molecule^{-1}$ is used for every component of CO$_2$. 
Note that while \cite{Gerakines95} measure a larger band strength for the CO$_2$ bending mode in a water-rich environment of 1.5$\times 10^{-17}\,\rm cm\,molecule^{-1}$, they also
state that a large uncertainty is associated with this measurement. The effect of using this band strength for CO$_2$ in a water-rich mixture would be
to decrease the CO$_2$ column density of the red component by 36\%.

\begin{table*}[h]
 
\caption{Ice column densities of the CO$_2$ components$^a\,^b$}
\begin{tabular}{llllllllll}
\hline
\hline
Source & total CO$_2$ & CO$_2$:H$_2$O & CO$_2$:CO $\sim$1:1  & CO$_2$:CO$\sim$1:25 & Pure CO$_2$ & shoulder & CO:CO$_2$ ratio & $\chi^2$  & H$_2$O$^c$ \vspace{0.1cm} \\

\hline
W3 IRS5        &  6.56  $\pm$	  0.20    &   4.41$\pm$  0.25	  &  0.93 $\pm$   0.20      & 0.02 $\pm$ 0.08	  &  0.83 $\pm$  0.12	 & 0.27 $\pm$	0.02   & 0.82$\pm$ 0.052 & 0.49  &  56.5  $\pm$    6.0    \\
L1448 IRS1     &  2.14  $\pm$	  0.06    &   1.46$\pm$  0.11	  &  0.32 $\pm$   0.10      & 0.06 $\pm$ 0.03	  &  0.19 $\pm$  0.06	 & 0.09 $\pm$	0.01   & $>$1.00           & 1.41  &   4.7  $\pm$    1.6    \\
L1448 NA       & 40.92  $\pm$	  0.35    &  32.76$\pm$  0.46	  &  3.79 $\pm$   0.23      & 0.00 $\pm$ 0.01	  &  4.21 $\pm$  0.12	 & 0.43 $\pm$	0.01   & $<$0.16           & 1.82  &  ...                   \\
L1455 SMM1     & 63.48  $\pm$	  4.43    &  45.63$\pm$  1.78	  &  5.72 $\pm$   1.68      & 0.17 $\pm$ 1.06	  &  7.96 $\pm$  0.49	 & 1.57 $\pm$	0.05   & $<$0.16           & 0.45  & 182.0  $\pm$   28.2    \\
RNO 15	       &  2.57  $\pm$	  0.05    &   2.10$\pm$  0.08	  &  0.43 $\pm$   0.04      & 0.02 $\pm$ 0.02	  &  0.00 $\pm$  0.05	 & 0.01 $\pm$	0.05   & $>$1.00           & 1.25  &   6.9  $\pm$    0.6    \\
IRAS 03254     &  8.86  $\pm$	  0.10    &   4.63$\pm$  0.18	  &  2.09 $\pm$   0.12      & 0.00 $\pm$ 0.01	  &  1.51 $\pm$  0.09	 & 0.55 $\pm$	0.01   & 0.62$\pm$ 0.030 & 1.32  &  40.5  $\pm$    3.7    \\
IRAS 03271     & 15.37  $\pm$	  0.09    &  10.65$\pm$  0.16	  &  2.56 $\pm$   0.11      & 0.04 $\pm$ 0.07	  &  1.68 $\pm$  0.06	 & 0.52 $\pm$	0.01   & 0.41$\pm$ 0.012 & 2.11  &  76.9  $\pm$   17.6    \\
B 1a	       & 20.85  $\pm$	  0.14    &  14.23$\pm$  0.25	  &  4.25 $\pm$   0.19      & 0.31 $\pm$ 0.10	  &  1.66 $\pm$  0.11	 & 0.37 $\pm$	0.01   & 0.65$\pm$ 0.009 & 1.31  & 104.0  $\pm$   23.0    \\
B 1c	       & 84.55  $\pm$	 15.70    &  68.10$\pm$  3.72	  & 13.95 $\pm$   2.02      & 0.00 $\pm$ 1.37	  &  0.00 $\pm$  0.10	 & 2.50 $\pm$	0.11   & $>$1.00           & 0.35  & 296.0  $\pm$   57.0    \\
IRAS 03439     &  3.32  $\pm$	  0.06    &   2.23$\pm$  0.11	  &  0.55 $\pm$   0.09      & 0.00 $\pm$ 0.03	  &  0.37 $\pm$  0.06	 & 0.17 $\pm$	0.01   & 0.76$\pm$ 0.082 & 1.63  &  10.1  $\pm$    0.9    \\
IRAS 03445     &  7.07  $\pm$	  0.09    &   5.31$\pm$  0.17	  &  1.75 $\pm$   0.12      & 0.00 $\pm$ 0.06	  &  0.00 $\pm$  0.07	 & 0.02 $\pm$	0.07   & 0.52$\pm$ 0.018 & 1.12  &  22.6  $\pm$    2.8    \\
L 1489	       & 16.20  $\pm$	  0.09    &  11.40$\pm$  0.16	  &  1.80 $\pm$   0.12      & 0.00 $\pm$ 0.07	  &  2.66 $\pm$  0.07	 & 0.04 $\pm$	0.07   & 0.44$\pm$ 0.036 & 1.95  &  47.0  $\pm$    2.8    \\
DG Tau B       &  5.40  $\pm$	  0.06    &   3.64$\pm$  0.11	  &  1.09 $\pm$   0.09      & 0.00 $\pm$ 0.01	  &  0.56 $\pm$  0.06	 & 0.12 $\pm$	0.07   & $>$1.00           & 0.86  &  26.3  $\pm$    2.6    \\
GL 989	       &  6.11  $\pm$	  0.07    &   3.97$\pm$  0.19	  &  1.37 $\pm$   0.13      & 0.00 $\pm$ 0.04	  &  0.51 $\pm$  0.09	 & 0.30 $\pm$	0.02   & 0.64$\pm$ 0.013 & 2.53  &  23.2  $\pm$    1.1    \\
HH 46	       & 21.58  $\pm$	  0.11    &  16.99$\pm$  0.20	  &  2.25 $\pm$   0.12      & 0.00 $\pm$ 0.09	  &  1.90 $\pm$  0.06	 & 0.49 $\pm$	0.01   & 0.37$\pm$ 0.016 & 1.34  &  77.9  $\pm$    7.3    \\
CED 110 IRS4   & 12.26  $\pm$	  0.12    &   8.34$\pm$  0.22	  &  2.10 $\pm$   0.17      & 0.01 $\pm$ 0.09	  &  1.49 $\pm$  0.10	 & 0.29 $\pm$	0.02   & 0.56$\pm$ 0.009 & 1.17  &  ...                   \\
B 35	       &  4.90  $\pm$	  0.15    &   3.48$\pm$  0.27	  &  0.38 $\pm$   0.20      & 0.09 $\pm$ 0.09	  &  0.56 $\pm$  0.14	 & 0.37 $\pm$	0.02   & 0.70$\pm$ 0.223 & 1.29  &  ...                   \\
CED 110 IRS6   & 14.30  $\pm$	  0.08    &  11.49$\pm$  0.14	  &  1.92 $\pm$   0.08      & 0.01 $\pm$ 0.07	  &  0.78 $\pm$  0.04	 & 0.07 $\pm$	0.04   & 0.28$\pm$ 0.005 & 1.90  &  47.0  $\pm$    6.0    \\
IRAS 12553     &  5.98  $\pm$	  0.09    &   4.84$\pm$  0.16	  &  1.07 $\pm$   0.11      & 0.00 $\pm$ 0.05	  &  0.04 $\pm$  0.07	 & 0.04 $\pm$	0.07   & 0.63$\pm$ 0.014 & 0.59  &  29.8  $\pm$    5.6    \\
ISO ChaII 54   &  1.81  $\pm$	  0.13    &   0.50$\pm$  0.25	  &  0.96 $\pm$   0.21      & 0.17 $\pm$ 0.07	  &  0.16 $\pm$  0.13	 & 0.22 $\pm$	0.03   & $>$1.00           & 1.23  &  ...                   \\
IRAS 13546     &  8.72  $\pm$	  0.12    &   6.22$\pm$  0.21	  &  2.00 $\pm$   0.16      & 0.01 $\pm$ 0.09	  &  0.30 $\pm$  0.09	 & 0.08 $\pm$	0.01   & 0.54$\pm$ 0.008 & 0.86  &  20.7  $\pm$    2.0    \\
IRAS 15398     & 52.16  $\pm$	  0.79    &  38.83$\pm$  0.90	  &  0.29 $\pm$   0.67      & 1.85 $\pm$ 0.59	  & 10.58 $\pm$  0.29	 & 0.80 $\pm$	0.02   & 0.16$\pm$ 0.000 & 1.34  & 148.0  $\pm$   39.5    \\
GSS 30 IRS1    &  3.28  $\pm$	  0.06    &   1.86$\pm$  0.10	  &  0.70 $\pm$   0.07      & 0.00 $\pm$ 0.05	  &  0.61 $\pm$  0.05	 & 0.09 $\pm$	0.05   & 0.77$\pm$ 0.078 & 1.04  &  15.3  $\pm$    3.0    \\
WL 12	       &  4.34  $\pm$	  0.05    &   2.72$\pm$  0.10	  &  1.36 $\pm$   0.08      & 0.04 $\pm$ 0.03	  &  0.03 $\pm$  0.05	 & 0.20 $\pm$	0.05   & $>$1.00           & 1.28  &  22.1  $\pm$    3.0    \\
GY 224	       &  2.69  $\pm$	  0.09    &   1.90$\pm$  0.17	  &  0.66 $\pm$   0.12      & 0.00 $\pm$ 0.01	  &  0.03 $\pm$  0.09	 & 0.15 $\pm$	0.04   & 0.65$\pm$ 0.057 & 1.28  & ...                    \\
WL 20S	       &  5.02  $\pm$	  0.06    &   3.86$\pm$  0.11	  &  0.75 $\pm$   0.08      & 0.00 $\pm$ 0.02	  &  0.28 $\pm$  0.05	 & 0.11 $\pm$	0.05   & 0.64$\pm$ 0.038 & 0.95  & ...                    \\
IRS 37	       &  4.05  $\pm$	  0.08    &   2.99$\pm$  0.14	  &  0.99 $\pm$   0.12      & 0.00 $\pm$ 0.05	  &  0.01 $\pm$  0.07	 & 0.01 $\pm$	0.05   & 0.65$\pm$ 0.012 & 0.85  &  36.5  $\pm$    5.0    \\
IRS 42	       &  4.49  $\pm$	  0.05    &   3.39$\pm$  0.11	  &  1.01 $\pm$   0.09      & 0.07 $\pm$ 0.03	  &  0.01 $\pm$  0.05	 & 0.09 $\pm$	0.05   & $>$1.00           & 1.01  &  19.5  $\pm$    2.0    \\
WL 6	       &  9.33  $\pm$	  0.08    &   6.86$\pm$  0.15	  &  2.17 $\pm$   0.10      & 0.00 $\pm$ 0.06	  &  0.12 $\pm$  0.06	 & 0.18 $\pm$	0.05   & 0.49$\pm$ 0.007 & 0.87  &  41.7  $\pm$    6.0    \\
CRBR 2422.8-3423& 10.54  $\pm$	  0.06    &   7.30$\pm$  0.11	  &  2.85 $\pm$   0.07      & 0.09 $\pm$ 0.04	  &  0.02 $\pm$  0.04	 & 0.36 $\pm$	0.05   & 0.48$\pm$ 0.003 & 1.52  &  45.0  $\pm$    5.0    \\
IRS 43         & 12.26  $\pm$	  0.12    &   8.11$\pm$	 0.23  	  &  1.91 $\pm$   0.16 	    & 0.00 $\pm$ 0.09     &  1.71 $\pm$  0.10    & 0.51 $\pm$   0.01   & 0.53$\pm$ 0.047 & 0.84  &  31.5  $\pm$    4.0    \\
IRS 44         &  6.92  $\pm$	  0.08    &   5.01$\pm$	 0.14  	  &  1.10 $\pm$   0.11 	    & 0.00 $\pm$ 0.01     &  0.50 $\pm$  0.08    & 0.34 $\pm$   0.01   & 0.87$\pm$ 0.040 & 1.17  &  34.0  $\pm$    4.0    \\
Elias 32       &  4.87  $\pm$	  0.09    &   2.89$\pm$  0.15	  &  1.38 $\pm$   0.12      & 0.07 $\pm$ 0.05	  &  0.55 $\pm$  0.07	 & 0.16 $\pm$	0.01   & 0.62$\pm$ 0.015 & 2.21  &  17.9  $\pm$    2.6    \\
IRS 46	       &  2.35  $\pm$	  0.12    &   1.64$\pm$  0.23	  &  0.43 $\pm$   0.20      & 0.02 $\pm$ 0.07	  &  0.09 $\pm$  0.12	 & 0.09 $\pm$	0.08   & $>$1.00           & 0.20  &  12.8  $\pm$    2.0    \\
VSSG 17        &  5.86  $\pm$	  0.11    &   3.46$\pm$  0.19	  &  2.20 $\pm$   0.14      & 0.00 $\pm$ 0.07	  &  0.01 $\pm$  0.09	 & 0.27 $\pm$	0.03   & 0.53$\pm$ 0.009 & 0.64  &  17.0  $\pm$    2.5    \\
IRS 51	       &  9.32  $\pm$	  0.07    &   5.44$\pm$  0.12	  &  3.30 $\pm$   0.09      & 0.27 $\pm$ 0.05	  &  0.07 $\pm$  0.05	 & 0.26 $\pm$	0.05   & 0.54$\pm$ 0.005 & 0.88  &  22.1  $\pm$    3.0    \\
IRS 63	       &  6.84  $\pm$	  0.05    &   4.49$\pm$  0.10	  &  2.26 $\pm$   0.05      & 0.01 $\pm$ 0.04	  &  0.00 $\pm$  0.05	 & 0.17 $\pm$	0.05   & 0.51$\pm$ 0.014 & 1.20  &  20.4  $\pm$    3.0    \\
L 1689 IRS5    &  3.37  $\pm$	  0.10    &   2.19$\pm$  0.17	  &  1.05 $\pm$   0.12      & 0.00 $\pm$ 0.06	  &  0.00 $\pm$  0.08	 & 0.22 $\pm$	0.01   & 0.58$\pm$ 0.015 & 1.26  & ...                    \\
RNO 91	       & 11.66  $\pm$	  0.16    &   5.94$\pm$  0.28	  &  2.37 $\pm$   0.20      & 0.00 $\pm$ 0.03	  &  2.70 $\pm$  0.14	 & 0.62 $\pm$	0.02   & 0.64$\pm$ 0.026 & 1.49  &  39.0  $\pm$    5.0    \\
W33A	       & 14.11  $\pm$	  0.29    &   9.96$\pm$  0.30	  &  1.70 $\pm$   0.22      & 0.02 $\pm$ 0.11	  &  1.13 $\pm$  0.13	 & 1.41 $\pm$	0.11   & 0.59$\pm$ 0.025 & 1.00  & 113.0	  28.3    \\
GL 2136        &  0.93  $\pm$	  0.03    &   0.42$\pm$	 0.03     &  0.26 $\pm$	  0.02      &  0.00$\pm$    0.00  &   0.15$\pm$    0.01  &  0.16$\pm$    0.02  & 0.57$\pm$  0.023&	0.69&	 47.2$\pm$	 4.7\\
S68N	       & 43.27  $\pm$	 16.54    &  30.99$\pm$  0.75	  &  5.38 $\pm$   1.18      & 0.09 $\pm$ 1.01	  &  6.81 $\pm$  0.38	 & 0.00 $\pm$	0.50   & $<$0.16           & 1.90  & ...                    \\
EC 74	       &  2.89  $\pm$	  0.08    &   2.00$\pm$  0.14	  &  0.79 $\pm$   0.07      & 0.00 $\pm$ 0.05	  &  0.00 $\pm$  0.01	 & 0.19 $\pm$	0.01   & 0.44$\pm$ 0.032 & 1.90  &  10.7  $\pm$    1.8    \\
SVS 4-5	       & 17.21  $\pm$	  0.10    &  12.37$\pm$  0.18	  &  3.48 $\pm$   0.13      & 0.00 $\pm$ 0.08	  &  0.68 $\pm$  0.07	 & 0.73 $\pm$	0.02   & 0.43$\pm$ 0.009 & 1.14  &  56.5  $\pm$   11.3    \\
EC 82	       &  2.54  $\pm$	  0.04    &   1.88$\pm$  0.08	  &  0.07 $\pm$   0.07      & 0.12 $\pm$ 0.02	  &  0.34 $\pm$  0.04	 & 0.08 $\pm$	0.05   & $>$1.00           & 2.18  &   3.9  $\pm$    0.7    \\
EC 90	       &  5.44  $\pm$	  0.05    &   3.53$\pm$  0.09	  &  1.67 $\pm$   0.07      & 0.12 $\pm$ 0.03	  &  0.01 $\pm$  0.05	 & 0.17 $\pm$	0.02   & 0.81$\pm$ 0.011 & 1.39  &  16.9  $\pm$    1.6    \\
SVS 4-10	       &  8.25  $\pm$	  0.05    &   5.37$\pm$  0.09	  &  2.35 $\pm$   0.07      & 0.07 $\pm$ 0.04	  &  0.12 $\pm$  0.04	 & 0.35 $\pm$	0.01   & 0.57$\pm$ 0.004 & 1.51  &  16.0  $\pm$    1.4    \\
CK 4	       &  1.98  $\pm$	  0.09    &   1.20$\pm$  0.16	  &  0.62 $\pm$   0.08      & 0.00 $\pm$ 0.05	  &  0.00 $\pm$  0.05	 & 0.15 $\pm$	0.00   & 0.75$\pm$ 0.099 & 1.00  &  15.6  $\pm$   15.6    \\
CK 2	       & 11.93  $\pm$	  0.21    &   9.02$\pm$  0.38	  &  2.29 $\pm$   0.23      & 0.01 $\pm$ 0.16	  &  0.26 $\pm$  0.15	 & 0.31 $\pm$	0.01   & 0.32$\pm$ 0.030 & 0.75  &  35.7  $\pm$    3.5    \\
RCRA IRS5      & 14.28  $\pm$	  0.13    &   8.38$\pm$  0.23	  &  4.52 $\pm$   0.17      & 0.01 $\pm$ 0.10	  &  0.80 $\pm$  0.10	 & 0.68 $\pm$	0.02   & 0.58$\pm$ 0.005 & 1.86  &  37.6  $\pm$    2.8    \\
RCRA IRS7A     & 19.64  $\pm$	  0.12    &  13.17$\pm$  0.21	  &  3.89 $\pm$   0.16      & 0.00 $\pm$ 0.09	  &  2.15 $\pm$  0.08	 & 0.52 $\pm$	0.01   & 0.58$\pm$ 0.006 & 1.20  & 109.0  $\pm$   19.2    \\
RCRA IRS7B     & 26.74  $\pm$	  0.22    &  19.02$\pm$  0.35	  &  2.38 $\pm$   0.24      & 0.00 $\pm$ 0.25	  &  4.94 $\pm$  0.11	 & 0.36 $\pm$	0.10   & $<$0.16           & 2.26  & 110.0  $\pm$   19.7    \\
IRAS 32        & 18.70  $\pm$	  0.21    &  10.21$\pm$  0.36	  &  4.00 $\pm$   0.28      & 0.13 $\pm$ 0.16	  &  3.04 $\pm$  0.16	 & 1.32 $\pm$	0.12   & 0.56$\pm$ 0.014 & 1.52  &  52.6  $\pm$   18.8    \\
S 140	       &  3.78  $\pm$	  0.07    &   1.17$\pm$  0.10	  &  1.06 $\pm$   0.07      & 0.00 $\pm$ 0.00	  &  1.06 $\pm$  0.04	 & 0.44 $\pm$	0.13   & 0.85$\pm$ 0.024 & 1.25  &  19.3  $\pm$    1.9    \\
NGC 7538 IRS 9 & 15.81  $\pm$	  0.13    &   9.41$\pm$  0.15	  &  3.03 $\pm$   0.12      & 0.01 $\pm$ 0.06	  &  2.51 $\pm$  0.08	 & 0.75 $\pm$	0.97   & 0.67$\pm$ 0.009 & 0.89  &  64.1  $\pm$    6.4    \\
IRAS 23238     & 32.51  $\pm$	  0.30    &  23.66$\pm$  0.42	  &  4.20 $\pm$   0.20      & 0.00 $\pm$ 0.02	  &  4.10 $\pm$  0.13	 & 0.40 $\pm$	0.01   & $<$0.16           & 3.02  & 130.0  $\pm$   22.6    \\

\hline			  
\end{tabular}
\begin{itemize}
\item[$^a$] All column densities are in $10^{17}\,\rm cm^{-2}$.
\item[$^b$] All uncertainties are statistical, and do not include systematic uncertainties from e.g. the continuum determination. 
\item[$^c$] For consistency, the new water ice column densities from Paper I are used, where available. They are generally consistent with the few published
values measured on the same, or similar, data sets, but with a few values differing by 15-20\%.  
\end{itemize}
\label{bigtable}
\end{table*}

\section{Relations of the CO$_2$ components}

\subsection{The abundance of CO$_2$ ice in low-mass YSO envelopes}
\label{CO2tot}
\begin{figure}
    
  \includegraphics[width=8cm]{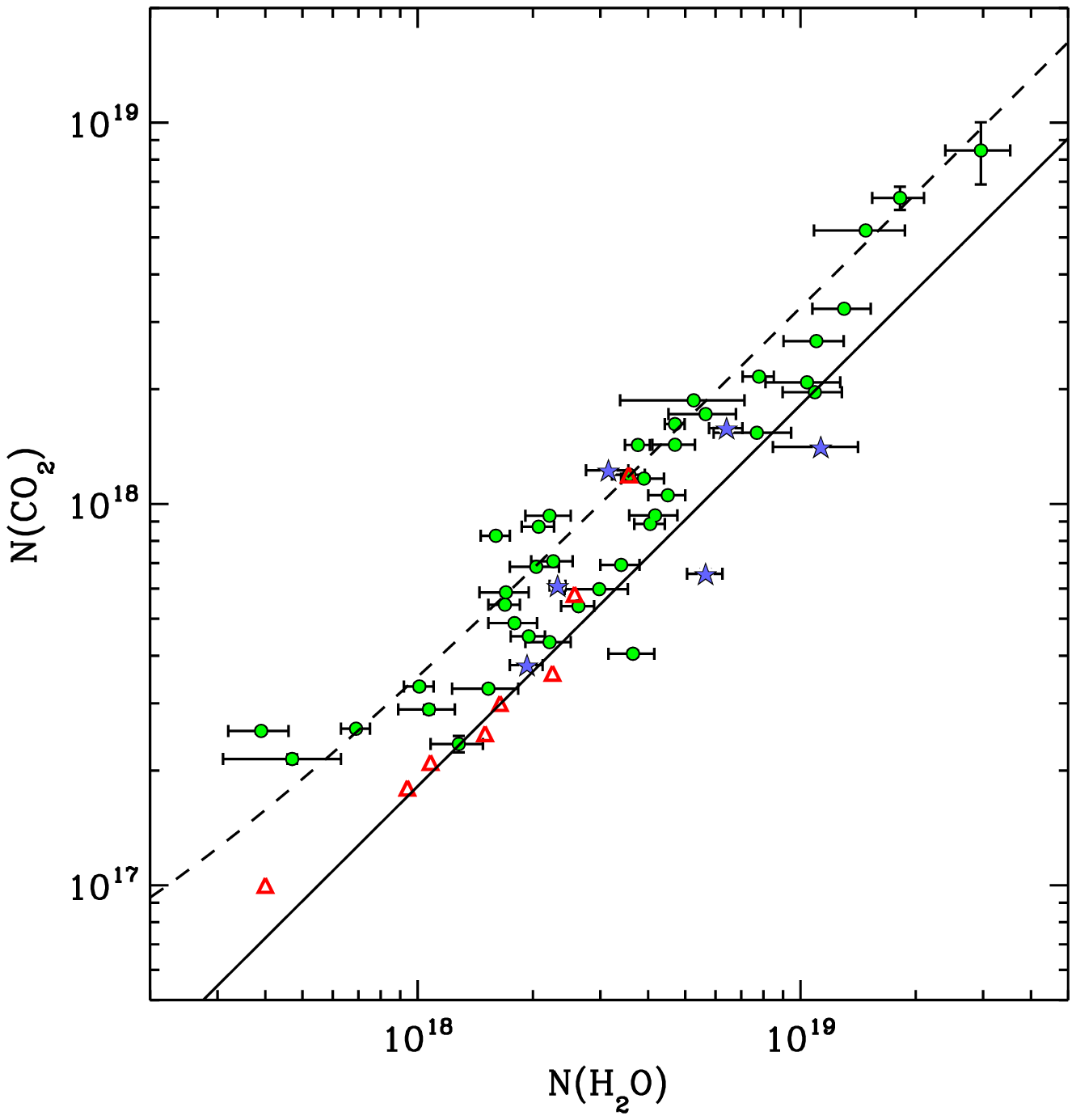}
  \caption{Relation between H$_2$O ice column density and CO$_2$ ice column density. The green points
are low-mass stars with luminosities of 0.1-100\,$\L_{\odot}$, while the blue stars indicate
the massive YSOs from the ISO sample with luminosities of $L=10^4-10^5\,L_{\odot}$. 
The red triangles are the background stars in Taurus from \cite{Whittet07}, as well as CK2 from \cite{Knez05}.The dashed line is
a linear least squares fit to the low-mass stars.  The
solid curve is the linear fit to the background stars. }
  \label{CO2_H2O}
\end{figure}

The relation between the observed H$_2$O ice column densities and the total CO$_2$ ice column densities are shown in Figure \ref{CO2_H2O}. 
Because it is difficult to measure the column density of H$_2$ gas along the line of sight, this is the relation typically used to define {\it ice abundance}
of a given solid-state species as a number fraction relative to water ice. The CO$_2$ ice abundance 
is remarkably constant, but does exhibit a scatter that is much larger than the uncertainties. A linear
fit to the low-mass stars reveals a number ratio of CO$_2$ to H$_2$O to be $0.32\pm0.02$, with a Pearson correlation coefficient of 96\%. There
is, however, a significant scatter in the relation, and a number of our sources show abundances between 0.2 and 0.3, relative to H$_2$O.  The relation
exhibits a slight tendency for higher CO$_2$ abundances at higher H$_2$O column densities. 

The CO$_2$ abundance derived here can be 
compared with that of 0.17 for the ISO sample of massive YSOs \citep{Gerakines99} and 0.18$\pm$0.04 for quiescent clouds as observed toward background stars
\citep{Whittet07}. Both these samples have been included in Figure \ref{CO2_H2O}. The points associated with the massive YSOs 
thus indicate a significantly lower CO$_2$ abundance. 
Inspection of the figure also reveals that the background stars generally
probe lower column densities than the YSOs. For these low column densities, the difference between the YSOs and the background stars is less significant. 
However, at water ice column densities higher than $\sim 2\times 10^{18}\,\rm cm^{-2}$, the difference in CO$_2$ abundance between background stars at low A$_V$ and 
low-mass YSOs is highly significant. This sudden change in the relation may
represent the activation of a new formation route to CO$_2$ in addition to that forming the H$_2$O:CO$_2$ mantle at lower A$_V$. 
An increase of CO$_2$ abundance in denser regions of a single cloud core was observed by \cite{Pontoppidan06}, and was also
discussed for the background stars in \cite{Whittet07}, based on a single star at high A$_V$.

\subsection{The CO$_2$:H$_2$O system}

\begin{figure}
    
  \includegraphics[width=8cm]{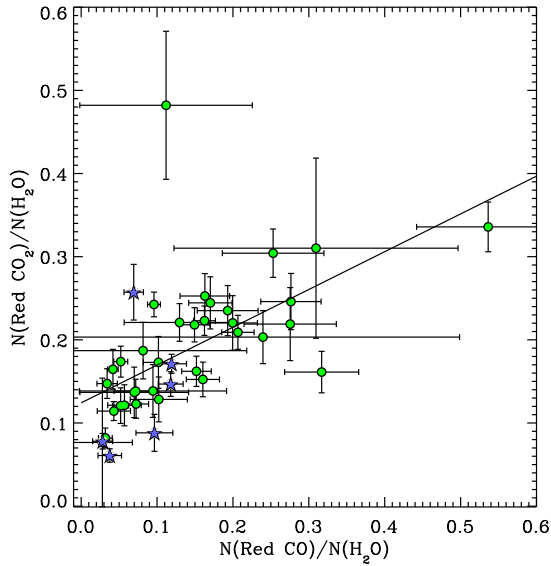}
  \caption{Correlation between the CO$_2$:H$_2$O component and the red CO component (CO:H$_2$O) from \cite{Pontoppidan03}.  The green
points are the low-mass stars, while the blue points are the high mass YSOs observed with ISO. The line is a linear fit to the low-mass YSOs.}
  \label{CO2:H2O}
\end{figure}

The CO$_2$:H$_2$O component dominates every CO$_2$ bending mode band, and the abundance variation of this component, or lack thereof, therefore mimics that of the total CO$_2$ band discussed in
section \ref{CO2tot}. In the following, ``abundance'' refers to an observed column density ratio averaged over the line of sight, while ``concentration'' is the 
local point number density of a species relative to the total number density of molecules in an ice film. The abundance of this component varies between 0.1 and 0.3 relative to water, with
a median value around 0.2. If the concentration equals the abundance, the relevant mixing ratio for a laboratory analog is CO$_2$:H$_2$O=1:($5^{+5}_{-2}$). The observed 
abundances of the CO$_2$:H$_2$O component are shown in Figure \ref{CO2:H2O} as a function of the corresponding CO:H$_2$O component of the CO stretching mode. 

The general value of this parameter is of some importance. \cite{Oberg07} showed that the band strengths of the various H$_2$O modes
are very sensitive to the CO$_2$ concentration, and they suggested that this may be an explanation for the discrepancy in 
observed band depths between the H$_2$O stretching and bending modes. The observed abundances of the CO$_2$:H$_2$O component suggest that 
the influence of CO$_2$ may explain a part of the H$_2$O bending/stretching mode discrepancy. This is discussed in greater detail in Paper I.

%\clearpage

\clearpage

\subsection{The CO$_2$:CO system}
\label{COCO2system}

\subsubsection{The ``blue'' component}
The component fit indicates that the CO$_2$:CO system forms a component separate from the CO$_2$ mixed with water. The evidence is
the higher abundance of the CO dominated component in dense, cold cloud regions as discussed in \cite{Pontoppidan06}, and as suggested by the offset of
low-mass protostellar envelopes relative to background stars and massive YSOs in Figure
\ref{CO2_H2O}, as well as the match of the profile of this blue CO$_2$ component with laboratory simulations of
CO$_2$:CO mixtures. An important parameter to determine is the concentration of CO$_2$ relative to CO in this component. There are at least two
ways of doing this. First, the concentration can be directly determined by comparing the observed strength of the blue CO$_2$ component with the corresponding
blue component of CO observed as part of the CO stretching mode band. Second, it can be indirectly inferred by the profile 
of the CO$_2$:CO component of the CO$_2$ bending mode, 
since this is sensitive to the concentration. 

\begin{figure}
    
  \includegraphics[width=8cm]{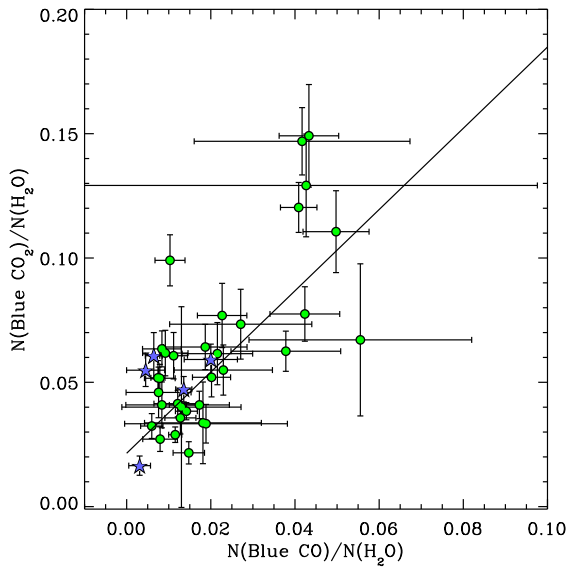}
  \caption{Correlation between the blue CO$_2$ component abundance and the blue CO component abundance from \cite{Pontoppidan03}, otherwise as Figure \ref{CO2:H2O}}
  \label{co_co2_1_1}
\end{figure}

The column density ratios of the CO$_2$ and CO ``blue'' are illustrated in Figure \ref{co_co2_1_1}. These components exhibit a fairly strong correlation
with a Pearson correlation coefficient of 0.70 and a slope of $N({\rm blue\,CO}_2)/N({\rm blue\,CO})=2.5\pm 0.2$, assuming a width of 3\,cm$^{-1}$ for
the blue CO component as empirically determined in \cite{Pontoppidan03}. The laboratory spectra of the CO stretching mode of CO$_2$:CO mixtures
are about twice as wide. Using the laboratory spectra instead to calculate the column densities of the ``blue'' CO component would decrease this ratio to 1.25$\pm$0.1. 
  
\begin{figure}
    
  \includegraphics[width=8cm]{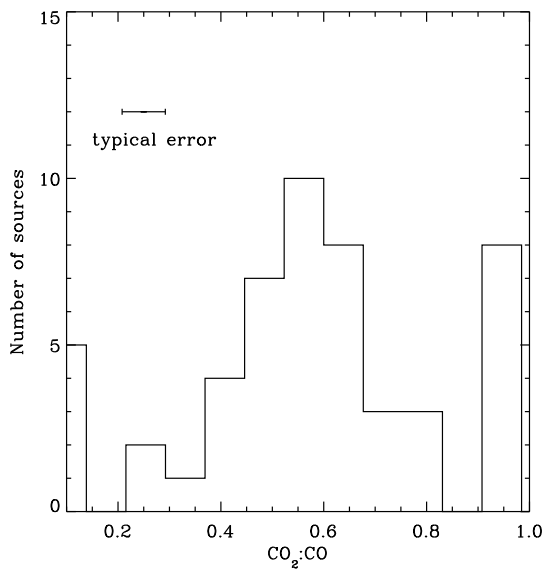}
  \caption{Distribution of CO$_2$:CO mixing ratios as determined from the CO$_2$ blue component profile fit. }
  \label{co_co2_histo}
\end{figure}

Conversely, Figure \ref{co_co2_histo} illustrates the indirect method of determining the concentration of CO$_2$ in CO from the 
profile of the ``blue'' CO$_2$ component. The figure shows the distribution of CO$_2$:CO mixing ratios as determined
by the component fit. It is seen that the mixing ratios are remarkably similar for most of the observed spectra, which
is also indicated by the direct correlation seen in Figure \ref{co_co2_1_1}. However, the median concentration determined using this
method is $N({\rm blue\,CO}_2)/N({\rm blue\,CO})=0.55\pm 0.2$, or a factor of 2-5 smaller than that determined using the directly measured column densities.

It is probably reasonable to assume that the direct method provides a better estimate of the concentration since the indirect method relies 
on an uncertain calibration of a set of laboratory experiments. However, it should be noted that the band strengths of both CO$_2$ and CO may depend on
concentration, which is an effect that is ignored here by assuming that the band strengths are constant. Variable band strengths may affect the direct method. 
Clearly, well-calibrated laboratory experiments are needed to resolve the issue. For this study, the ``blue'' CO components, as determined by
the profile of the blue CO$_2$ band, are divided by a factor 3 to 
better match the CO bands, as dictated by the direct concentration measurement. Note that \cite{Pontoppidan03} in their study of the CO stretching vibration
band at 4.67\,$\mu$m found that the available laboratory spectra of CO$_2$:CO mixtures were 
generally much too wide to reproduce the blue wing of the CO bands, as confirmed by \cite{Broekhuizen06}, which
led them to consider alternatives to explain the ``blue'' component of the CO band. Here it is concluded, based on the clear correlation between the
blue CO$_2$ and CO components, that they indeed represent the same component, but that since their detailed profiles do not 
fit well to the laboratory analogs, the structure of this component is not yet fully understood. 

\subsubsection{The ``dilute'' component}
\label{dilute}

\begin{figure}
    
  \includegraphics[width=8cm]{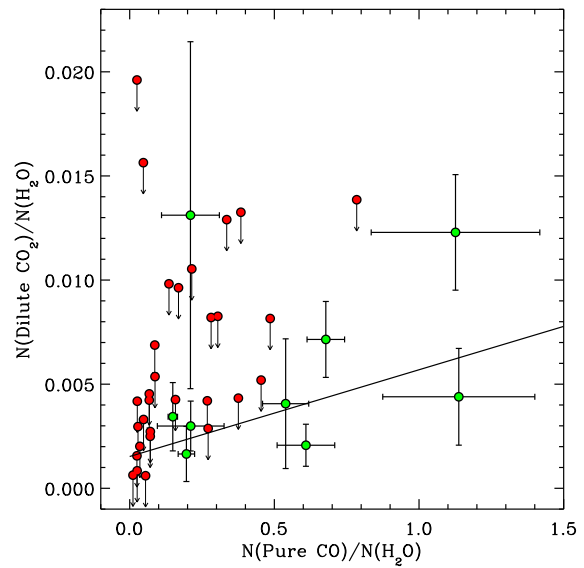}
  \caption{Relation between the dilute CO$_2$ component abundance and the middle (pure) CO component abundance from \cite{Pontoppidan03}. The green points are detections, while
the red points indicate upper limits on the dilute CO$_2$ component. The line is a linear fit to the detections.}
  \label{co_co2_dilu}
\end{figure}

Some CO$_2$ bending mode profiles exhibit a very narrow single peak at 15.15\,$\mu$m (660\,cm$^{-1}$), as indicated in Figure \ref{co2_components_sketch}. 
The most obvious sources with this property are IRS 51 and CRBR 2422.8-3423, as well
as the background star CK 2. The profile of this component corresponds 
closely to that of a CO$_2$ ice very dilute in CO with a concentration less than 10\%; the profile of the band does not change appreciably at lower
concentrations. The ``dilute'' component typically appears toward sources that also have very large column densities of ``pure'' CO (the ``middle'' component of \cite{Pontoppidan03}). The
relation between these two components is shown in Figure \ref{co_co2_dilu}, where it is seen that typical concentrations of CO$_2$ in the CO is 1:100-250.
This indicates that there are vastly different mixing ratios of CO$_2$ to CO along each line of sight, possibly even on each grain.

\subsection{Relation with CH$_3$OH}

\begin{figure}
    
  \includegraphics[width=8cm]{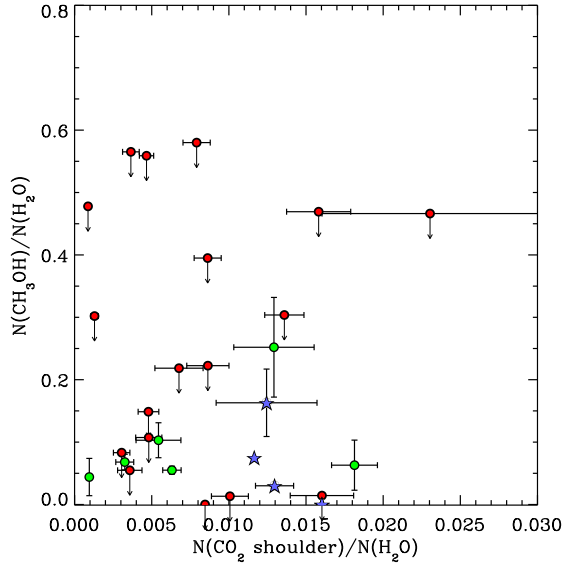}
  \caption{Relation between the abundance of the CO$_2$ ``shoulder'' component and the CH$_3$OH abundance from Paper I. }
  \label{CH3OH}
\end{figure}

Because CH$_3$OH has been related to the shoulder on the red side of the CO$_2$ bending mode \citep{Dartois99}, it is natural to estimate whether
there is a connection with the direct measurement of CH$_3$OH abundances from Paper I. The relation, shown in Figure \ref{CH3OH}, does
not exhibit an obvious relationship between the CO$_2$ shoulder and the CH$_3$OH abundance. 
This does not necessarily indicate that the shoulder is not related to interactions
with CH$_3$OH if the concentration of CO$_2$ in the CH$_3$OH varies significantly. Keeping the assumption that the band strength
of the CO$_2$ shoulder is that of pure CO$_2$, the abundance CO$_2$:CH$_3$OH varies between 1:20 and 1:3. It is therefore likely
that the CO$_2$ is highly dilute in the CH$_3$OH ice. 

\subsection{Upper limits on C$_3$O$_2$}
\label{C3O2}
\cite{Jamieson06} found that carbon suboxide (C$_3$O$_2$) is formed in abundance along with CO$_2$ during electron irradiation of a CO ice. It
has prominent bands around 17-19\,$\mu$m \citep{GerakinesMoore01}, but with exact central positions that vary considerably in the literature, presumably as a result
of different ice mixtures. Some of the IRS
spectra do have weak structure in the general wavelength region, but nothing that resembles a consistently recurring absorption band at a single
wavelength. It is therefore concluded that there is no obvious evidence for absorption from C$_3$O$_2$ at the 5\% level. 

\subsection{The 17\,$\mu$m feature}
Some spectra show a clear excess absorption feature centered on 17\,$\mu$m, most notably toward IRS 42, EC90, VSSG 17 and ISO ChaII 54. The origin of this feature is
unclear, and only its observed properties are reported here. It appears that there is some relation between the presence of the feature
and sources that show contamination by silicate emission from the central disk. Thus the feature may not be a real absorption signature, but emission from
the 18\,$\mu$m silicate band affecting the continuum determination. However, if real, the feature has a typical width of 35\,$\rm cm^{-1}$, and optical depths
of ~0.1, where detected. 

\section{Origin and evolution of the CO$_2$ components}

\subsection{Formation routes to CO$_2$ in the interstellar medium}

CO$_2$ ice in protostellar envelopes as observed here is widely believed to be formed 
by surface reactions as opposed to a gas-phase route followed by freeze-out \citep{Tielens82}. Pure gas-phase chemical
models of typical cold, dark clouds predict CO$_2$ abundances of 10$^{-9}$ relative to H$_2$ \citep{Bergin95}, making it
unlikely that there is any contribution from directly from gas-phase chemistry to the
observed solid state CO$_2$. There is strong observational
evidence that CO$_2$ forms readily in cold quiescent molecular clouds, and it does not appear that ``extra'' photo-processing of the ice is required, beyond
what can be explained by cosmic ray induced UV photons and photons from the interstellar radiation field managing
to penetrate to $A_V$s of 3-5 magnitudes \citep{Whittet98}.
In fact, the relatively constant fraction of CO$_2$ relative to water ice (15-40\%) under a wide range of conditions suggests that it forms under
``common'' quiescent conditions of densities of $10^4-10^5\,\rm cm^{-3}$, temperatures near $10\,$K and a standard cosmic ray field. 
Thus far, no line of sight has been observed to have CO$_2$ abundances of $N_{\rm CO_2}/N_{\rm H_2O}\lesssim 0.15$. 
However, the exact chemical pathway remains controversial.  Possible routes to the formation of CO$_2$ are via the reactions:
\begin{equation}
\rm{CO}+\rm{OH}\rightarrow CO_2+H 
\label{OHroute}
\end{equation}
\begin{equation}
\rm{O}+\rm{HCO}\rightarrow CO_2+H 
\label{HCOroute}
\end{equation}
\begin{equation}
\rm{CO}+\rm{O}\rightarrow CO_2 
\label{Oroute}
\end{equation}

Route \ref{Oroute} is often included as an important grain surface reaction \citep{Tielens82,Stantcheva04}, but
has also been found in at least one study to possess a prohibitively large barrier \citep{Grim86}. However, 
a similar experiment by \cite{Roser01} finds that the reaction proceeds with no or little barrier. It should
be noted that it is expected that the barrier to route \ref{Oroute} is sensitive to 
the electronic state of the oxygen atoms, such that $O(^1 S)$ may react much easier with CO than oxygen
in the ground state $O(^3 P)$ \citep{Fournier79}. The energy difference between these two states correspond to a red photon (6300\,\AA),
which will penetrate much deeper into dark clouds than the UV photons normally considered for photolysis reactions.  

In some grain surface models, route \ref{OHroute} is used as a dominant reaction \citep{Chang07}.

It is also known that CO$_2$ can form with a low or non-existing activation barrier through an electronically excited state of CO:

\begin{equation}
\rm CO^*+CO\rightarrow CO_2 + C
\label{COroute}
\end{equation}

Obviously, this reaction requires that the CO molecule is excited, and was studied extensively in the context of 
UV photolysis \citep[e.g.][]{Gerakines96,Loeffler05}. However, \cite{Oberg07_letter} finds {\it no} formation of CO$_2$ from CO 
in a ultra-high vacuum (UHV) UV irradiation experiment, 
which is much less contaminated by H$_2$O than the previous high vacuum experiments. While this argues against route \ref{COroute}
as an effective pathway to CO$_2$ ice, electron irradiation does produce CO$_2$ from pure CO under UHV conditions \citep{Jamieson06}. 

Consequently, the rates of most solid-state reactions leading to CO$_2$ are still controversial, and theoretical models have struggled to consistently reproduce the observed abundances. 
Based on existing observations, any model for the formation of CO$_2$ should be required to reproduce both the absolute abundance of CO$_2$ of
15-40\% relative to water ice, {\it as well as the apparent universality} of this abundance. Additionally, the separate molecular environments
of the CO$_2$ ice should also be explained, in particular the presence of CO$_2$ in both H$_2$O-rich and CO-rich environments.

It is worth mentioning that CO accreted from the gas-phase is not the only potential source for the carbon in CO$_2$. \cite{Mennella04} found that
the carbon in hydrogenated carbon grains will form both CO and CO$_2$ when covered with a water ice mantle and subjected to cosmic rays. This
is potentially a route to forming CO$_2$ embedded in a water ice matrix. However, having a different source of carbon for CO$_2$ formation
than gas-phase CO must explain why the $^{12}$C/$^{13}$C ratios of gas-phase CO, solid state CO and solid state CO$_2$ are all so similar to the Solar value of 89 
\citep[50-100][]{Boogert00, Boogert02, Pontoppidan03}. 
In contrast, pre-solar carbonaceous grains have highly variable $^{12}$C/$^{13}$C ratios with a tendency toward
ratios a few to 100 times that of the Sun, although some have ratios as low as $\sim$1 \cite[e.g.][]{Lin02,Croat05}. 
The question is whether such scatter is reflected in the ice isotopologue ratios if the carbonaceous grains are the source of carbon.

In the following, it is explored how the data presented here can help constrain the formation and evolution of CO$_2$ in the interstellar medium 
and in protostellar envelopes in particular. 

\subsection{Formation of the CO$_2$:CO system}
Having established its existence, how can the presence of large amounts of CO$_2$ within the CO-dominated mantle be explained?
Is there a connection
with the freeze-out of CO at densities higher than 10$^5\,\rm cm^{-3}$?  
There is evidence from laboratory experiments that CO and CO$_2$ do not mix upon warmup of separately deposited layers \citep{Broekhuizen06}, 
and the possibility that the CO$_2$ is formed directly as part of the CO mantle using the carbon of the CO accreted 
from the gas-phase is therefore explored. While a thick mantle of CO ice certainly offers
a significant reservoir of carbon and oxygen for the formation of CO$_2$, the breaking of the CO triple bond to form CO$_2$ directly from CO requires a significant
energy input. The most direct way of providing a high input of energy in
dense molecular clouds is via cosmic rays. The cosmic rays can hit the grains directly, but are also the
dominant source of UV photons through their interaction with hydrogen molecules. 
A number of laboratory experiments have been performed simulating and 
comparing the formation of CO$_2$ from a pure CO ice through UV and cosmic ray irradiation \citep{Gerakines01,Loeffler05,Jamieson06}. 
\cite{Jamieson06} showed that CO$_2$ can be formed from a pure CO ice layer during irradiation with a 5 keV electron beam simulating
heavier and more energetic cosmic rays. The experiment converted 0.49\% of the CO to CO$_2$ with a deposited energy of 
$5.8\times 10^{16}\,\rm MeV\,cm^{-3}$, or 3.4\,$\rm eV\,molecule^{-1}$, assuming a CO ice density of 0.8\,$\rm g\,cm^{-3}$.

This value can be put into the context of dense molecular clouds by estimating the time scale for depositing the same amount of 
energy to a CO ice mantle with a standard cosmic ray field. Following the approach of \cite{Shen04}, the total deposited energy by the 
cosmic ray field per CO molecule per second is:

\begin{equation}
  E_{\rm dep}(CR) = \frac{m_{\rm CO}}{\rho_{\rm CO}} \sum_Z 4\pi \int_{E_{\rm min}(Z)}^{E_{\rm max}} \frac{{\rm d}Q}{{\rm d}s}\frac{{\rm d}n}{{\rm d}E} f_Z{\rm d}E,
\end{equation}
where $m_{\rm CO}$ is the mass of a CO molecule and $\rho_{\rm CO}$ is the density of CO ice. ${\rm d}Q/{\rm d}s$ is
the energy loss as the cosmic ray traverses a length, d$s$, through the CO ice and ${\rm d}n/{\rm d}E$ is the cosmic ray flux spectrum, and finally, 
$f_Z$ is the fraction of the energy deposited that actually remains with the grain the rest being ejected in highly energetic
electrons. \cite{Leger85} estimate that $f_Z\sim 0.6$.  
The minimum cosmic ray energy, $E_{\rm min}$, is determined by the various factors that may drain energy from the particles. These include
interactions with dust grains, with the molecular gas itself, as well as drag from magnetic fields. The importance of these effects
depend sensitively on $Z$ as well as the cosmic ray energy. \cite{Leger85} showed that interactions with the gas only affects the cosmic ray spectrum
at $A_V > 50\,$mag. However, dust can effectively stop low energy ($< 100\,$MeV for iron) cosmic rays at $A_V<10\,$mag, and that determines 
$E_{\rm min}$. 

For these assumptions, the energy deposited by cosmic rays in a CO ice mantle is 
$8\times 10^{-14}\,\rm eV\,molecule^{-1}\,s^{-1}$, which means the \cite{Jamieson06} experiment
corresponds to roughly 350\,yrs of cosmic ray irradiation or, assuming a constant rate of CO$_2$ formation, 
$\sim 3.5\times 10^4\,$yrs to convert 50\% of the CO mantle to CO$_2$. 
Although the input cosmic ray flux spectrum is very uncertain, especially at low energies, the conclusion is that it is plausible that
cosmic rays can provide the necessary energy input to form the observed CO$_2$:CO component. 

In this context, what is the implication of the presence of the ``dilute'' component discussed in Section \ref{dilute}? One possibility is
that the CO ice with a low concentration of CO$_2$ is younger than the CO$_2$:CO$\sim$1:1 component. This would happen if the conversion of 
CO to CO$_2$ occurs at a constant rate. 

\cite{Jamieson06} predicts the presence of a range of carbon oxide species in
addition to CO$_2$, of which the most abundant is C$_3$O$_2$. The quality of the Spitzer-IRS spectra allows a sensitive
search for this molecule in the solid state through its modes at $\sim 18.5\,\mu$m, yet it is not clearly
detected in any of the spectra presented here (see Section \ref{C3O2}).

\subsection{CO$_2$ as a temperature tracer}

CO$_2$ has been suggested to be a tracer of strong heating based on simulated annealing experiments in the laboratory to 100\,K \citep{Gerakines99}.
The proposed mechanism is that the CO$_2$ segregates out of the hydrogen-bonding mixture with water and possibly CH$_3$OH to produce inclusions
of pure CO$_2$. These inclusions in turn produce the characteristic double peak observed in many high mass YSOs. While
the very high temperatures required for the segregation process in a laboratory setting probably correspond to somewhat lower temperatures on
astronomical time scales, they are still well above the temperatures of 10--40\,K that dominate
the column densities through protostellar envelopes around low-mass stars. Thus, it seems surprising that many of the surveyed low-mass
stars show a double peak. The decomposition and visual inspection of the spectra reveals that a pure CO$_2$ component is clearly detected in
18 of the 48 low-mass stars, or almost 40\%.  

\begin{figure}
    
  \includegraphics[width=8cm]{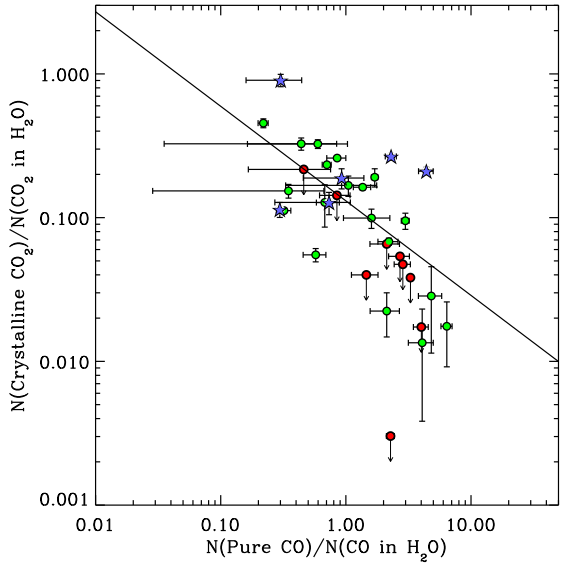}
  \caption{Relation between the CO and CO$_2$ temperature tracers. The CO$_2$ temperature tracer is the ratio of the column density in the double peaked, pure CO$_2$ component
relative to CO$_2$ in water. The CO temperature tracer is the amount of pure CO relative to CO trapped in water ice. Green points are low-mass YSOs, blue points are high-mass
YSOs, and red points are upper limits on the CO$_2$ temperature tracer. The line is a fit to the low-mass YSOs. }
  \label{CO2_temptracers}
\end{figure}

At least one other mechanism to produce a pure CO$_2$ ice component exists. The existence of a ubiquitous CO$_2$:CO component has been suggested before and
is strengthened by the sample presented here. This component may also produce pure CO$_2$ through distillation.
Upon warmup of a CO$_2$:CO mixture the CO will desorb leaving the CO$_2$ behind, but will do so at much
lower temperatures \citep[20-30\,K][]{Broekhuizen06}. 

One way of distinguishing this process with the formation of pure CO$_2$ via segregation is to calculate the 
fraction, $P$, of thermally processed icy material at temperatures above a certain critical temperature, $T_{\rm crit}$, along a given line of sight through a 
protostellar envelope. Clearly, $T_{\rm crit}$ is expected to be much lower for the
distillation process than for the segregation process. Assuming $T$ is a monotonically decreasing function of radius, $P$ is given by:

\begin{equation}
P=\frac{ \int_{R_{\rm crit}}^{R_{\rm sub}}n_{\rm CO_2}(r)dr }{ \int_{\infty}^{R_{\rm sub}}n_{\rm CO_2}(r)dr },
\end{equation}
where $R_{\rm crit}$ is the radius where $T=T_{\rm crit}$ and $R_{\rm sub}$ is the radius where the processed CO$_2$ ice sublimates. $n_{\rm CO_2}$ 
is the density of the CO$_2$ component that is transformed to pure, crystalline CO$_2$ upon heating to $T_{\rm crit}$.
The use of a critical temperature assumes that the process forming pure CO$_2$ is a thermal 
process governed by some activation energy and described by an Arrhenius relation. 

The first step is to estimate the value of $T_{\rm crit}$ for the segregation process in water-rich ice. 
Because of the long time scales and low pressures in the interstellar medium compared to the short time
time scales and high pressures of a laboratory experiment, it is not appropriate to apply the critical temperatures from the laboratory directly to
an astrophysical problem. For a process not dependent on pressure, such as the segregation of CO$_2$ from a water matrix, 
the critical temperatures in the two settings are related via:

\begin{equation}
\frac{\tau_{\rm astro}}{\tau_{\rm lab}} = \exp\left[ E_{\rm a}\left(\frac{1}{T_{\rm astro}}-\frac{1}{T_{\rm lab}}\right)\right],
\end{equation}
where $E_a$ is the activation energy in Kelvin, while $\tau_{\rm astro}$ and $\tau_{\rm lab}$ are
the e-folding time scales of a given process in the interstellar medium and in the laboratory, respectively. 
Because $E_a$ is unknown, measurements of the laboratory time scale at two different temperatures are
required. 

Unfortunately, the kinetics of the segregation process are not well known.
From the experiment of \cite{Ehrenfreund99}, a rough estimate can be made of $\tau_{\rm lab}\sim 1\,$hr at 100\,K and $\tau_{\rm lab}\sim 1\,$min at
120\,K. However, they use a tertiary CO$_2$:H$_2$O:CH$_3$OH=1:1:1 mixture, which has a concentration of methanol much higher than that found in 
typical low-mass protostellar envelopes. Conversely, recent experiments with a binary CO$_2$:H$_2$O=1:4 mixture by \cite{Oberg07} show that 
the CO$_2$ bending mode double peak has formed already at 75\,K on laboratory time scales of hours. 
Using a time scale of $10^5\,$yr, a typical time for the collapse front to reach the outer boundary of the protostellar core, 
the Ehrenfreund et al. values give $E_a\sim 4900\,$K and $T_{\rm crit}\equiv T_{\rm astro}\sim 70\,$K. \cite{Boogert00} find
$E_a=4900$ and $T_{\rm crit}=77\,$K with similar assumptions.
If it is instead assumed that $T_{\rm lab}=75\,$K for an e-folding time scale of 1 hour, as indicated by \cite{Oberg07}, but 
the activation energy of 4900\,K is retained, $T_{\rm crit}\sim 57\,$K.
It is stressed that these values for $T_{\rm crit}$ are educated guesses at best, and that quantitative kinetic laboratory
experiments are needed to measure the actual value. In conclusion, $T_{\rm crit}$ is taken in the range 60 to 80\,K. 
 
It is also important to note that the time a dust grain can be expected to spend at temperatures between, say, 70 and 90\,K is much less than 
$10^5$ years. In the simplest physical 1-dimensional model of an infalling envelope \citep{Shu77}, a dust grain at the radii corresponding to such temperatures 
will be in free fall. The time scale for it passing through this region for a typical 1\,$M_{\odot}$ young star is 75 years, which in turn will increase $T_{\rm crit}$
for segregation to 78\,K for the CH$_3$OH-rich mixture. Using a 2-dimensional infall model that takes rotation into account will likely increase the infall time scale somewhat.
The confidence of the value of $T_{\rm crit}$ can obviously be improved significantly with a quantitative laboratory simulation coupled with a more detailed infall model.
 
A value for the desorption temperature, $T_{\rm sub}$, of the pure, crystalline CO$_2$ component formed by the segregation process is also needed. It is reasonable
to expect the segregated CO$_2$ to be in the form of inclusions embedded in the water ice. The question is whether
the CO$_2$ is trapped in the water or will be able to escape at temperatures lower than the interstellar water ice desorption temperature of 110\,K \citep{Fraser01}.
While the Ehrenfreund experiments retain the CO$_2$ inclusions until the water ice desorbs at 150\,K, the {\"O}berg et al. experiments find that bulk of the
CO$_2$ ice desorbs at temperatures much lower than the water ice. This is consistent with the result of \cite{Collings04}, who classified CO$_2$
as a molecule that is not easily trapped in a water ice matrix. The fact that the tertiary mixture Ehrenfreund experiment appears to retain the CO$_2$
to higher temperatures than the binary H$_2$O mixtures may be related to the CH$_3$OH changing the trapping properties of the matrix. 
along with  In any case, to match the different experiments, an interstellar CO$_2$ desorption temperature of 110\,K is assumed for the Ehrenfreund experiment, 
60\,K for the {\"O}berg experiment and 50\,K for the pure CO$_2$ layer produced by the distillation process of CO$_2$:CO. The physical interpretation is
that the methanol-rich ice traps the CO$_2$ ice until the water desorbs. The water-rich binary mixture only traps CO$_2$ to temperatures
slightly higher than the desorption temperature of pure CO$_2$ in accordance with \cite{Collings04}, while CO$_2$:CO mixture obviously does not trap CO$_2$
at all. 

The next step is to choose a radial temperature and density structure that can be used to calculate $P$. While $P$ is
sensitive to a range of structural parameters for the envelope, the dominant one is the luminosity of the central source. It is beyond the scope of this paper to 
explore the parameter space of the structures of protostellar envelopes, but it is instructive to construct an example. For simplicity a static, one-dimensional 
power-law envelope ($\rho(R)\propto (R/R_0)^{-1.5}$) with A$_V$ of 50 mag and $R_0=100-300\,$AU is assumed. Furthermore, it is assumed that the envelope is empty within $R_0$. 
This is actually an envelope structure that favors a large $P$. More evolved envelopes dominated by infalling material to large radii will have a shallower 
density profile and thus more of the line of sight column density at larger radii. In this sense, the model $P$ curves are upper limits.
The dust temperature is calculated using the Monte Carlo code RADMC \cite{Dullemond04} coupled with a dust opacity constructed to
fit the extinction curve as measured using the c2d photometric catalogues (Pontoppidan et al., in prep.). The resulting dust temperature at 100\,AU varies from
40 to 1200\,K for source luminosities of 0.1 to $10^5\,L_{\odot}$.

\begin{figure*}
    
  \includegraphics[width=8cm,angle=90]{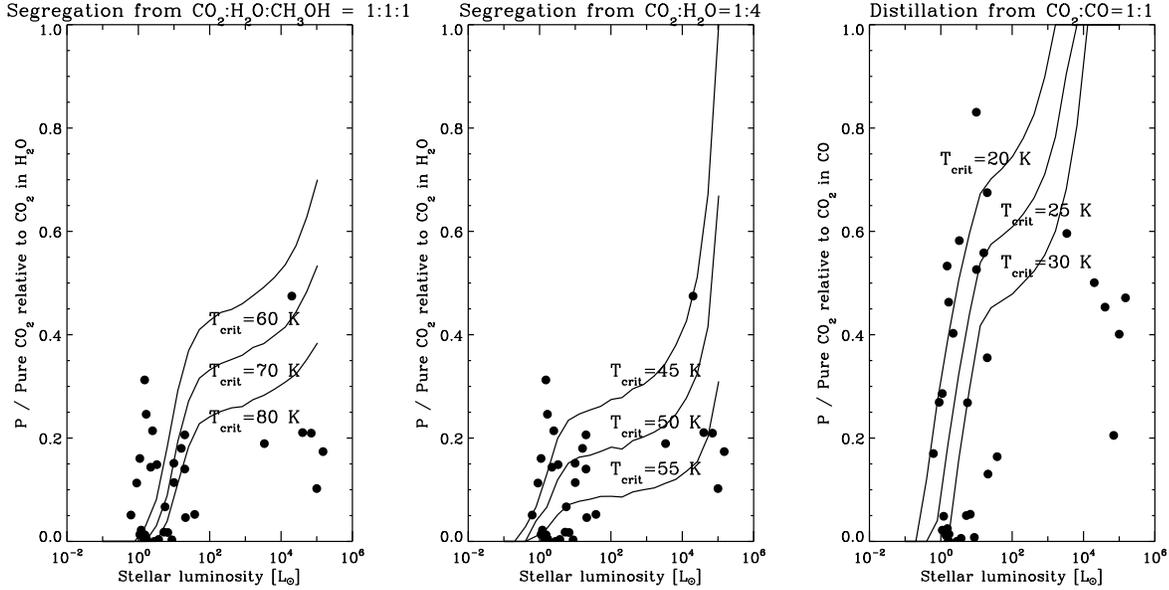}
  \caption{Fraction of pure CO$_2$ compared to models for different critical temperatures relevant for segregation (left) and CO desorption (right). 
In the models on the left and middle panels, an inner edge at 100\,AU is used, while the models on the right panel fit better with an inner edge at 300\,AU. The
dots are the observed values for the subset of our sample that have measured luminosities from \cite{Berrilli89,Ladd93,Chen95,Saraceno96,Bontemps01,Larsson00,Kaas04,diskshadow}.}
  \label{Pfracs}
\end{figure*}

Figure \ref{Pfracs} shows the observed values of $P$ as a function of source luminosity compared to model curves for different values of $T_{\rm crit}$ under the assumptions that the 
pure CO$_2$ originates either in the CO$_2$:H$_2$O component or in the CO:CO$_2$ component through the processes discussed above. The $P$-curves for
both the Ehrenfreund et al. and {\"O}berg et al. experiments are shown, as well as the curves expected for the formation of pure CO$_2$ through distillation of the CO$_2$:CO component.

First, it is noted that $P$ is not necessarily a monotonic increasing function with luminosity. This is because the evaporation of the pure CO$_2$ ice component
in the innermost regions of the envelope where $T>T_{\rm sub}$ competes with the formation of pure CO$_2$ at temperatures $T_{\rm crit}<T<T_{\rm sub}$. 
This is seen in Figure \ref{Pfracs} as a turnover in the models as the luminosity increases. For higher power law indices of the envelope, the $P$-curves may
even decrease with increasing luminosity.

Comparing the data points with the model curves, the results are as follows: Assuming a complete transformation of CO$_2$ mixed with water
ice to pure CO$_2$ inclusions, the observed points are consistent with a critical temperature for this process of 50-70\,K, depending on
whether the Ehrenfreund et al. or the {\"O}berg et al. experiments are considered.
Conversely, assuming
a conversion of CO$_2$ mixed with CO to pure CO$_2$, a critical temperature of at most 25\,K explains the highest observed $P$ values toward
low-mass stars. 
The measured values of $P$ are
very sensitive to the presence of cold foreground clouds contributing an unrelated column density, as well as the detailed structure of the inner envelope, in
particular the arbitrary location of an inner edge at 100-300\,AU. This has important consequences for the use of the
the CO$_2$ bending mode as an astrophysical tracer. For instance, the models show that the double peak should have a 
roughly constant relative strength for sources with luminosities between a few and at least $10^3\,L_{\odot}$. Therefore, 
if a source within that luminosity range shows no sign of a double peak, it is an indication of the presence of a significant
contribution to the extinction from foreground material, unrelated to the protostellar envelope. 

In conclusion, the splitting of the CO$_2$ bending mode toward low-mass protostars can be explained by segregation in strongly heated water-rich ices as 
described in \cite{Gerakines99} only for protostellar envelopes with steep density density profiles extending all the way to 100 AU from the central star, 
and for concentrations of CH$_3$OH much higher than the observed abundances.
The experiments of \cite{Oberg07} may also explain the data for such envelopes, but $T_{\rm crit}$ is not well-defined for this experiment.

\begin{figure*}[p]
    
  \includegraphics[width=13cm,angle=0]{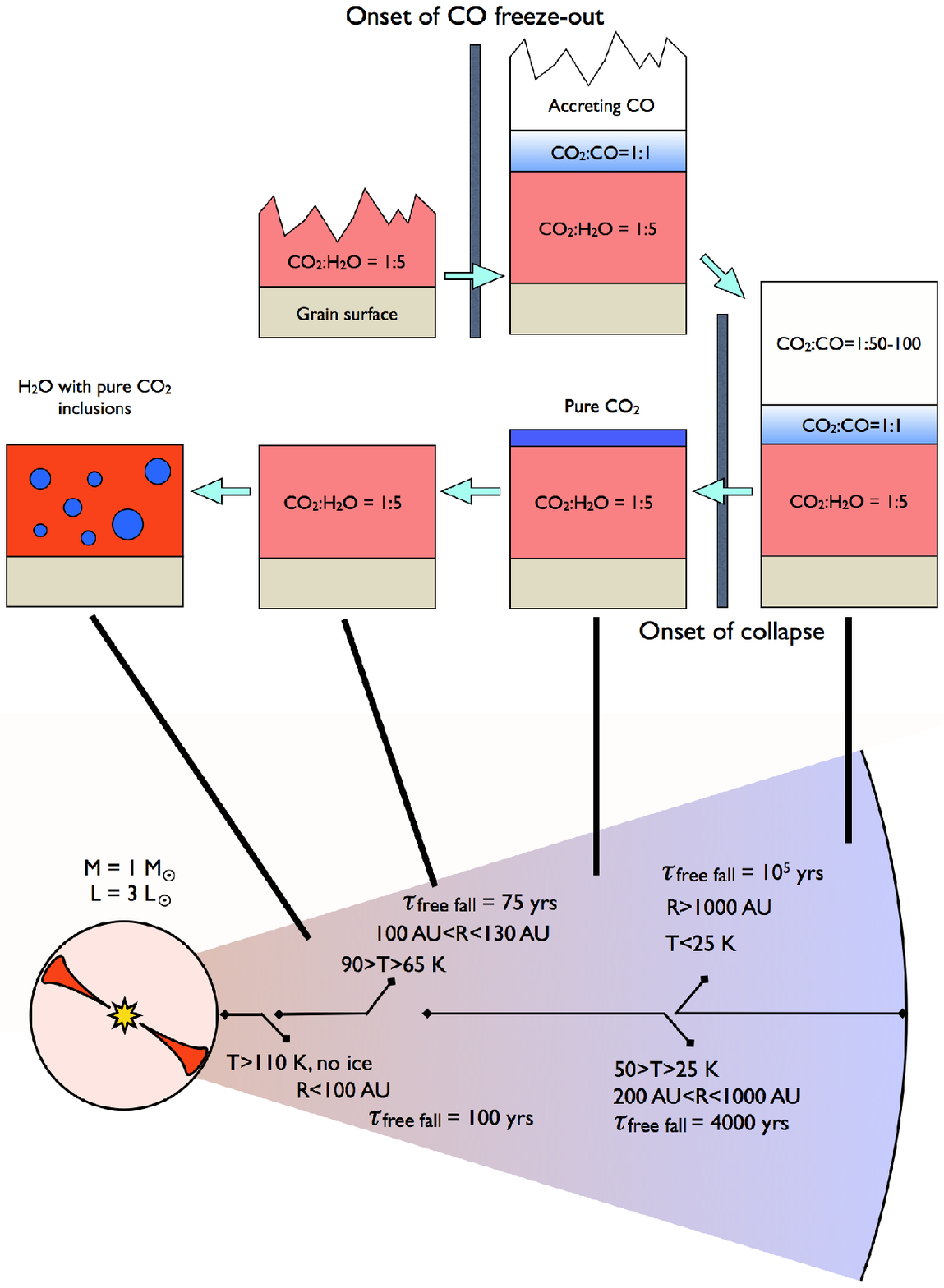}
  \caption{Sketch of the CO$_2$ ice structure in a typical low-mass protostellar envelope. The upper section of the sketch shows the
suggested evolution of the CO:CO$_2$:H$_2$O system on a single dust grain from the formation of the protostellar core until the
grain is accreted on a proto-planetary disk surrounding the central star. The lower panel indicates where in a 
collapsing envelope the various stages of ice mantle evolution can be found.}
  \label{env_sketch}
\end{figure*}

The resulting distribution of CO$_2$ ice in the different environments
is sketched in Figure \ref{env_sketch}.

\section{Discussion and conclusion}

In this paper, a picture is presented in which the dominant ice components, H$_2$O, CO$_2$, CO and, in a limited number of cases, CH$_3$OH appear to constitute an
intimately connected system; the observational characteristics of each species are directly affected by the presence of the others. 
Other molecules act as trace constituents with abundances that are too low to significantly
affect the profiles of the dominant species, i.e. other species did not have to be invoked to explain the observed band profiles.
This is in contrast to the very complex 5-8\,$\mu$m region discussed in Paper I.  

The survey has shown that the CO$_2$ bending mode profiles toward low-luminosity ($0.1<L<100\,\rm L_{\odot}$), solar-type YSOs
do not differ strongly in their basic structure from those observed toward massive, luminous YSOs ($\rm L> 1000\,L_{\odot}$). 
All protostellar envelopes as well as quiescent molecular clouds probed by lines of sight toward background stars are dominated
by CO$_2$ mixed with water with abundances of 20\% relative to water ice, but with a significant additional contribution of 10\% of CO$_2$ mixed with CO. The other components needed
to explain the CO$_2$ bending mode are present with much smaller abundances, or a few \% each with respect to water ice. 

While the CO$_2$:H$_2$O component might form from CO and OH via reaction (\ref{OHroute}), 
it is probable that, because of the lack of residual atomic hydrogen, the CO$_2$:CO component forms through a different chemical route. 
Additionally, since the band profiles indicate that the CO$_2$ in this component can have concentrations of anything from 2:1 to less than 1:100 relative to CO, 
a variable effect highly sensitive to either the environment or age of the ice must play a role. Based on observations of a sample of background stars, 
\cite{Bergin05} suggest that the formation of the CO$_2$:CO component is related to
the abundance of atomic oxygen relative to atomic hydrogen in the gas-phase and predict that it forms in low-density regions of the cloud.
Observationally, this would produce CO and CO$_2$ profiles dominated by the CO$_2$:CO$\sim$1:1 component in low-density regions. It would 
also result in the CO$_2$:CO layer being placed below the water-rich layer on each grain. In this case, the formation of pure CO$_2$ by the desorption
of CO as proposed here would not work.  

A different scenario is suggested in which the CO$_2$:CO component is connected to the rapid freeze-out of pure CO at higher densities. 
This scenario is practically independent on the gas-phase chemistry, but requires a mechanism for forming CO$_2$ directly from CO.
Such a mechanism will most likely involve a highly energetic input from cosmic rays; UV irradiation with $\lambda > 1200\,\AA$ will not work in the absence
of water \citep{Oberg07_letter}. While the energy input from a standard cosmic ray field is likely sufficient, this process would also
tend to form more complex carbon oxides in the ice. While these are are not detected in our Spitzer spectra, their absence is
not strongly constraining, given that the molecular properties of their bands, such as strengths, are not well known. 

It is confirmed that the CO$_2$ ice profile is an excellent ice temperature tracer. Comparison with the stretching vibration band of solid CO shows that the
CO$_2$ correlates well with the pure CO to CO:H$_2$O ratio, an established ice temperature tracer. 
The observed differences between CO$_2$ bending mode profiles are consistent with differences in the fraction of the sight lines that have been 
heated above a certain threshold temperature, $T_{\rm crit}$, to produce a double-peaked profile. 
More luminous YSOs have generally had a larger fraction of their ice column densities at temperatures above $T_{\rm crit}$, but the prevalence of the characteristic double-peak shows that 
even the low-luminosity YSOs have thermally processed inner envelopes. 

There are several possibilities for the value of $T_{\rm crit}$. 
 For the massive YSOs, strong annealing of methanol-rich ices to temperatures
higher than 100\,K in a laboratory setting was invoked to 
explain the observed abundance of pure CO$_2$ ice by \cite{Gerakines99}. It is suggested that this corresponds to $T_{\rm crit}=50-80\,$K under
the conditions in collapsing protostellar envelopes. These are high temperatures that puts restrictions on the density structure of
the protostellar envelopes in order to reproduce the observed abundances of pure CO$_2$.
Laboratory experiments measuring the kinetics of the annealing process in methanol-poor CO$_2$:H$_2$O ice mixtures will be needed to use this process in quantitative 
modeling of ice mantle processing. 

It is therefore argued that the presence of a significant fraction of pure CO$_2$
toward so many of the low-mass YSOs in the survey indicates that another, lower temperature process may play a role. Our survey, as well as others, measure
a significant fraction of the CO$_2$ ice mixed with CO, rather than with water. This mantle component will form a layer of pure CO$_2$ upon very moderate heating
to $T_{\rm crit}=$20-30\,K as the CO desorbs, leaving the CO$_2$ behind. This process has therefore the potential to create pure CO$_2$ by {\it distillation} rather than
{\it segregation}. Note that the new observations presented here do not rule out that the segregation process is responsible for a significant fraction of the pure
CO$_2$ in the sample of {\it massive} YSOs originally discussed in \cite{Gerakines99}, in particular since less CO will be frozen out in the 
warm envelopes surrounding massive YSOs. The simple protostellar models presented here seem to indicate that both mechanisms for 
producing CO$_2$ in general contribute to the double-peak in low-mass young stellar objects.

\acknowledgements{Support for KMP was provided by NASA through Hubble Fellowship grant 1201.01
awarded by the Space Telescope Science Institute, which is operated by the Association of Universities 
for Research in Astronomy, Inc., for NASA, under contract NAS 5-26555.
Astrochemistry in Leiden is supported by a SPINOZA grant of the Netherlands 
Organization for Scientific Research (NWO) Support for this work, part of the Spitzer Space 
Telescope Legacy Science Program, was provided by NASA through Contract Numbers 1224608 and 
1230779 issued by the Jet Propulsion Laboratory, California Institute of Technology under NASA contract 1407.
Some of the data presented herein were obtained at the W.M. Keck Observatory, which is operated as a 
scientific partnership among the California Institute of Technology, 
the University of California and the National Aeronautics and Space Administration. 
The Observatory was made possible by the generous financial support of the W.M. Keck Foundation. 
The authors wish to recognize and acknowledge the very significant cultural role and reverence 
that the summit of Mauna Kea has always had within the indigenous Hawaiian community.  
We are most fortunate to have the opportunity to conduct observations from this mountain.
}

\bibliographystyle{apj}
\bibliography{ms}

\end{document}